\documentclass{jpp}
\usepackage{graphicx,natbib}
\usepackage{epstopdf, epsfig}
\usepackage{amsfonts,amsmath,amssymb}
\usepackage[pdftex,colorlinks,citecolor=blue]{hyperref}
\usepackage{bm}
\usepackage{xcolor}
\newcommand{\mockalph}[1]{}

\newcommand\bb[1]{\mbox{\boldmath{$#1$}}}
\newcommand\grad{\bb{\nabla}}
\newcommand\bcdot{\,\bb{\cdot}\,}
\newcommand\btimes{\,\bb{\times}\,}
\newcommand{\D}[2]{\frac{{\rm d} #2}{{\rm d} #1}}
\newcommand{\pD}[2]{\frac{\partial #2}{\partial #1}}
\newcommand{\DD}[2]{\frac{{\rm d}^2 #2}{{\rm d} {#1}^2}}
\newcommand{\eb}{\hat{\bb{b}}}
\newcommand{\ez}{\hat{\bb{z}}}
\newcommand{\ex}{\hat{\bb{x}}}
\newcommand{\ey}{\hat{\bb{y}}}
\newcommand{\dbldot}{{\,\bb{:}\,}}
\newcommand{\erm}{{\rm e}}
\newcommand{\imag}{{\rm i}}

\ifCUPmtlplainloaded \else
  \checkfont{eurm10}
  \iffontfound
    \IfFileExists{upmath.sty}
      {\typeout{^^JFound AMS Euler Roman fonts on the system,
                   using the 'upmath' package.^^J}%
       \usepackage{upmath}}
      {\typeout{^^JFound AMS Euler Roman fonts on the system, but you
                   dont seem to have the}%
       \typeout{'upmath' package installed. JPP.cls can take advantage
                 of these fonts, if you use 'upmath' package.^^J}%
       \providecommand\upi{\upi}%
      }
  \else
    \providecommand\upi{\upi}%
  \fi
\fi

\ifCUPmtlplainloaded \else
  \checkfont{msam10}
  \iffontfound
    \IfFileExists{amssymb.sty}
      {\typeout{^^JFound AMS Symbol fonts on the system, using the
                'amssymb' package.^^J}%
       \usepackage{amssymb}%
         
       \let\ge=\geqslant  
      }{}
  \fi
\fi

\ifCUPmtlplainloaded \else
  \IfFileExists{amsbsy.sty}
    {\typeout{^^JFound the 'amsbsy' package on the system, using it.^^J}%
     \usepackage{amsbsy}}
    {\providecommand\boldsymbol[1]{\mbox{\boldmath $##1$}}}
\fi

\title[Kinetic Magnetorotational Instability]{Pressure-anisotropy-induced nonlinearities in the kinetic magnetorotational instability}

\author[J.~Squire and others]%
{J.~Squire$^{1,2}$
 \thanks{Email address for correspondence: jsquire@caltech.edu},
E.~Quataert$^{3}$ 
and M.~W.~Kunz$^{4,5}$}

\affiliation{$^1$TAPIR, Mailcode 350-17, California Institute of Technology, Pasadena, CA 91125, USA\\[\affilskip]
$^2$Walter Burke Institute for Theoretical Physics, Pasadena, CA 91125, USA\\[\affilskip]
$^3$Astronomy Department and Theoretical Astrophysics Center, University of California, Berkeley, CA 94720, USA\\[\affilskip]
$^4$Department of Astrophysical Sciences, Princeton University, Peyton Hall, Princeton, NJ 08544, USA\\[\affilskip]
$^5$ Princeton Plasma Physics Laboratory, PO Box 451, Princeton, NJ 08543, USA}

\pubyear{2017}
\volume{}
\pagerange{}
\date{?; revised ?; accepted ?. - To be entered by editorial office}
\begin{document}

\maketitle

\begin{abstract}
In collisionless and weakly collisional plasmas, such as hot accretion flows onto compact objects, the magnetorotational instability (MRI) can differ significantly from the standard (collisional) MRI. In particular, \emph{pressure anisotropy} with respect to the local magnetic-field direction can both change the linear MRI dispersion relation and cause nonlinear modifications to the mode structure and growth rate, even when the field and flow perturbations are very small. This work studies these pressure-anisotropy-induced nonlinearities in the weakly nonlinear, high-ion-beta regime, before the MRI saturates into strong turbulence. Our goal is to better understand how the saturation of the MRI in a low collisionality plasma  might differ from that in the collisional regime. We focus on two key effects: (i) the direct impact of self-induced pressure-anisotropy nonlinearities on the evolution of an MRI mode, and (ii) the influence of pressure anisotropy on the ``parasitic instabilities'' that are suspected to cause the mode to break up into turbulence. Our main conclusions are: (i) The mirror instability regulates the pressure anisotropy in such a way that the linear MRI in a collisionless plasma is an approximate nonlinear solution once the mode amplitude becomes larger than the background field (just as in MHD). This implies that differences between the collisionless and collisional MRI become unimportant at large amplitudes. (ii) The break up of large amplitude MRI modes into turbulence via parasitic instabilities is similar in collisionless and collisional plasmas. Together, these conclusions suggest that the route to magnetorotational turbulence in a collisionless plasma may well be similar to that in a collisional plasma, as suggested by recent kinetic simulations. As a supplement to these findings, we offer guidance for the design of future kinetic simulations of magnetorotational turbulence.
\end{abstract}

\begin{PACS}
\end{PACS}

%
%
\section{Introduction}

Across a wide variety of accreting astrophysical systems, the inflow of matter is thought to 
rely on turbulent angular momentum transport driven by the magnetorotational instability (MRI; \citealp{Balbus:1991fi,Balbus:1998tw}). 
The majority of works studying the  MRI, in particular its 
saturation into turbulence (e.g., \citealp{Hawley:1995gd,2001ApJ...554L..49H,2017ApJ...840....6R}),
 have been based on magnetohydrodynamics (MHD). They
thus implicitly assume that the collisional mean free path of gas particles is 
small in comparison to the scales of all fluid motions. However, this assumption
can be far from valid in many accreting systems. For instance, in  radiatively 
inefficient accretion flows onto supermassive black holes (RIAFs; see \citealp{1998tbha.conf..148N,2002ApJ...573..738H,2003ANS...324..435Q,2014ARA&A..52..529Y}), a large
portion of 
 the gravitational potential energy of the infalling gas
 is converted directly into thermal energy, suggesting ion temperatures $T_{i}\sim 10^{12}~{\rm K}$ with 
 corresponding ion collisional mean free paths that are orders of magnitude larger than the system size. 

As shown by  \citet{Quataert:2002fy} (hereafter \defcitealias{Quataert:2002fy}{Q02}\citetalias{Quataert:2002fy}), \citet{Sharma:2003hf}, and \citet{Balbus:2004hg}, 
the linear magnetorotational instability still exists in collisionless and weakly collisional plasmas. This kinetic MRI (KMRI) has seen subsequent theoretical attention. As well as extensions to the original linear analyses using either fully kinetic treatments \citep{Sharma:2003hf,Heinemann:2014cl,Quataert:2015dv} or fluid models \citep{2007ApJ...662..512F,Rosin:2012ej},
various works have explored turbulent transport in the fully nonlinear regime, generally
finding behaviour that bears a strong similarity to that seen in standard resistive MHD. \citet{Sharma:2006dh} (hereafter \defcitealias{Sharma:2006dh}{S06}\citetalias{Sharma:2006dh}) was the first 
to study MRI turbulence in this regime using a kinetically motivated fluid closure, an approach that 
has been followed in a variety of works since (e.g., \citealp{Sharma:2007cr,Chandra:2015bh,Foucart:2015gy}). Recently, it has become possible to study 
the MRI using truly kinetic particle-in-cell (PIC) methods, both in 2D \citep{Riquelme:2012kz,2013ApJ...773..118H,Kunz:2014hm} and 3D \citep{2015PhRvL.114f1101H,2016PhRvL.117w5101K}. Most notably, in \citet{2016PhRvL.117w5101K}, a fully collisionless plasma
was seen to develop into MRI  turbulence with strong similarities to that seen in comparable MHD calculations 
(or, even more so, similarities to the  model of \citetalias{Sharma:2006dh}), providing a fascinating example
of a fully collisionless plasma behaving as a collisional fluid. 

In this paper, we explore the regime between fully nonlinear turbulence  and
the linear KMRI. Our purpose is to move towards understanding the nonlinear saturation of 
the KMRI, in particular the similarities with, and differences to, the standard MRI.
Our philosophy is to examine the simplest (and, hopefully, the most significant)  modifications
to MHD. With this in mind, we consider both the kinetically motivated Landau-fluid (LF) model used by \citetalias{Sharma:2006dh} and  ``Braginskii'' MHD \citep{Braginskii:1965vl}, which is valid in the weakly collisional regime.
 We study two interlinked  effects, each of which  could
   have a strong influence on how the KMRI saturates into turbulence. The first effect is the \emph{pressure anisotropy} ($\Delta p$) driven by growing KMRI 
 modes. This can nonlinearly affect the modes' evolution (even in 1D)
 at amplitudes far smaller -- by a factor $\sim$$\beta$, the ratio of thermal to magnetic pressure -- than occurs in a standard compressible gas with isotropic pressure. It is important to understand the 
 effect of this nonlinearity, because it is these anisotropy-modified KMRI modes 
 that will be disrupted at large amplitudes and excite  turbulence. 
 The second effect that we study is the nonlinear disruption of KMRI modes by \emph{parasitic modes}, which are thought to govern the transition  into strong turbulence \citep{Goodman:1994dd,Pessah:2009uk,Latter:2009br,Latter:2010iz,Longaretti:2010ha}. MHD parasitic modes are Kelvin-Helmholtz 
 and tearing modes that feed off  strong gradients in the large-amplitude MRI ``channel'' mode\footnote{The azimuthally and radially constant
 MRI modes are often termed ``channel modes'' because they are able to 
 survive unmodified to very large amplitudes.}. 
A significant difference in parasitic-mode growth rates in a collisionless plasma (compared to MHD) would suggest that the saturation of the KMRI into turbulence  would also be significantly modified, perhaps with important implications for KMRI-driven turbulence.

Our main results are twofold. First,  once the KMRI channel mode amplitude $\delta B$ surpasses the strength of the background field $B_{0}$, its evolution always reverts to MHD-like behaviour. In particular, because of the pressure-anisotropy-limiting effects of kinetic microinstabilities \citep{Schekochihin:2008en,Kunz:2015gh} and the specific form of  MRI modes, the pressure anisotropy has very little effect 
on mode evolution once $\delta B\gtrsim B_{0}$. The MRI modes are then approximate nonlinear solutions of the Landau-fluid or Braginskii models until they reach very large amplitudes.
However, at moderate mode amplitudes $\delta B\lesssim B_{0}$, the effect of pressure anisotropy can be significant;
for example, it causes strong modifications to the KMRI in the presence of a background azimuthal field\footnote{This is sometimes termed the {magnetoviscous instability} (MVI), following \citet{Balbus:2004hg}.} 
 at amplitudes well below where it would saturate into turbulence. Our second result is that  there is \emph{not} a strong difference in parasitic-mode growth rates between the kinetic and MHD models, which indicates that modes can grow to similar amplitudes before being disrupted in collisionless and collisional systems.
Together these conclusions suggest that the saturation of MRI modes into turbulence in high-$\beta$ 
collisionless and weakly collisional regimes will be similar to what occurs in a collisional (MHD) plasma. This appears to be the case in the simulations that have been run up to now, including those that do not rely on fluid closure schemes  \citep{Riquelme:2012kz,2013ApJ...773..118H,2015PhRvL.114f1101H,2016PhRvL.117w5101K}.

In some ways, the results of this work will primarily be of interest for understanding and designing
future 3-D fully kinetic simulations of MRI turbulence. Such simulations are the only clear
method available to explore the collisionless accretion flows without \emph{ad hoc} assumptions, 
but are very demanding computationally. The primary difficulty arises from the enormous scale separation in RIAFs between the ion gyrofrequency $\Omega_{i}$ (which must be resolved in a kinetic code)
and the disk rotation frequency $\Omega$. Simulations are necessarily limited to  
modest values of $\Omega_{i}/\Omega$, and it is thus crucial to understand some of
the basic differences between the MRI and KMRI in designing and analyzing simulations, so
as to ensure that observed effects are not an artifact of limited scale separation. 
To add to these difficulties, our understanding of the processes governing even the simplest MHD MRI 
turbulence remains somewhat limited (e.g., see \citealt{Fromang:2007cy,Lesur:2007bh,2009A&A...507...19F,Simon:2012dq,Blackman:2012wd,Meheut:2015it,Squire:2016iq}). 

The remainder of the paper is organized as follows. In \S\ref{sec:equations etc.} we introduce 
the models used throughout our work and the numerical methods used to solve them (\S\ref{sub: numerics}). Because the effects that arise due to pressure anisotropy may be unfamiliar to many readers, 
we provide a brief account in \S\ref{sub: general comparison} of the primary differences between each model and MHD. One-dimensional nonlinearities arising due to self-generated 
pressure anisotropies are then treated in \S\ref{sec:1D}, starting with an overview of the linear 
physics, and then treating the pure-vertical-field KMRI (\S\ref{sec: 1D vertical}) and azimuthal-field 
KMRI (\S\ref{sec: 1D azim}) separately. An overview of these results is given in \S\ref{sub:1-D conclusions}. We then consider the evolution of parasitic modes
 in \S\ref{sec:3D}, starting with linear calculations on sinusoidal background 
profiles (\S\ref{sub:Parasitics}) and then showing fully nonlinear calculations using the {\sc Zeus} code used by  
\citetalias{Sharma:2006dh} (\S\ref{sub: zeus sims}). This section is deliberately kept brief, due to the null result 
that nonlinear saturation is not strongly affected by the pressure anisotropy. A discussion of kinetic effects neglected in our model is given in \S\ref{sub: extra kinetics}. 
We then combine our results with those from previous kinetics simulations of mirror and magnetorotational instabilities to provide guidance for the design of future simulations of KMRI turbulence (\S\ref{sec:implications}). Finally, we conclude with a summary in \S\ref{sec:discussion}. 

%
%
\section{Governing equations: the effects of pressure anisotropy}\label{sec:equations etc.}

Our philosophy throughout this 
work is to consider the simplest and most general modifications to  MRI evolution on the largest (MHD) scales due to kinetic physics. We anticipate (though cannot prove) that these
are the most important kinetic modifications to the MRI. We thus focus on
the development of a \emph{gyrotropic} pressure anisotropy -- i.e., a pressure
tensor that differs in the directions parallel ($p_{\parallel}$) and perpendicular ($p_{\perp}$) to the magnetic-field lines, but that is unchanged by rotations about the field line. 
This pressure anisotropy, $\Delta p \equiv p_{\perp}-p_{\parallel}$, causes an additional stress in the momentum equation, which 
can nonlinearly affect the MRI modes at much lower amplitudes than
occurs in standard MHD.
The gyrotropic approximation is generally valid when the magnetic field varies on spatial and temporal scales much larger than the ion gyroradius and inverse gyrofrequency. 

\subsection{Basic equations and closure models}\label{sec:eqns}

Our equations are obtained as follows. A small patch of an accretion disc, co-orbiting with a fiducial point $R_0$ in the mid-plane of the unperturbed disc at an angular velocity $\bm{\Omega} = \Omega \ez$, is represented in Cartesian coordinates with the $x$ and $y$ directions corresponding to the radial and azimuthal directions, respectively. Differential rotation is accounted for by including the Coriolis force and by imposing a background linear shear flow, $\bm{U}_0 = -S x \ey$, where $S \equiv -{\rm d}\Omega/{\rm d}\ln R_0 > 0$ is the shear frequency; Keplerian rotation yields $S = (3/2)\Omega$. The evolutionary equations for the first three moments 
of the plasma distribution function are then \citep{CGL:1956,Kulsrud:1980tm,Schekochihin:2010bv},
\begin{gather}
\D{t}{\rho} = -\rho \grad\bcdot\bm{u} ,\label{eq:KMHD rho}\\*
\rho \left( \D{t}{\bm{u}} + 2\Omega \ez\btimes\bm{u} - 2S\Omega x\ex \right) = -\grad \left( p_\perp + \frac{B^2}{8\upi} \right) + \grad\bcdot\left[ \eb\eb \left( \frac{B^2}{4\upi} + \Delta p \right) \right] ,\label{eq:KMHD u}\\*
\D{t}{\bm{B}} = \bm{B}\bcdot\grad\bm{u} - \bm{B}\grad\bcdot\bm{u} ,\label{eq:KMHD B} \\*
\D{t}{p_{\perp}} = - \grad \bcdot (q_{\perp}\hat{\bm{b}}) - q_{\perp} \grad \bcdot \hat{\bm{b}} + p_{\perp} \hat{\bm{b}}\hat{\bm{b}}\dbldot \grad \bm{u} - 2p_{\perp} \grad \bcdot \bm{u}  -  \nu_{c}\Delta p,\label{eq:KMHD pp}\\*
\D{t}{p_{\parallel}} = - \grad \bcdot (q_{\parallel}\hat{\bm{b}}) + 2q_{\perp} \grad \bcdot \hat{\bm{b}} -2 p_{\parallel} \hat{\bm{b}}\hat{\bm{b}} \dbldot \grad \bm{u} - p_\parallel \grad\bcdot\bm{u} +2\nu_{c}\Delta p,\label{eq:KMHD pl}
\end{gather}
%
where ${\rm d}/{\rm d}t \equiv \partial/\partial t + \bm{u}\bcdot\grad$ is the convective derivative. The velocity
$\bm{u}$ in \eqref{eq:KMHD rho}--\eqref{eq:KMHD pl} includes both the background shear flow $\bm{U}_{0}$ and perturbations on top of this $\delta\bb{u}$ (e.g., the MRI). The other symbols have their usual meanings: $\rho$ is the mass density, $\bm{B}$ is the magnetic field, $\nu_{c}$ is the particle collision frequency, and $B\equiv \left| \bm{B} \right|$ and $\hat{\bm{b}}\equiv\bm{B}/B$ denote the magnetic-field strength and direction (note that $\grad \bcdot (\hat{\bm{b}}\hat{\bm{b}}B^{2})$ is simply $\bm{B}\bcdot \grad \bm{B}$). 
The pressures perpendicular and parallel to $\hat{\bm{b}}$ are $p_{\perp}$ and $p_{\parallel}$ respectively, while $q_{\perp}$ and $q_{\parallel}$  denote the fluxes 
of perpendicular and parallel heat in the  direction parallel to $\hat{\bm{b}}$.
Note that $p_{\perp}$ and $p_{\parallel}$ in \eqref{eq:KMHD u} should  in principle be summed over both particle species (with separate pressures for each species), while $\rho$ and $\bm{u}$ in \eqref{eq:KMHD rho}--\eqref{eq:KMHD B} 
are the ion density and flow velocity. For simplicity, in  this work we solve only the ion 
pressure equations (i.e., $p_{\perp} = p_{\perp,i}$, $p_{\parallel} = p_{\parallel,i}$), which 
is justified in the limit of cold electrons (as expected in RIAFs, e.g., \citealt{Sharma:2007cr}). 
We have also neglected nonideal corrections
to the induction equation \eqref{eq:KMHD B} (e.g., the Hall term), which 
is appropriate given our neglect of finite Larmor radius (FLR) effects in \eqref{eq:KMHD u} (we will, however, include a hyper-resistivity term in equation \eqref{eq:KMHD B} for numerical  reasons; see \S\ref{sub: numerics}).
For convenience, we define the dimensionless anisotropy $\Delta \equiv \Delta p/p_{0}$, where $p_{0} = (p_{\perp}+2p_{\parallel})/3$ is the total thermal pressure, as well as the ratio of thermal to magnetic pressure $\beta \equiv 8\upi p_{0}/B^{2}$, the Alfv\'en 
 speed $v_{A} = B/\sqrt{4\upi \rho}$ and its $\hat{\bm{z}}$ component $v_{Az}=B_{z}/\sqrt{4\upi \rho}$,  the sound speed $c_{s} = \sqrt{p_{0}/\rho}$, and the parallel sound speed $c_{s\parallel} = \sqrt{p_{\parallel}/\rho}$. The double-dot notation used in \eqref{eq:KMHD pp}--\eqref{eq:KMHD pl} and throughout this work is $\hat{\bm{b}}\hat{\bm{b}}\dbldot\grad\bm{u} \equiv \sum_{i,j} b_i b_j \partial_i u_j$.

In their present form, equations \eqref{eq:KMHD rho}--\eqref{eq:KMHD pl} are not closed, due to the presence of the unspecified heat fluxes, $q_{\perp}$ and $q_{\parallel}$. These 
must be either  specified  using a closure scheme, neglected, or solved for using the full kinetic equations. 
In this work we consider three closures for $q_{\perp}$ and $q_{\parallel}$ (or equivalently, three approximations to \eqref{eq:KMHD pp}--\eqref{eq:KMHD pl}). These will be 
  seen to lead to quite different behaviour in solutions of \eqref{eq:KMHD rho}--\eqref{eq:KMHD pl}. They are:
  \begin{description}
\item[{Collisionless Landau fluid closure}: ] {Landau fluid (LF) closures have been used extensively 
in the fusion community \citep{Snyder:1997fs,Hammett:1990,Hammett:1992} and, to a lesser degree,  
for astrophysical applications (\citetalias{Sharma:2006dh}; \citealt{Sharma:2007cr}). They are particularly well suited for 
modeling collisionless ($\nu_{c} = 0$) plasmas. In the LF closure, the 
heat fluxes,
\begin{gather}
q_{\perp} = -\frac{2 c_{s\parallel}^{2}}{\sqrt{2 \upi }c_{s\parallel} |k_{\parallel}|+\nu_{c}} \left[ \rho  \nabla_{\parallel} \left(\frac{p_{\perp}}{\rho}\right)  - p_{\perp}\left(1-\frac{p_{\perp}}{p_{\parallel}} \right)\frac{\nabla_{\parallel} B}{B}  \right], \label{eq:GL heat fluxes qp}\\ 
q_{\parallel} = - \frac{8 c_{s\parallel}^{2}}{\sqrt{8 \upi }c_{s\parallel} |k_{\parallel}|+(3\upi-8)\nu_{c}} \,\rho \nabla_{\parallel} \left(\frac{p_{\parallel}}{\rho}\right),\label{eq:GL heat fluxes ql}
\end{gather} 
are constructed to replicate the effects of linear Landau damping. Here $\nabla_{\parallel} \equiv \hat{\bm{b}}\bcdot \grad $  denotes the gradient parallel to the field and
$|k_{\parallel}|$ denotes the wavenumber parallel to the field, which must 
be considered as an operator because it appears in the denominator of \eqref{eq:GL heat fluxes qp}--\eqref{eq:GL heat fluxes ql}. The forms of the heat fluxes in \eqref{eq:GL heat fluxes qp}--\eqref{eq:GL heat fluxes ql} accurately reproduce the true kinetic growth rates and frequencies for
a variety of large-scale (MHD) modes, including the MRI (\citetalias{Quataert:2002fy}; \citealt{Sharma:2003hf}). We refer the reader to \citet{Snyder:1997fs} and \citetalias{Sharma:2006dh} for more information.
}
\item[{Weakly collisional ``Braginskii'' closure: }]{In the Braginskii regime, $|\grad \bm{u}| \ll \nu_{c}$ \citep{Braginskii:1965vl}, the pressure anisotropy is strongly influenced by collisional relaxation. We thus neglect 
${\rm d}_{t}p_{\perp}$ and ${\rm d}_{t}p_{\parallel}$ in \eqref{eq:KMHD pp}--\eqref{eq:KMHD pl} and balance the double-adiabatic production of pressure anisotropy (the $\hat{\bm{b}}\hat{\bm{b}}\dbldot\grad\bm{u}$ and $\grad\bcdot\bm{u}$ terms) against its collisional relaxation (the $\nu_{c}\Delta p$ terms) to find
\begin{equation}
\Delta \approx \frac{1}{\nu_{c}}\left(\hat{\bm{b}}\hat{\bm{b}}\dbldot\grad \bm{u} - \frac{1}{3}\grad \bcdot \bm{u} \right) = \frac{1}{\nu_{c}} \D{t}{} \ln \frac{B}{\rho^{2/3}} ,\label{eq:Brag closure}
\end{equation}
where we have also used the fact that  $\nu_{c}/|\grad\bm{u}|\gg 1$ implies $\Delta p \ll p_{0}$. (For $\beta \gg 1$, the $\grad\bcdot \bm{u}$ term can also be neglected). When inserted into the momentum equation \eqref{eq:KMHD u}, equation \eqref{eq:Brag closure} has the form of an anisotropic viscous stress, and is thus referred to as ``Braginskii viscosity'' (or Braginskii MHD for the full set of equations). Note that we have neglected heat fluxes in arriving at \eqref{eq:Brag closure}, a simplification that is rigorously obtained if $\nu_{c}/|\grad\bm{u}|\gg \beta^{1/2}$ (the ``high-collisionality'' regime). On the other hand, if $\nu_{c}/|\grad\bm{u}|\ll \beta^{1/2}$ (the ``moderate-collisionality'' regime), the heat fluxes are strong over the time scales of the motion \citep{Mikhailovskii:1971}, and their contribution to the (ion) pressure anisotropy must be retained. (See Appendix \ref{app:sub: Braginskii} for further discussion.) In this case, there is no simple closure that can be devised (e.g., see appendix B of \citealt{Squire:2016ev2}) and it is usually easier to consider the full LF system.
}
\item[{Double-adiabatic closure: }]{The double-adiabatic, or Chew-Goldberger-Low (CGL),
closure \citep{CGL:1956} simply involves setting $q_{\perp} = q_{\parallel} = 0$. 
This approximation is far from justified for subsonic motions in the high-$\beta$ plasmas considered
here; however, the closure is useful for comparison with the LF closure 
by virtue of its relative simplicity. It has also been employed in a variety of
previous computational studies (e.g., \citetalias{Sharma:2006dh}; \citealt{Kowal:2011iy,SantosLima:2014cn}), and so it is worthwhile to diagnose the model's successes and limitations. 
}
\end{description}

An important caveat for each of these approximations to \eqref{eq:KMHD pp}--\eqref{eq:KMHD pl} relates to plasma microinstabilities. For our purposes, given the focus of this work on the high-$\beta$ regime, the most
 significant of these are the firehose instability  \citep{Rosenbluth:1956,Chandrasekhar:1958,Parker:1958,Yoon:1993of}, which is excited if
 \begin{equation}
\Delta \lesssim -\frac{2}{\beta},\label{eq:FH lim}
\end{equation}
and the mirror instability \citep{1969PhFl...12.2642H,sk93,Hellinger:2007}, which is excited if
 \begin{equation}
\Delta \gtrsim \frac{1}{\beta}.\label{eq:MI lim}
\end{equation}
(There are corrections to these $\beta$-dependent thresholds that arise from particle resonances and depend on the specific form of the distribution function; see, e.g.,  \citealt{2015PhPl...22c2903K}). Important aspects of these instabilities (e.g., their regularization at small scales or particle
scattering in their nonlinear evolution) are not captured by the closures we employ here, and
 kinetic calculations are needed to correctly understand their saturation.
There have been a variety of recent works in this vein \citep{Schekochihin:2008en,Hellinger:2008hd,Rosin:2011er,Kunz:2014kt,Hellinger:2015en,Rincon:2015mi,2015ApJ...800...27R,2015ApJ...800...88S,2016ApJ...824..123R,Melville:2015tt},
which have shown that these microinstabilities generally act to 
pin the pressure anisotropy at the marginal stability limits. 
Interestingly, when $\Delta$ is driven beyond the stability boundaries,
 the microinstabilities achieve this in two stages: first, while the
microscale fluctuations are growing secularly, by increasing (if ${\rm d}_t B <0$) or decreasing  (if ${\rm d}_t B >0$) 
the magnetic-field strength $B$ (in other words, the small-scale fluctuations contribute to $B$); second,
as the instabilities saturate, by  enhancing the scattering of particles and thus increasing $\nu_{c}$ in \eqref{eq:KMHD pp}--\eqref{eq:KMHD pl}.
While the time for the firehose instability at moderate $\beta$ to saturate is essentially set by gyro-scale physics, and so might be considered as instantaneous in a fluid model,  the 
mirror instability saturates on a time scales comparable to the turnover time of the large-scale motions driving the anisotropy (see \S\ref{sub: firehose mirror beta constraints} and \citealt{Kunz:2014kt,Rincon:2015mi,2015ApJ...800...27R,Melville:2015tt}).  We also note that there are also various other kinetic 
 instabilities that could be important, for instance, the ion-cyclotron instability, or electron instabilities. We  do not consider these in detail because the mirror and firehose instabilities are thought to be the most relevant to the high-$\beta$, ion-dominated regime that is the focus of this work (see \S\ref{sub: extra kinetics} for further discussion).

In practice, because the primary effect in both  the secular and scattering regimes is to limit $\Delta$ at the threshold boundaries, we model these effects as a ``hard wall'' limit
on $\Delta$, following prior work (\citetalias{Sharma:2006dh}; \citealt{Sharma:2007cr,SantosLima:2014cn,Chandra:2015bh,Foucart:2015gy}). This simply limits $\Delta$ to $1/\beta$ or $-2/\beta$ if the dynamics  drive $\Delta$ across these boundaries.\footnote{Within the context of the LF closure \eqref{eq:KMHD rho}--\eqref{eq:GL heat fluxes ql}, the only obvious  place where the difference between saturation via 
 unresolved small-scale fields and saturation via particle scattering may be important is in the heat fluxes.
 These are significantly reduced at high collisionality (see \eqref{eq:GL heat fluxes qp} and \eqref{eq:GL heat fluxes ql}), and likely also modified by microscale mirror or firehose fluctuations \citep{2016MNRAS.460..467K,2016ApJ...824..123R,2017arXiv170803926R}. We have experimented with 
 either including $\nu_{c}$ in $q_{\perp,\parallel}$ at the mirror
 and firehose boundaries or not, and this difference does not appear to strongly affect the results presented herein. }
One should, however, be careful with this simple ``limiter'' method not to  inadvertently 
remove interesting physics from the model. For instance, the parallel firehose instability 
is destabilized at the same point $\Delta = -2/\beta$ as where the magnetic tension is nullified 
by $\Delta p$ (indeed, this is the cause of the instability), which can have a strong influence 
on the largest scales \citep{Squire:2016ev,Squire:2016ev2} and is captured in even our simplest 1-D models.  For this reason, in considering azimuthal-field KMRI modes, we have run calculations
both with and without a firehose limiter, seeing very similar dynamics in each case. Finally, we note that with finite scale separations between $\Omega_{i}$ and $S$, as 
is the case in numerical simulations \citep{Riquelme:2012kz,2015PhRvL.114f1101H,2016PhRvL.117w5101K}, there
can be significant overshoot of the pressure anisotropy beyond the limits \eqref{eq:FH lim} and \eqref{eq:MI lim} \citep{Kunz:2014kt}. This overshoot may be 
important for the large scale evolution (see \S\ref{sec:1D}) but is probably not representative of what happens in real 
systems, which usually have a very large dynamic range between $\Omega_{i}$ and $S$. These 
effects are discussed in detail in \S\ref{sec:implications}.

%
%
\subsection{Computational methods}\label{sub: numerics}

A number of different numerical methods are used to solve \eqref{eq:KMHD rho}--\eqref{eq:GL heat fluxes ql}. For investigating the 1-D evolution of a channel mode, 
we use a pseudo-spectral method, with standard dealiasing 
and hyper-diffusion operators used to remove the energy 
just above the grid scale. A very similar numerical method, albeit on a 3-D Fourier grid, is used for the studies of parasitic modes. For studies of the fully nonlinear 3-D evolution, we use a
modified version of the finite-difference code {\sc Zeus}, as described in \citetalias{Sharma:2006dh}.
For simulations in the weakly collisional regime, we solve the full 
LF system  \eqref{eq:KMHD rho}--\eqref{eq:GL heat fluxes ql}, so as to correctly capture the
effects of the heat fluxes in the moderate- and high-collisionality  regimes (see Appendix \ref{app:sub: Braginskii}).

A few words are needed regarding the numerical treatment of heat fluxes in the LF model \eqref{eq:GL heat fluxes qp}--\eqref{eq:GL heat fluxes ql}. 
In particular, the $1/|k_{\parallel}|$ operator is numerically awkward, because it is not diagonal in either Fourier space 
or real space. 
We thus use the prescription of \citetalias{Sharma:2006dh} and replace this
by a pre-chosen $k_{L}$ for  the  {\sc Zeus} implementation and the 
parasitic mode studies, while for the 1-D collisionless calculations we use $1/|k_{\parallel}|=1/|k_{z}|$ (once the mode reaches larger amplitude this may somewhat underestimate the
heat fluxes, since $k_{\parallel} < k_{z}$ if the field lines are not straight).
Following \citetalias{Sharma:2006dh}, we have checked that varying the choice of $k_{L}$ within a reasonable range, or using the choice $|k_{\parallel}|=|k_{z}|$,  does not significantly affect the dynamics.
 
We use the methods detailed in appendix A3 of \citetalias{Sharma:2006dh} to limit the pressure
anisotropy: $\nu_{c}$ is instantaneously increased in \eqref{eq:KMHD pp}--\eqref{eq:KMHD pl} whenever 
$\Delta$ passes the limits \eqref{eq:FH lim} or \eqref{eq:MI lim}. 
We do not make any distinction between the ``secular growth'' and ``particle scattering'' phases of 
microinstability evolution with this method (see discussion above, around \eqref{eq:MI lim}, and \S\ref{sub: firehose mirror beta constraints}), and more 
study is needed to better understand the successes and limitations of this simple limiter approach.

%
%
\subsection{General comparison of kinetic models} \label{sub: general comparison}

Before commencing with our analysis of the KMRI, we highlight in this section some of the key similarities and differences between the LF, Braginskii, and CGL models, as well as that of standard MHD.

First, irrespective of the closure details, the general effect of pressure anisotropy is to modify the Lorentz force, either by enhancing ($\Delta p>0$) or reducing ($\Delta p <0$) the effective magnetic tension (see \eqref{eq:KMHD u}). 
The same \emph{relative} pressure anisotropy $\Delta = \Delta p/p_{0}$ will thus have a greater dynamical effect as $\beta $ increases, because $p_{0}$ increases compared to $B$. In contrast, the generation of $\Delta$ in a changing magnetic field,
\begin{equation}
\D{t}{\Delta} \sim   \hat{\bm{b}}\hat{\bm{b}}\dbldot\grad \bm{u} \sim \D{t}{\ln B} 
\end{equation}
(or $\Delta \sim \nu_{c}^{-1} \hat{\bm{b}}\hat{\bm{b}}\dbldot\grad \bm{u}$ for Braginskii),
does not depend on $\beta$. Thus, the dynamics of higher-$\beta$ plasmas
are in general more strongly influenced by self-generated pressure anisotropies than are the dynamics of  lower-$\beta$ plasmas.

Secondly, the spatial form of $\Delta p$ generated in a given time-dependent $B$ differs significantly between the LF, CGL, and Braginskii models. This $\Delta p$ is important for
 the nonlinear behaviour of the collisionless MRI. 
 The influence of a spatially constant $\Delta $ can be considered from an essentially linear standpoint: it simply acts to enhance or reduce the Lorentz force by the factor $(1+\beta \Delta/2)$.
In contrast,  
when the spatial variation in $\Delta$ is similar to its magnitude, its effect is inherently nonlinear. For the 
same reason, detailed conclusions about the expected change in the spatial shape of an MRI 
mode 
due to pressure anisotropy \emph{do} depend on the regime of interest (and model) -- for example, 
collisionless vs.~Braginskii \citep{Squire:2016ev}, or low vs.~high $\beta$ -- rather than 
being generic consequences of any self-generated pressure anisotropy. 
For our purposes, one can consider high-collisionality Braginskii MHD (equation \eqref{eq:Brag closure}) as the
limiting model that develops large spatial variation in $\Delta$ (since $\Delta$ is
tied directly to ${\rm d}_t B$), while 
the high-$\beta$ limit of the LF model is the opposite, developing large
$\Delta$ with very little spatial variation.\footnote{The double-adiabatic 
model and moderate-collisionality Braginskii regime lie somewhere between these limits; see \eqref{eq:pp CGL expansion}.} This is because collisions generically 
act to reduce the magnitude of $\Delta$ without affecting the 
spatial variation in $p_{\perp}$ and $p_{\parallel}$, whereas 
heat fluxes act to reduce spatial variation in $p_{\perp}$ and $p_{\parallel}$
without affecting the spatial average  of $\Delta$.

This last statement warrants further explanation, given the complexity of 
\eqref{eq:KMHD pp}--\eqref{eq:GL heat fluxes ql}. 
The effect of the LF heat fluxes \eqref{eq:GL heat fluxes qp}--\eqref{eq:GL heat fluxes ql} can be clarified if we assume $\nu_{c}=0$ and $\Delta p \ll p_{0}$ (the latter is always valid at
high $\beta$), so that
\begin{equation}
q_{\parallel} \approx - \sqrt{\frac{8 }{\upi }} \rho c_{s}\frac{\nabla_{\parallel}}{| k_{\parallel}|} \left(\frac{p_{\parallel}}{\rho}\right), \quad q_{\perp} \approx - \sqrt{\frac{2 }{\upi }} \rho c_{s}\frac{\nabla_{\parallel}}{| k_{\parallel}|} \left(\frac{p_{\perp}}{\rho}\right). \label{eq:simp heat fluxes}
\end{equation}
Then, assuming $\hat{\bm{b}}\bcdot \grad q_{\perp,\parallel} \gg q_{\perp,\parallel} \grad \bcdot \hat{\bm{b}} $, which is valid when the perturbation 
to the background field is small (i.e., when the field lines are nearly straight), 
the contribution to the $p_{\perp}$ and $p_{\parallel }$ evolution equations has the form 
$\partial_{t} p_{\perp} \sim -\rho c_{s}|k_{\parallel}| (p_{\perp}/\rho) $ and $\partial_{t} p_{\parallel}\sim -\rho c_{s}|k_{\parallel}| (p_{\parallel}/\rho) $, respectively. These terms, which model the effect of Landau damping \citep{Snyder:1997fs},
act as a ``scale-independent'' diffusion of $p_{\perp}$ and $p_{\parallel}$, damping inhomogeneities\footnote{The heat-flux induced damping is technically 
of $p_{\perp}/\rho = T_{\perp}$ and $p_{\parallel}/\rho = T_{\parallel}$, rather
than  of $p_{\perp}$ and $p_{\parallel}$ themselves. Because the spatial   variation in 
$\rho$ will generally be similar to that of $p_{\perp,\parallel}$, the effective damping 
is less than what it would be if the variation in $\rho$ were ignored (see Appendix \ref{app:sub: LF}, equations \eqref{eq:pp Landau expansion}--\eqref{eq:pl Landau expansion}). However, this variation in $\rho$ can never completely cancel the variation in $p_{\perp,\parallel}$ and preclude 
 damping: there is always some $T_{\perp,\parallel}$ variation induced by the changing magnetic-field strength.} over the
 sound-wave timescale $|k_{\parallel}| c_{s}$.
This is important because $|k_{\parallel}| c_{s} \sim \beta^{1/2} \gamma_{\mathrm{MRI}}$ (where 
$\gamma_{\mathrm{MRI}}$ is the MRI growth rate), showing
that for $\beta \gg 1$, the heat fluxes will rapidly erase spatial variation in $\Delta p$ on the 
timescale that the MRI grows. As a result, the spatial variation in $\Delta p$ will be dwarfed by its mean; i.e., 
$\Delta p$ will be  nearly spatially constant. 
A more thorough discussion of these ideas is given in Appendix~\ref{app: MRI nonlinearity}, where we 
solve explicitly for the $\Delta p$ that arises in an exponentially growing, spatially varying magnetic 
perturbation. This shows that the double-adiabatic model for $p_{\perp,\parallel}$ generates a $\Delta p $ with spatial variation on the order of its mean, while the addition of LF heat fluxes decreases the spatial variation
of $\Delta p$ by a factor $\beta^{1/2}$  while leaving the mean $\Delta p$ unchanged.

%
%
\section{One-dimensional evolution}\label{sec:1D}

In this section, we discuss various nonlinear effects that occur 
due to the pressure anisotropy that develops in a growing KMRI mode.  These 
effects are one-dimensional (i.e., unrelated 
to ``parasitic'' modes and turbulence, which is discussed in \S\ref{sec:3D}) and occur at very low mode amplitudes: when $\delta B \sim \beta^{-1/2}B_{0}$ in a purely vertical field, or when $\delta B_{y} \sim \beta^{-2/3}B_{0}$ and $\delta B_{x} \sim \beta^{-1/3}B_{0}$ with an azimuthal field (where $\delta B$ denotes the mode amplitude).
This provides an interesting counterpoint 
to MHD MRI channel modes, which are nonlinear solutions of the incompressible MHD 
equations \citep{Goodman:1994dd}, and  only exhibit notable nonlinear modifications 
as the mode amplitude approaches the sound speed, $\delta B \sim  \beta^{1/2} B_{0}$. 
However, we will also see that despite this early (low-amplitude) nonlinear modification, once a KMRI mode's amplitude starts to dominate over the background field ($\delta B\gtrsim B_{0}$), it \emph{reverts} to being an approximate nonlinear solution, because of the pressure-anisotropy-limiting behaviour of the mirror instability. Thus 
a KMRI mode behaves very similarly to an MHD channel mode as $\delta B$ approaches $\beta^{1/2} B_{0}$.

We begin by outlining the basic linear physics of the KMRI (\S\ref{sec:linear MRI}), which will be relevant to its nonlinear evolution. We then examine 
various stages in the evolution of a 1-D collisionless KMRI mode 
in  vertical (\S\ref{sec: 1D vertical}) or mixed azimuthal-vertical (\S\ref{sec: 1D azim}) background magnetic fields,  before 
the mode saturates into turbulence. We also discuss 
how these stages are modified in the weakly collisional (Braginskii) regime (\S\S\ref{sec: 1D brag}, \ref{subsub: Braginskii with By}). Throughout this section 
we denote the MRI perturbation velocity and magnetic field as $\delta\bm{u}$ and $\delta\bm{B}$ respectively (their magnitudes are $\delta u$ and $\delta B$), the background magnetic field as 
$\bm{B}_{0} = B_{0y}\hat{\bm{y}}+B_{0z}\hat{\bm{z}}$, and $\beta_{0} = 8\upi p_{0}/B_{0}^{2}$   is defined with respect to the background field (we also use $\beta_{0z} = 8\upi p_{0}/B_{0z}^{2}$).
For the reader interested in a general overview of results, the summary in \S\ref{sub:1-D conclusions} should be understandable without a careful reading of \S\S\ref{sec:linear MRI}--\ref{sec: 1D azim}.

%
%
\subsection{Linear KMRI}\label{sec:linear MRI}

Before discussing any nonlinear effects, it is helpful to first review aspects of the linear KMRI. We consider only
the simplest case of purely vertical wavenumbers $k_{z}\neq 0$, $k_{x}=k_{y}=0$; i.e., $(\bm{k}\parallel \bm{B} \parallel \bm{\Omega})$.
This choice is motivated by the stabilizing influence of
a nonzero radial wavenumber $k_{x}$  
 (see discussion in \S 4 of \citetalias{Quataert:2002fy}), meaning that $k_{x}=0$ modes should dominate if growing from small amplitudes, while treating $k_{y}\neq 0$ modes 
 requires a global and/or time-dependent method \citep{Balbus:1992du,Johnson:2007wo,Squire:2014es}. We also 
neglect the possibility of a radial background magnetic field since this leads to  a
 time-dependent background azimuthal field. More thorough discussion and detailed derivations can be found in \citetalias{Quataert:2002fy}; 
\citet{Sharma:2003hf,Balbus:2004hg,Rosin:2012ej,Heinemann:2014cl,Quataert:2015dv}, as well 
in Appendix~\ref{app: AKMRI linear properties}, where we derive properties of the  KMRI 
in a mixed vertical-azimuthal field.

%
%
\begin{figure}
\begin{center}
\includegraphics[width=0.7\columnwidth]{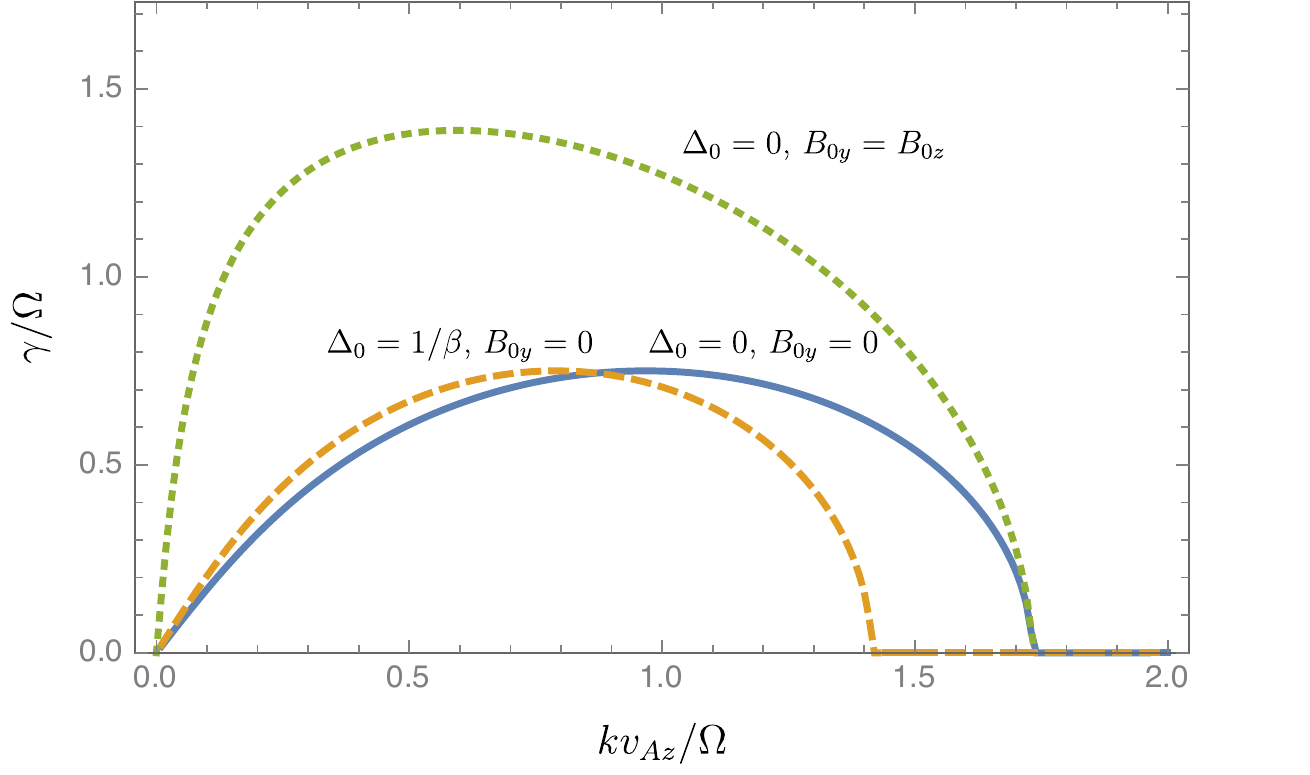}
\caption{Dimensionless linear growth rate $\gamma/\Omega$ of the KMRI at $\beta_{0z}=8\upi p_{0}/B_{0z}^{2}=400$ and $S/\Omega = 3/2$, plotted as a 
function of dimensionless vertical wavenumber $k_{z} v_{Az}/\Omega$
(with $k_{x}=k_{y}=0$). The solid blue curve shows the case with a purely vertical $\bm{B}_{0}$ and no background pressure anisotropy $\Delta_{0}$, for which the dispersion relation is identical to the standard MRI. The orange dashed line
shows the case with $B_{0y}=0$ and $\Delta_{0}=1/\beta_{0}$, which is approximately 
the anisotropy at which the mirror limit is first reached in the growing mode. Finally, the green dotted 
line shows the growth rate in the case with an azimuthal field $B_{0y}=B_{0z}$ (with $\beta_{0z}=400$, $\beta_{0}= 200$),
where the growth rate is strongly enhanced compared to the MHD MRI 
(which is unaffected by the azimuthal field for $k_{x}=k_{y}=0$).}
\label{fig:linear}
\end{center}
\end{figure}

We linearize \eqref{eq:KMHD rho}--\eqref{eq:GL heat fluxes ql} with $\nu_{c}=0$ and $S/\Omega=3/2$, then insert the 
Fourier ansatz $\delta f (z,t)=\delta f e^{i k z - i\omega t}$ for each variable ($f = \rho$, $\bm{u}$, $\bm{B}$ etc.).
Solution of the resulting polynomial equation for $\omega$ yields the linear KMRI growth rates, $\gamma/\Omega= \Im({\omega})/\Omega$, as shown
in figure \ref{fig:linear} for several relevant cases.
For the case of purely vertical field and no background pressure
anisotropy (solid line), the KMRI dispersion relation 
is identical to the collisional MRI. This occurs because when $\bm{B}_{0}=B_{0}\hat{\bm{z}}$
the MRI does not linearly perturb the pressure 
(because $\partial_{z}(\bm{B}_{0}\bcdot \delta \bm{B})=0 $), and thus is
ignorant of the equation of state. With a nonzero 
radial wavenumber $k_{x}\neq 0$, this is no longer the case (since then $\delta B_{z}\neq 0$), and the growth rate is lower than for the
standard MRI. 

The addition of a background positive pressure anisotropy, $\Delta_{0}>0$, 
shifts the MRI to larger wavelengths (smaller $k$; dashed curve in figure \ref{fig:linear}). This is because the anisotropic-pressure stress
in the momentum equation has a form identical to that of the Lorentz force (see \eqref{eq:KMHD u}). 
Thus the only difference in comparison to the standard MRI dispersion relation is the replacement of  $k v_{A}$  with $k v_{A}(1+\beta_{0} \Delta_{0}/2)^{1/2}$, which decreases 
the wavenumber that maximizes $\gamma$ from $\sim\! \Omega v_{A}^{-1}$ to 
$\sim\! \Omega v_{A}^{-1}(1+\beta_{0} \Delta_{0}/2)^{-1/2}$, while keeping the 
maximum itself constant. This  is relevant to the nonlinear 
behaviour of the KMRI, since the mode generates a pressure anisotropy
as it evolves. 

Finally, the addition of a background azimuthal field causes the KMRI 
dispersion relation to differ significantly  from that of the standard MRI (\citetalias{Quataert:2002fy}; \citealp{Balbus:2004hg}; dotted curve in figure \ref{fig:linear}), increasing the growth rate  and moving
the instability to  larger wavelengths. This differs from the standard (MHD) MRI, which is
unaffected by $B_{0y}\neq 0$ when $k_{y}=0$. Due to the different  
physical processes that lead to the large growth rates at low $k$, this instability is
also known as the \emph{magnetoviscous instability} (MVI; \citealt{Balbus:2004hg}). Unlike 
the standard MRI, the growth mechanism relies on the azimuthal pressure force $\hat{\bm{b}}_{0}\hat{\bm{b}}_{0} \delta \Delta p$, which is destabilizing and
dominates over the magnetic tension by a factor of $\beta_{0}^{1/2}$ (see \citetalias{Quataert:2002fy}). 
As shown in Appendix~\ref{app: AKMRI linear properties}, for $\beta_{0}\gg 1$ the KMRI 
growth rate approaches $\gamma = (2S\Omega)^{1/2}$, with a maximum growth 
rate at wavenumber $k_{\mathrm{max}}v_{Az}/\Omega\approx 1.8\,\beta_{0z}^{-1/6}$ when $B_{0y}\approx B_{0z}$ and $S/\Omega=3/2$ (see \eqref{eq:app: max k MVI}). In this fastest-growing mode, the
relative amplitudes of the various components are $\beta_{0}^{-2/3}\delta B_{x}/B_{0}\sim \beta_{0}^{-1/3}\delta B_{y}/B_{0}\sim \beta_{0}^{-5/6}\delta u_{x}/v_{A}\sim \beta_{0}^{-5/6}\delta u_{y}/v_{A}\sim \delta p_{\perp,\parallel}/p_{0}$ (unlike the standard MRI, where $\delta B_{x}/B_{0}\sim \delta B_{y}/B_{0}\sim \delta u_{x}/v_{A}\sim \delta u_{y}/v_{A}$). While we shall use these scalings below in our discussion of the nonlinear behaviour of MRI modes, we caution that they only apply at rather high $\beta_{0}$, a problem that is exacerbated if $B_{0y}/B_{0z}\neq 1$. At more moderate $\beta_{0}$, as feasible for simulations, $k_{\mathrm{max}}$ tends 
towards its value for  standard MRI, $k_{\mathrm{max}}v_{Az}/\Omega\approx 1$    (see figure \ref{fig: maximizing k for MVI}). It is also worth noting that the dispersion relation, $\gamma(k)$, varies only slowly around $k=k_{\mathrm{max}}$ (see e.g., dotted curve in figure~\ref{fig:linear}). This implies that, if starting from 
random initial conditions, a long time would be required before the fastest-growing mode dominates, suggesting
that when nonlinear effects become important there will likely still be several modes of similar amplitudes. 
With these caveats duly noted, in our discussion of nonlinear effects below (\S\ref{sec: 1D azim}),  we will consider only the  fastest-growing KMRI mode (i.e., set $k=k_{\mathrm{max}}$) and use the above scalings, rather than keep  $k$ arbitrary. 


%
%
%
\subsection{Nonlinear KMRI: Vertical magnetic field}\label{sec: 1D vertical}

In this section, we consider the nonlinear evolution of the MRI 
in a vertical background field.  In this case the linear dispersion relation 
is identical to that obtained in ideal MHD. However, we shall show that 
the modes are (modestly) nonlinearly modified by the pressure anisotropy at low amplitudes $\delta B\sim\beta^{-1/2}B_{0}$. To do so, we first describe the evolution
of a truly collisionless mode using the LF closure with $\nu_{c} = 0$ (\S\ref{sec: 1D LF}), and then examine the weakly collisional Braginskii case in \S\ref{sec: 1D brag}.

\subsubsection{Collisionless (LF) regime}\label{sec: 1D LF}


%
%
\begin{figure}
\begin{center}
\includegraphics[width=1\columnwidth]{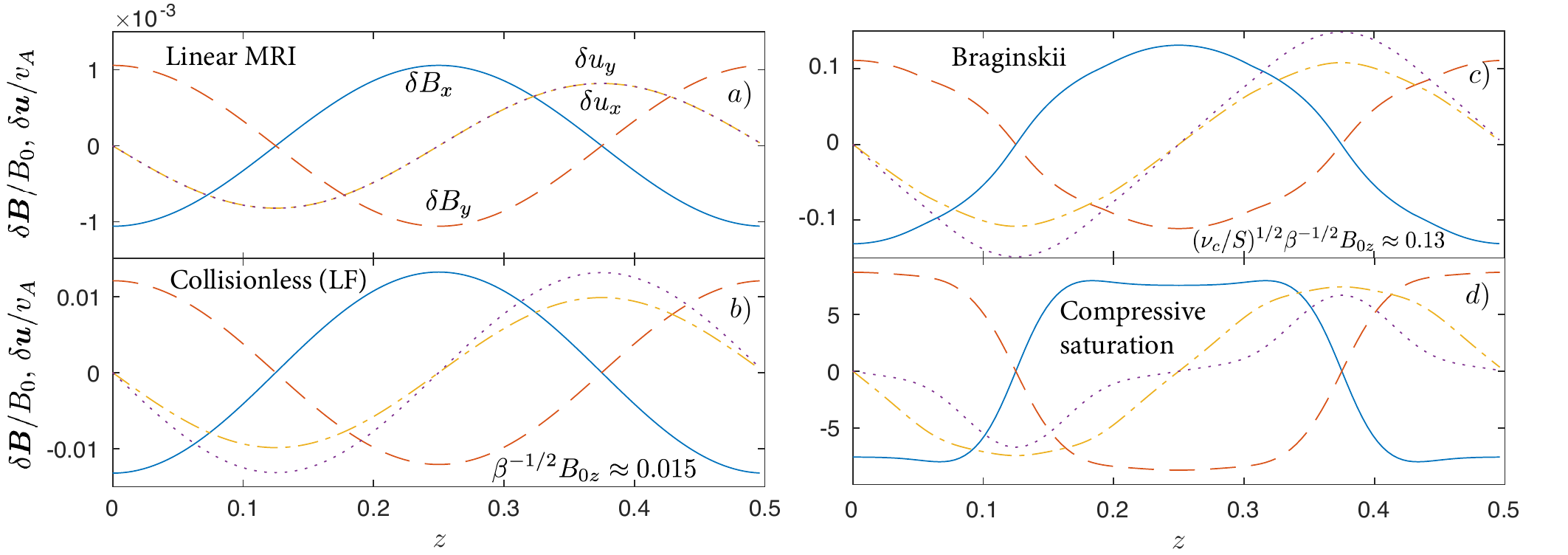}
\caption{The structure of a kinetic MRI mode evolving in a vertical background field $B_{0z}$ in various regimes, as computed from the 1-D LF model (with $\nu_{c}\neq 0$ in panel c). We take $\beta_{0}=337$  in a domain with $\Omega L_{z}/c_{s}=1$, such 
that the peak of the MRI dispersion relation (i.e., the maximum $\gamma$) is at $k=2\times 2\upi/L_{z}$ (each figure shows only half of a scale height).
Each plot illustrates $\delta B_{x}$ (blue solid line), $\delta B_{y}$ (red dashed line), $\delta u_{x}$ (yellow dot-dashed line), and $\delta u_{y}$ (purple dotted line).  The various
subfigures show: (a) the linear KMRI mode (this is the initial conditions for each simulation), which is identical in structure to an MHD MRI mode at 
these parameters; (b) the collisionless MRI mode when $\Delta p$ reaches the mirror
limit  ($\delta B \sim \beta^{-1/2}B_{0z}\approx 0.015 $), which remains very nearly sinusoidal because the heat 
fluxes make $\Delta p$ spatially uniform; (c) a mode in the high-collisionality Braginskii  regime (with ${\nu}_{c}/S= \beta_{0}^{3/4}$) when $\Delta p$ reaches 
the mirror limit (at $\delta B \sim (\nu_{c}/S)^{1/2}\beta^{-1/2}B_{0z}\approx 0.13 $), which is nonsinusoidal because of the $\mathcal{O}(1)$ spatial variation in $\Delta p$; (d) the MRI mode at very large amplitudes, when compressibility becomes important. The structure of this final compressible stage of evolution is the same across all models, including standard (collisional) MHD.}
\label{fig:Bz MRIs}
\end{center}
\end{figure}

A maximally unstable MRI mode satisfies 
\begin{subequations}\label{eq:channel profile}
\begin{gather}
\delta u_{x} = \delta u_{y} = - \sqrt{\frac{3}{5}} \frac{\delta B_{0}}{\sqrt{4\upi \rho}} \erm^{\gamma t} \sin(k z), \\*
-\delta B_{x} = \delta B_{y} = \delta B_{0} \erm^{\gamma t} \cos(k z),
\end{gather}
\end{subequations}
where $\gamma = S/2$ is the growth rate and $\delta B_{0}$ is the initial 
mode amplitude. Because of the opposing signs between $\delta B_{x} \delta u_{x}$ and $\delta B_{y}\delta u_{y}$ in the
mode, only $-S \hat{b}_{x} \hat{b}_{y}$ contributes to $\hat{\bm{b}}\hat{\bm{b}}\dbldot \grad \bm{u} \approx 
-S \delta B_{x}\delta B_{y}/B_{0z}^{2}$.
 As shown in
Appendix \ref{app:sub: LF}, at high $\beta$ in a collisionless plasma (using the LF closure), the mean of the pressure anisotropy dominates
over its spatial variation by a factor of $\beta^{1/2}$ (see \eqref{eq: pressure 2 is smooth}--\eqref{app:eq: LF delta MRI}), and $\Delta$ is approximately spatially constant:
\begin{equation}
\Delta   \sim \frac{3}{2} \frac{  \delta B_{0}^{2} }{  B_{0z}^{2}} \erm^{2\gamma t} .\label{eq:DeltaLF}
\end{equation}
The mirror threshold is reached at $\Delta = 1/\beta $, when $\delta B/B_{0z}\sim \beta_{0}^{-1/2}$. As discussed in \S\ref{sec:linear MRI}, a positive pressure 
anisotropy modifies the MRI mode  by effectively increasing the magnetic tension, and
the mode will be stabilized if $k v_{A} \sqrt{1+\Delta(t)\beta_{0}/2} \ge \sqrt{2S\Omega}$.
Fortunately, for a mode near the peak growth rate, $ k v_{A} = \sqrt{ S(\Omega - S/4)}$,
this stabilization does not occur, because the fast-growing mirror fluctuations limit $\sqrt{1+\Delta(t)\beta_{0}/2}$ to values at or below $\sqrt{3/2}$.
As shown in figure \ref{fig:Bz MRIs}{\it b}, this pressure anisotropy causes a rather minor modification to the shape of the MRI mode because $\Delta p$ is almost spatially constant, 
decreasing $u_{x}$ and $B_{y}$ relative to  $u_{y}$ and $B_{x}$ in the same way as for an
MHD MRI mode that is not at the fastest growing wavelength.\footnote{Note that if there is insufficient scale separation between 
the gyrofrequency and the MRI growth rate, the pressure 
anisotropy may grow far enough beyond the mirror limit to stabilize the
mode, which occurs if  $\Delta \ge 2\beta_{0}^{-1}[2S \Omega /( kv_{A})^{2}-1]$. In this case, the mode can move to longer
wavelengths and continue to grow if there is sufficient space in the box. This occurs in 
the simulations of \citetalias{Sharma:2006dh}, where the chosen mirror 
limit ``hard wall''  is $\Delta = 3.5/\beta$, large enough that
pressure anisotropy would stabilize the fastest growing MRI mode before $\Delta$ is artificially limited; see \S\ref{sec:3D} and \S\ref{sec:implications} for further discussion.}

As the mode continues to evolve to larger amplitudes $\delta B/B_{0z}\gtrsim \beta_{0}^{-1/2}$, the pressure
anisotropy remains limited by mirror fluctuations. Leaving aside, for the moment, questions related to how the mode disrupts and becomes turbulent, 
there are two other amplitudes of interest: (i) when the mode amplitude surpasses the background field at $\delta B/B_{0z}\sim 1$, and (ii) when compressibility becomes important at $\delta B/B_{0z}\sim \beta_{0}^{1/2}$ ($\delta u\sim c_{s}$). 
Interestingly, if the pressure anisotropy is efficiently limited by mirror fluctuations, $\Delta p=B^{2}/8\upi$, then the pressure anisotropy nonlinearity has little effect:
\begin{equation}
\grad\bcdot (\Delta p \hat{\bm{b}}\hat{\bm{b}})=\frac{\grad\bcdot (B^{2}\hat{\bm{b}}\hat{\bm{b}})}{8\upi} = \frac{\bm{B}\bcdot \grad\bm{B}}{8\upi} = \frac{\bm{B}_{0}\bcdot \grad\delta \bm{B} + \delta \bm{B}\bcdot \grad\delta \bm{B}}{8\upi} = \frac{\bm{B}_{0}\bcdot \grad\delta \bm{B}}{8\upi},\label{eq: pressure anisotropy dB}
\end{equation}
since $\delta \bm{B}\bcdot \grad\delta \bm{B}\approx 0$ for an MRI channel mode \citep{Goodman:1994dd}. Thus, once $\delta B_{x}\sim \delta B_{y}\gtrsim B_{0}$, the effect of the pressure-anisotropy nonlinearity on the mode remains identical to when $\delta B_{x}\sim \delta B_{y}\lesssim B_{0}$, even though the pressure anisotropy has a  large ($\Delta p \sim \delta B^{2}$) variation in space (i.e., 
the change is simply a modified magnetic tension, as shown in  figure \ref{fig:Bz MRIs}{\it b}). In other words, 
because $\Delta p\propto B^{2}$, there is no significant change to the mode as the mode amplitude surpasses the background field strength. 

The final phase of  evolution, once $\delta B\sim \beta^{1/2}B_{0}$ (i.e., $\delta u\sim c_{s}$), is then very similar to standard MHD, and the mode develops 
a rather distinctive shape, which we illustrate in figure \ref{fig:Bz MRIs}{\it d}. Because the nonlinearity is dominated at this point in the evolution by  
inhomogeneities in the  density, this mode shape appears generically in all of the models studied here (including compressible MHD)
across a wide range of parameters \citep[cf.][]{Latter:2009br}.


%
%
%
\subsubsection{Weakly collisional (Braginskii) regime}\label{sec: 1D brag}

In the Braginskii regime, which is valid  for $ \overline{\nu}_{c}\equiv \nu_{c}/S\sim \nu_{c}/\gamma \gg 1$ and is relevant (i.e., represents a potentially  significant
correction to standard MHD) when $\overline{\nu}_{c} \ll \beta_{0}$, 
there are two subregimes. In the moderate-collisionality regime, $\overline{\nu}_{c} \ll \beta_{0}^{1/2}$, the heat fluxes play a very significant role and smooth the pressure anisotropy spatially, 
as occurs in the collisionless case described in \S\ref{sec: 1D LF}. 
In the high-collisionality regime, $\overline{\nu}_{c} \gg \beta_{0}^{1/2}$, the heat fluxes are  
sub-dominant and do not play a significant dynamical role, leading to $\Delta p $ profiles that vary significantly in space. For more information, see Appendix \ref{app:sub: Braginskii} and appendix B of \citet{Squire:2016ev2}.

In the moderate-collisionality regime, up to $\overline{\nu}_{c} \lesssim  \beta_{0}^{1/2}$, 
the behaviour of the mode at   amplitudes $\delta B\lesssim (\beta_{0}/\overline{\nu}_c)^{1/2}B_{0}$ (see below)
is effectively identical to the collisionless regime discussed in \S\ref{sec: 1D LF}. In particular, the pressure anisotropy that develops from the growing mode with $\delta B\ll B_{0}$ is
\begin{equation}
\Delta   \approx \frac{1}{\nu_{c}}\langle \hat{\bm{b}}\hat{\bm{b}}:\grad\bm{u}\rangle \sim \frac{S}{\nu_{c}}\frac{  \delta B_{0}^{2}} {  B_{0z}^{2}} \, \erm^{2 \gamma t}=\frac{1}{\overline{\nu}_{c}}\frac{  \delta B_{0}^{2}} {  B_{0z}^{2}} \, \erm^{2 \gamma t},
\end{equation}
because the heat fluxes are smoothing $\Delta p$ faster than it is being created (by some factor between 
$\beta_{0}^{1/2}$ and $1$, depending on $\overline{\nu}_{c}\,\beta_{0}^{-1/2}$; see \citealt{Squire:2016ev2}).
Thus, by the time that $\Delta p $ reaches the mirror limit (i.e., when $\delta B\sim (\overline{\nu}_{c}/\beta_{0})^{1/2}B_{0}$), 
it is nearly smooth in space, implying that the same conclusions reached in the collisionless limit also apply here. The same is then true as the mode continues growing to $\delta B\gtrsim B_{0}$ (but $\delta B\lesssim (\beta_0/\overline{\nu}_{c})^{1/2 } B_{0}$); it forces $\Delta p$ to the 
mirror limit everywhere in space, implying the mode is not strongly affected by the pressure-anisotropy nonlinearity (see \eqref{eq: pressure anisotropy dB}).

In the high-collisionality regime,  $\overline{\nu}_{c} \gg  \beta_{0}^{1/2}$, the heat fluxes do not
significantly smooth spatial variation in $\Delta p$ and we must modify various aspects of the conclusions
from the previous section.
The pressure anisotropy that develops from the growing mode (for $\delta B \ll B_{0}$) is now spatially inhomogenous:
\begin{equation}
\Delta =\frac{1}{\nu_{c}} \hat{\bm{b}} \hat{\bm{b}}\dbldot\grad \bm{u} \sim \frac{1}{\overline{\nu}_{c}}\frac{\delta B_{0}^{2}}{B_{0z}^{2}} \, \erm^{2 \gamma t }\cos^{2}(kz). 
\end{equation}
 This implies that, as the pressure anisotropy first reaches the mirror limit in some regions of space (near 
 the antinodes of the mode, where ${\rm d}_t \ln B$ is largest), it also changes the shape of the mode, {\it viz.}, 
 it couples different Fourier components of $\delta \bm{B}$ and $\delta \bm{u}$. This causes minor modifications
 to the shape of the mode, which are shown in figure \ref{fig:Bz MRIs}{\it c} for $\overline{\nu}_{c}=\beta_{0}^{3/4}$ (i.e., in the middle of the high-collisionality regime; the shape changes at other $\overline{\nu}_{c}$ and $\beta_{0}$ are generally similar to this). As the mode
 grows further, $\Delta p$ becomes limited by the mirror instability ($\Delta p\propto B_{0}^{2}+ \delta B^{2}$) across a larger region of space, 
 implying (by the arguments above) that the mode regains its sinusoidal shape (albeit briefly, see next paragraph).
 
 There is one final effect, not included in the collisionless discussion,  which  occurs in both the moderate-collisionality and high-collisionality Braginskii regimes at large mode amplitudes, $\delta B\gtrsim (\beta_{0}/\overline{\nu}_{c})^{1/2}B_{0}$.
 The difference compared to the collisionless case arises because, in the Braginskii  regime, $\Delta p$ is proportional to the \emph{current}
 value of $\hat{\bm{b}}\hat{\bm{b}}\dbldot\grad\bm{u}$ rather than to its time history. Once $\delta B\gtrsim B_{0}$, this value $\hat{\bm{b}}\hat{\bm{b}}\dbldot\grad\bm{u}\approx - S\delta B_{x}\delta B_{y}/\delta B^{2}$ becomes
 constant in time, despite the fact that $B^{2}\approx \delta B^{2}$ is growing. Thus, $\Delta p$ moves 
 back \emph{below} the mirror limit when $\delta B^{2}\gtrsim p_{0}/\overline{\nu}_{c} $, {\em viz.}, when $\delta B\gtrsim (\beta_{0}/\overline{\nu}_{c})^{1/2}B_{0}$. This occurs before compressibility affects the mode (at $\delta B\sim \beta_{0}^{1/2}B_{0}$), and causes the pressure anisotropy to vary in space, which in 
 turn modifies the shape of the mode. These modifications are very minor, even at very high $\beta_{0}\sim 10^{5}$ and high $\overline{\nu}_{c}$ (not shown), and so it seems unlikely that they should modify mode saturation in 3D in any significant way. As the  amplitude approaches $\delta B\sim \beta_{0}^{1/2}B_{0}$, the mode is affected by compressibility in exactly the same way as is the  collisionless (or MHD) MRI (see figure \ref{fig:Bz MRIs}{\it d}).

%
%
\begin{figure}
\begin{center}
\includegraphics[width=0.52\columnwidth]{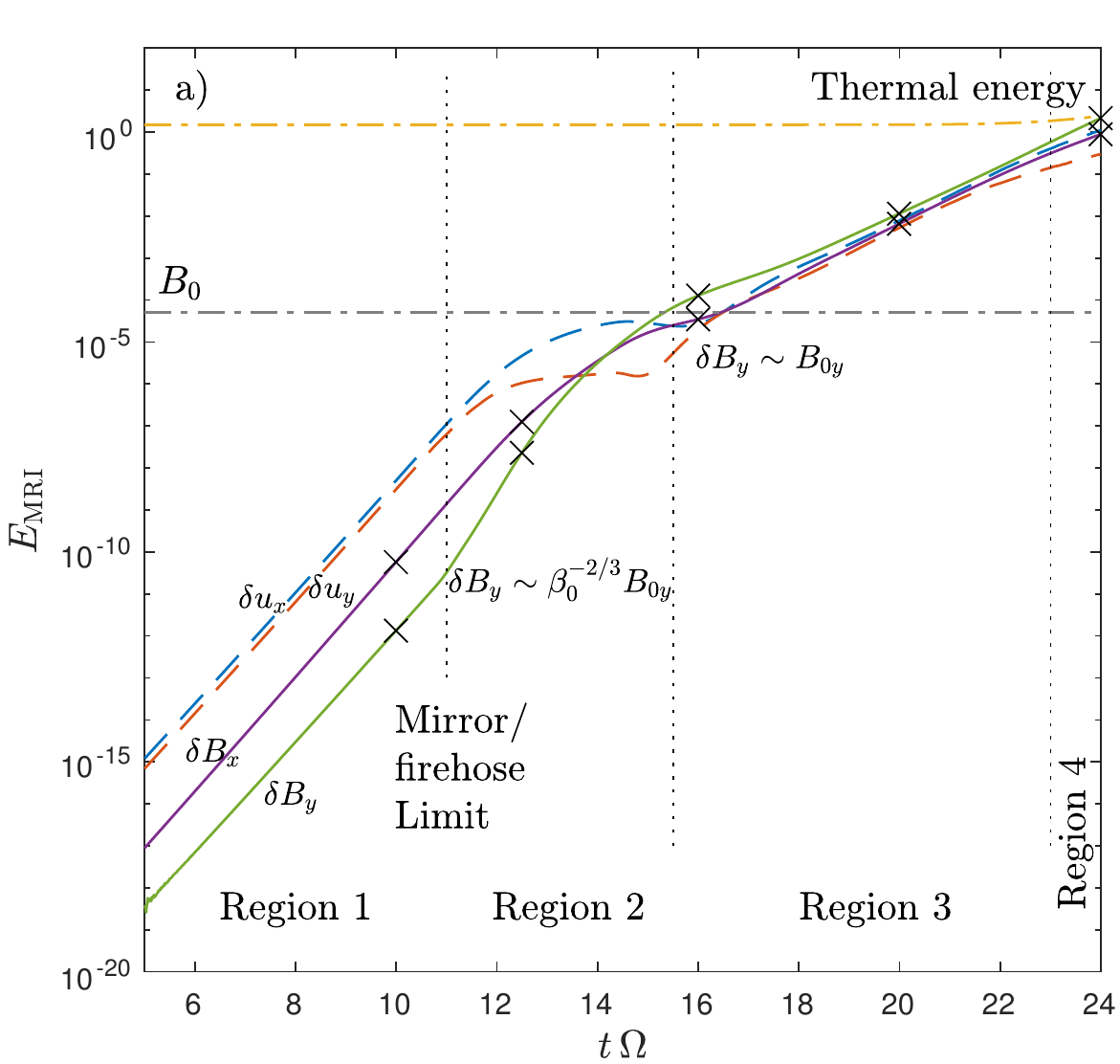}\includegraphics[width=0.48\columnwidth]{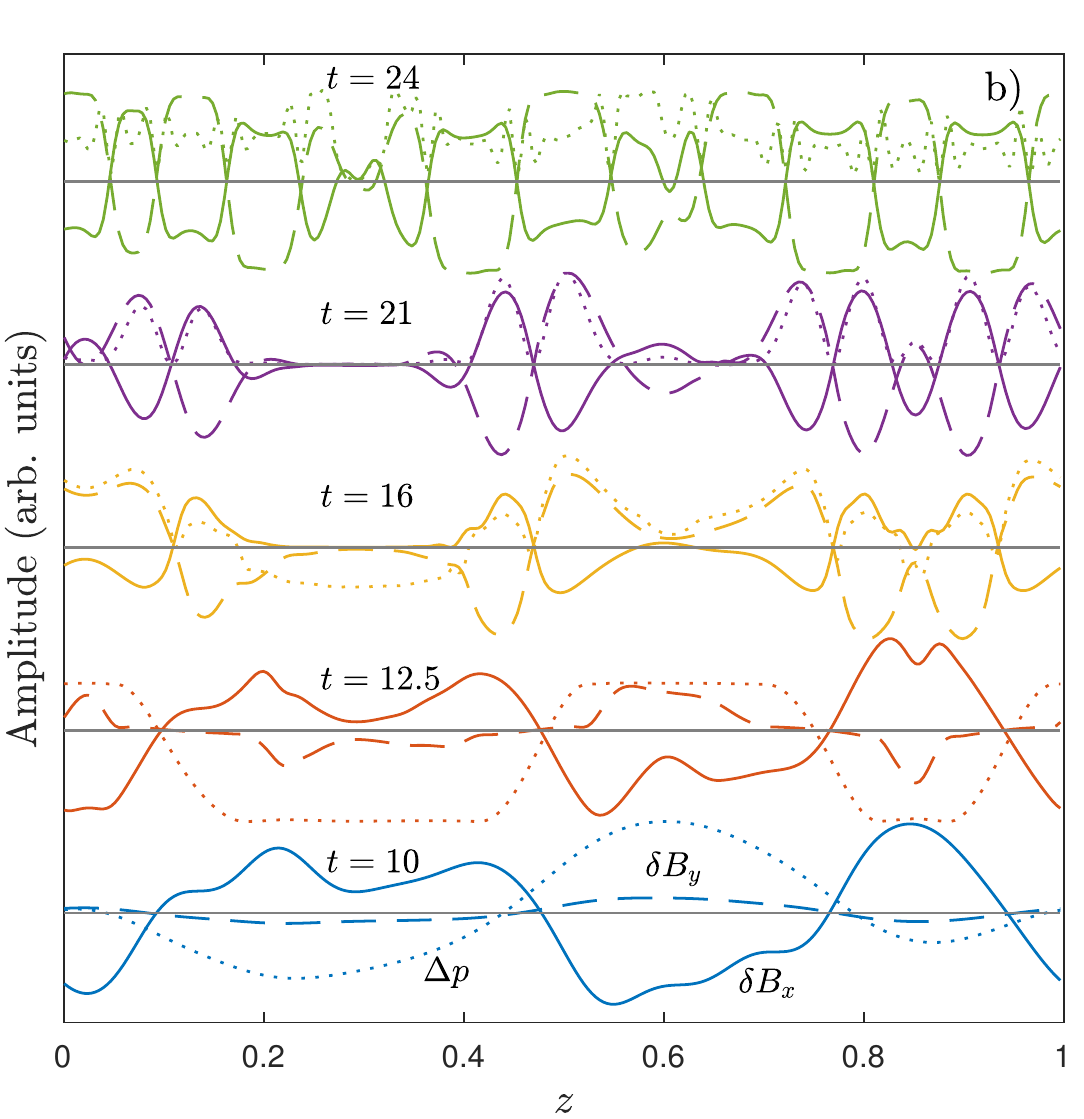}
\caption{(a) Energy evolution of each component of the growing KMRI mode in a mixed azimuthal-vertical field with $B_{0y}=B_{0z}$, at $\beta_{0} = 5000$ ($\beta_{0z}=10000$) in a domain such that $\Omega L_{z}/c_{s}=1$. We show $\delta {B}_{x}$ (solid purple line; $E_{\mathrm{MRI}}=\int dz\, \delta B_{x}^{2}/8\upi$), $\delta {B}_{y}$ (solid green line; $E_{\mathrm{MRI}}=\int dz\, \delta B_{y}^{2}/8\upi$),  $\delta {u}_{x}$ (dashed blue line; $E_{\mathrm{MRI}}=\int dz\, \rho \delta u_{x}^{2}/2$), and $\delta {u}_{y}$ (dashed red line; $E_{\mathrm{MRI}}=\int dz\, \rho \delta u_{y}^{2}/2$). The calculation, which uses the 1-D LF model \eqref{eq:KMHD rho}--\eqref{eq:GL heat fluxes ql} with $\nu_{c}=0$, is initialized with random Fourier amplitudes, scaled by $k^{-2}$ (initial phase of evolution not shown for clarity). For comparison, we also show the thermal energy (yellow dot-dashed line) and the energy of the background magnetic field (grey dot-dashed line). 
Following the linear  phase with large growth rate (Region 1), the linearly perturbed
pressure anisotropy reaches the mirror and firehose limits when $\delta B_{y}\sim \beta_{0}^{-2/3} B_{0y}$, $\delta B_{x}\sim \beta_{0}^{-1/3} B_{0y}$. There follows a transition phase (Region 2) in which the perturbed pressure 
anisotropy can no longer contribute to the instability and the mode moves
to the much shorter wavelengths characteristic of the standard MRI. Then, 
once $\delta B_{y}>B_{0y}$, the mode grows similarly to the vertical-field KMRI (Region 3) with $\Delta$ at 
the mirror limit, until finally it is affected by compressibility in the same way as illustrated in 
figure \ref{fig:Bz MRIs}{\it d} (Region 4). (b) Spatial structure of the azimuthal-field KMRI mode at a variety of times corresponding to ``$\times$'' markers in panel (a), which 
are chosen to illustrate the different phases of evolution. At each time, offset on the vertical axis for clarity with times listed in units of $\Omega^{-1}$, we show $\delta B_{x}/\max(\delta B)$ with solid lines,  $\delta B_{y}/\max(\delta B)$ with dashed lines, and  $ \Delta p/\max(|\Delta p|)$ with dotted lines  (the grey lines show $0$ to more clearly separate each curve). The mode transitions 
(around $t\approx 16\Omega^{-1}$) from structures characteristic of the azimuthal-field KMRI with $\Delta p$ both 
positive and negative, to those characteristic of the MHD-like vertical-field MRI, with the pressure anisotropy everywhere positive and at the mirror limit. Although  less clean than the single-mode case studied in figure \ref{fig:Bz MRIs}, the structures at very late times ($t=24$) are again affected 
by compressibility in the same way (cf., $\delta B_{x}$ and $\delta B_{y}$  with those shown in figure \ref{fig:Bz MRIs}{\it d}).}
\label{fig:By NL}
\end{center}
\end{figure}

\subsection{Nonlinear KMRI: Azimuthal-vertical magnetic field}\label{sec: 1D azim}

In this section, we consider the effect of an additional background azimuthal magnetic field. We focus mainly
on the collisionless KMRI (LF model; \S\ref{subsub: LF with By}), briefly mentioning the Braginskii version (for which the conclusions are  similar) in \S\ref{subsub: Braginskii with By}.
As shown in figure \ref{fig:linear}, the MRI (or MVI) under such conditions is significantly different
from the vertical background field case, growing fastest
at long wavelengths $k_{\mathrm{max}} v_{Az}/\Omega \ll 1$ with growth rates exceeding $S/2$.
This situation is arguably more relevant astrophysically than is the pure vertical-field case: at high $\beta_0$, even small azimuthal fields significantly modify the dispersion 
relation (see Appendix \ref{app: AKMRI linear properties}). 

\subsubsection{Collisionless (LF) regime}\label{subsub: LF with By}

As with the vertical-field KMRI, nonlinear effects become important in the collisionless case at very low amplitudes, specifically
 when $\delta B_{y} \sim \beta_{0}^{-2/3}B_{0}$ (or when $\delta B_{x}\sim \beta_{0}^{-1/3}B_{0}$, see \eqref{eq:app: Bx mode amp MVI}--\eqref{eq:app: uy mode amp MVI}). 
Once the perturbed field becomes larger than the background
 field,   the magnetic field is dominated by the mode itself and 
 the instability  again
 becomes similar to the MHD MRI (or, equivalently, the vertical-field KMRI at large amplitudes).

We now describe how such a mode transitions through four distinct stages in its nonlinear
evolution, which is  illustrated schematically in figure \ref{fig:By NL}.  Let us consider
each stage of evolution separately, assuming $B_{0y}\approx B_{0z} = B_{0}/\sqrt{2}$ for the sake of simplicity:
\begin{description}
\item[1. Linear evolution ] {Unlike in the vertical-field case, the KMRI mode linearly produces a pressure anisotropy,
\begin{equation}
\Delta \approx \sqrt{2\upi}\frac{1}{c_{s} k} \frac{\partial_{t}\delta B_{y}}{B_{0}} \approx \sqrt{2\upi} \frac{\gamma}{c_{s} k} \frac{\delta B_{y}}{B_{0}} ,\label{eq:By linear dP}
\end{equation}
because $\delta \bm{B}\bcdot \bm{B}_{0}=\delta B_{y}B_{0y}\neq 0$  (here we have set $B_{0y}\approx B_{0z}$; see \citetalias{Quataert:2002fy} and Appendix~\ref{app: AKMRI linear properties}). This
implies that, as the mode evolves, it pushes the plasma towards both the mirror limit, in regions where 
$\delta \bm{B}\bcdot \bm{B}_{0}>0$ (i.e., $\delta B_{y} B_{0y}>0$), and the firehose limit, in regions where $\delta \bm{B}\bcdot \bm{B}_{0}<0$ ($\delta B_{y} B_{0y}<0$).
Note that the factor $\gamma/(c_{s}k)^{-1}\sim \beta_{0}^{-1/2} (v_{A}k/\gamma)^{-1}$ in equations \eqref{eq:By linear dP} arises due to the smoothing effect of the heat fluxes  and reduces $\Delta$ significantly
(for example, in the CGL model where $q_{\perp}=q_{\parallel}=0$, $\Delta\approx 3\delta B_{y}/B_{0}$ is much larger than \eqref{eq:By linear dP}).
The linear  phase ends as $\Delta$ approaches the microinstability limits $|\Delta|\sim \beta^{-1}$ and becomes flattened by growing mirror and firehose fluctuations. 
Assuming the mode grows at the scale that maximizes the growth rate, $k_{\mathrm{max}} v_{A}/\Omega \sim \beta_{0}^{-1/6}$ (see \eqref{eq:app: max k MVI} and figure~\ref{fig: maximizing k for MVI}), these limits are reached when  $\delta B_{y} \sim \beta^{-2/3} B_{0}$, or when $\delta B_{x} \sim \beta^{-1/3} B_{0}$ (since $\delta B_{x}\sim \beta_{0}^{1/3}\delta B_{y}$ for these fastest growing modes; see \eqref{eq:app: Bx mode amp MVI}--\eqref{eq:app: uy mode amp MVI}).  At this 
point, the perturbation of $B$ due to $\delta B_{x}$ is similar to that due to $\delta B_{y}$, meaning that both components contribute  to the  pressure anisotropy\footnote{There is a minor ambiguity 
here because, while the perturbation to $B$ due to $\delta B_{y}$  perturbs $\Delta p$ in both 
the positive (mirror) and negative (firehose) directions, that due to $\delta B_{x}$  perturbs 
$\Delta p$ only in the positive direction (it is proportional to $\delta B_{x}^{2}$). A priori, it is thus unclear  whether 
the system will always reach the firehose limit, or whether the decrease in $B_{y}$ can be offset by the increase in $B_{x}$. However, it seems that once the nonlinearity starts becoming important, the rate of change of $\delta B_{y}$ 
increases sufficiently fast  (while that of $\delta B_{x}$ slows; see figure~\ref{fig:By NL}{\it a} around $t\Omega\approx12$) so as to  cause the contribution from $\delta B_{y}$ to dominate and $\Delta p$ to reach the firehose. Various tests,  similar to that shown in figure~\ref{fig:By NL} but with different $\beta_{0}$ and $B_{0y}/B_{0z}$, have confirmed that this  picture holds so long as $\beta_{0}$ is sufficiently large. }.
}
\item[2. Pressure anisotropy limited, with $\bm{\delta B_{y}< B_{0y}}$ ] {In the limit 
that the mirror and firehose fluctuations efficiently constrain the growing $|\Delta|$, the pressure
profile will  develop a step function profile  in space
between the mirror limit, $\Delta \approx 1/\beta$, and the firehose limit $\Delta \approx -2/\beta$ (see figure \ref{fig:By NL}{\it b} at $t=12.5\Omega^{-1}$).
A key effect of these limits is that they  suppress the
influence of the $\delta p_{\perp}$ and $\delta p_{\|}$ perturbations on the mode evolution. Without 
such pressure perturbations, the MRI reverts back to standard, MHD-like behaviour characteristic 
of the vertical-field MRI (this can be seen, for example, by artificially suppressing $\delta p_{\perp,\|}$ perturbations
in a calculation of the $B_{0y}\neq 0$ KMRI dispersion relation). 
Specifically, 
the growth rate approaches zero for $k v_{A}\lesssim \Omega$ and  peaks at smaller scales $k v_{A}\sim  \Omega$.
Thus, the long-wavelength linear modes are significantly
stabilized (i.e., $\gamma$ at low $k$ is  small) and the mode moves to 
shorter wavelengths.  Such behaviour is expected intuitively because the azimuthal pressure
force, which is the cause of the enhanced low-$k$ linear growth rate (\citetalias{Quataert:2002fy}; \citealp{Balbus:2004hg}), is limited by the mirror and firehose fluctuations. 
Because the standard MRI 
grows with the perturbed 
energies in approximate equipartition, $\delta B_{y}$   
must ``catch up'' to $\delta B_{x}$, and both of these must
catch up to the velocity perturbations (which grow linearly with $\delta u_{x}\sim \delta u_{y}\sim \beta_{0}^{1/2}\delta B_{y}$) during this phase  of evolution (in other words, $\delta \bm{u}$ grows more slowly than $\delta \bm{B}$).
  This picture is confirmed by 1-D numerical calculations, with the growth 
of $\delta \bm{u}$ decreasing significantly as shorter wavelengths take over; see figure \ref{fig:By NL}{\it a}, Region 2. 
We also see smaller-scale perturbations growing on top of the longer-wavelength mode in figure \ref{fig:By NL}{\it b} at $t\Omega = 12.5$ (e.g., around $z=0.2$ and $z=0.85$), particularly in those regions at the firehose limit where the MRI preferentially grows at 
smaller scales because $\Delta p<0$.
Note that the mode cannot be stabilized completely  during this phase because
the MRI growth rate on a background $\Delta p$, though small, is nonzero as $k\rightarrow 0$. }
\item[3. $\bm{\delta B_{y} > B_{0y}}$ ]{As the amplitude of $\delta B$ grows larger than the background
$B_{0y}$ field, the mode enters a second phase of nonlinear evolution. With the field-line direction
dominated by the perturbation, the presence of a background $B_{0y}$ loses its 
dynamical importance and the mode behaves similarly to the  vertical-field case, growing at $k v_{A}\sim \Omega$. In addition, because the magnitude of the magnetic
field is now growing everywhere in space, $\Delta$ becomes everywhere positive  
and will be limited only by the mirror instability; see figure \ref{fig:By NL}{\it b} at $t=21\Omega^{-1}$. As with the 
vertical-field MRI, when the perturbation amplitude dominates the total $B$, the pressure anisotropy $\Delta p \approx (\delta B_{x}^{2}+\delta B_{y}^{2})/8\upi$ does not cause significant nonlinear modifications, because $\delta \bm{B}\bcdot \grad \delta \bm{B} = 0$ for an MRI channel mode.}
\item[4. Compressibility effects ]{There is no 1-D
mechanism to halt the growth of the mode until $\delta u$ approaches the
sound speed ($\delta {B}_{x}\sim\delta B_{y}\sim \beta^{1/2} {B}_{0}$). Thus, the mode behaves similarly to the vertical-field MRI, with large variations in $\rho$.  The profiles that develop have the same 
characteristic shape as those seen in figure \ref{fig:Bz MRIs}{\it d} (cf., figure \ref{fig:By NL}{\it b} at $t\Omega=24$).}
\end{description}

\vspace{1ex}
Each of the four stages discussed above can be seen in the mode
energy evolution, 
shown in figure \ref{fig:By NL}{\it a}. In contrast to the calculation of the
vertical-field MRI evolution (figure \ref{fig:Bz MRIs}), we 
begin from random large-scale initial conditions
at much higher $\beta_{0} = 5000$ (with $B_{0y}=B_{0z}$), so as
to clearly distinguish between the different regions of evolution and
allow the smaller scale modes to grow after $\Delta$ reaches the microinstability limits. 
While the details of the process described above will be modified depending on a variety of factors 
-- e.g., the mode wavelength in comparison to the domain, the spectrum of the initial conditions, and the value of $B_{0y}/B_{0z}$ -- 
the basic concepts and phases of evolution should be generally applicable.
It is also worth noting that, at the large values of  $\beta$ for which our arguments are most applicable, 
the mode will likely collapse into turbulence before the  final nonlinear stage where compressibility is important (see \S\ref{sec:3D}).

\subsubsection{Weakly collisional (Braginskii) regime}\label{subsub: Braginskii with By}

As with the vertical-field MRI in the Braginskii regime (\S\ref{sec: 1D brag}), there are two different
Braginskii sub-regimes: (i) if $\overline{\nu}_{c}\equiv \nu_{c}/S \ll \beta^{1/2}$, the instability grows at a similar rate
to the collisionless instability (and the heat fluxes play an important dynamical role), or (ii) if $\overline{\nu}_{c}\ \gg \beta^{1/2}$, the growth rate of the MVI is reduced, reaching the collisional (standard MHD) regime when $\overline{\nu}_{c}\sim \beta$ (see figures 1--3 of \citealt{Sharma:2003hf}).
To understand this latter point -- that the Braginskii MRI becomes MHD-like when $\overline{\nu}_{c}\gtrsim \beta$ -- we compare  the Lorentz force, $\bm{B}_{0}\bcdot \grad \delta \bm{B}=\grad\bcdot (\bm{B}_{0}\delta \bm{B})$, to the pressure-anisotropy force  
$\grad\bcdot (\hat{\bm{b}}\hat{\bm{b}}\Delta p)$ that arises from the (linear) $\Delta p$ induced by the KMRI mode,
\begin{equation}
 \Delta p \sim \frac{p_{0}}{\nu_{c}}\frac{1}{B} \D{t}{B} \sim \frac{p_{0}}{\nu_{c}}\frac{1}{B_{0}} \D{t}{\delta B} \sim \frac{ \beta}{{\nu}_{c}}  \, \gamma B_0 \delta B(t) , \label{eq: Brag By Dp}
\end{equation}
where $B_{0}^{2}=B_{0y}^{2}+B_{0z}^{2}$ and we have assumed $B_{0y}\sim B_{0z}$. Note 
that we have neglected the heat fluxes in \eqref{eq: Brag By Dp}, which act to reduce $\Delta p$ (see \eqref{eq:By linear dP}), as is appropriate for the
high-collisionality regime (see appendix \ref{app:sub: Braginskii}).
We see from \eqref{eq: Brag By Dp} that the pressure-anisotropy stress  is larger than the Lorentz force (and thus important for the mode's evolution) only if $\overline{\nu}_{c}\lesssim  \beta$, as expected \citep{Sharma:2003hf}.

Numerical experiments (not shown) have revealed that, in the moderate-collisionality regime $\overline{\nu}_{c}\equiv \nu_{c}/S \ll \beta^{1/2}$, the Braginskii KMRI behaves 
similarly to a truly collisionless mode (this should be expected based on the arguments above and in \S\ref{sec: 1D brag}). As described in \S\ref{sec: 1D brag}, at very large amplitudes, $\delta B>B_{0}$  (Region 3), the evolution differs from a  collisionless mode once $\delta B\gtrsim (\beta_{0}/\overline{\nu}_c)^{1/2}B_{0}$ (the growing mode can no longer sustain $\Delta p$ at the mirror limit). 
In the high-collisionality regime,  $\overline{\nu}_{c}\gtrsim \beta^{1/2}$, the linear mode itself transitions 
back to the standard MHD MRI, and much of the discussion above loses its relevance. In particular, the
most-unstable mode moves to larger scales as $\overline{\nu}_{c}$ increases (see figure 2{\it b} of \citealt{Sharma:2003hf}). This implies  that the sudden reduction in the scale of the mode when $\Delta p \sim B^{2}$ (see figure \ref{fig:By NL}{\it b} between $t\approx 12.5$ and $t\approx 21$) is much less relevant. The mode 
 causes $\Delta p$ to reach the mirror and firehose limit  at some amplitude $\delta B\lesssim B_{0}$ (with the exact point depending on $\overline{\nu}_{c}/\beta$) and then transition to the behaviour discussed in \S\ref{sec: 1D brag} once $\delta B\gtrsim B_{0}$.

\subsection{General discussion of  1-D nonlinearities}\label{sub:1-D conclusions}

With the diverse assortment of nonlinear effects outlined in the previous sections, 
it seems prudent to conclude with a discussion of some overarching ideas.
As is generally the case in high-$\beta$ collisionless plasma
dynamics \citep[e.g.][]{Schekochihin:2006tf,Squire:2016ev}, nonlinearity can be important for perturbation amplitudes far below what one might naively expect.
For the MRI in the collisionless regime, this occurs a factor of $\sim$$\beta_{0}$ (or $\sim$$\beta_{0}^{7/6}$ for $\delta B_{y}$ with $B_{0y}\neq 0$)
below where nonlinear effects become important  in standard MHD. 
However, we have also seen that, in all cases
considered, the KMRI (or MVI)  always reverts to MHD-like evolution at large amplitudes due to the anisotropy-limiting response of the mirror and firehose instabilities \citep[e.g.][]{Kunz:2014kt}. This behaviour 
has also been seen in fully kinetic 2-D  simulations \citep{Riquelme:2012kz}. More specifically, this  
arises because  of the form of the pressure anisotropy, {\em viz.}~$\Delta p = B^{2}/8\upi$, when it is limited at the mirror threshold. If
$B$ is dominated by the mode itself ($B^{2}\approx \delta {B}^{2}$), then the anisotropic stress 
in the momentum equation (\ref{eq:KMHD u}), 
\begin{equation}
\grad \bcdot (\Delta p \hat{\bm{b}}\hat{\bm{b}}) = \frac{1}{8\upi}\grad \bcdot (B^{2} \hat{\bm{b}}\hat{\bm{b}}) \approx \frac{1}{8\upi} \delta \bm{B}\bcdot \grad \delta \bm{B},
\end{equation}
acts like an additional Lorentz force, which  is zero for the MRI channel mode \citep{Goodman:1994dd}. This  implies that the pressure-anisotropy effects that are critical to the difference between the KMRI and the MRI in linear theory become unimportant once $\delta B\gtrsim B_{0}$. Due to this, a large-amplitude collisionless KMRI channel mode is an approximate nonlinear solution (of the LF equations), as in MHD, until its amplitude approaches the sound speed (there is a minor complication in the Braginskii regime; see  \S\ref{sec: 1D brag}). Thus, the effect of compressibility on the large-amplitude mode structure  -- if this occurs before breakdown into turbulence --
looks nearly identical in the MHD and kinetic models, with either a  Braginskii or LF closure,  and with or without azimuthal fields (this structure is shown in figure \ref{fig:Bz MRIs}{\it d}). Very similar structures are also seen at large amplitudes in fully kinetic simulations \citep{Kunz:2014hm}. 

In contrast, the effect of the nonlinearity at intermediate amplitudes $\delta B\lesssim B_{0}$ differs between plasma regimes (collisionless and
Braginskii) and background field configurations.
For modes growing in a purely vertical background field, the nonlinear modification of the
mode is  modest, so long as the mirror fluctuations 
limit $\Delta$ to be close to $1/\beta$. However, even moderate overshoot past the mirror limit -- for example, as might occur in numerical simulations due to insufficient scale separation between $\Omega_{i}$ and $\Omega$, see \S\ref{sec:implications} -- can cause quite strong modifications and/or cause the mode to move to longer wavelengths.
Modes evolving in a mixed vertical-azimuthal background field behave very differently, 
because  the pressure perturbation participates directly in the linear instability \citepalias{Quataert:2002fy}. Upon
reaching the mirror and firehose microinstability limits, the pressure perturbation can no longer contribute to the linear growth and
the mode moves towards the smaller scales characteristic of the standard MRI
once $\delta B_{y} \gtrsim \beta^{-2/3}B_{0}$ (or $\delta B_{x}\gtrsim \beta^{-1/3}B_{0}$). This transition is illustrated graphically in figure \ref{fig:By NL}.

\section{Saturation into turbulence}\label{sec:3D}

In the previous section we have argued that kinetic physics does 
not offer alternate routes for the saturation of MRI modes in one dimension. In particular, 
in all cases considered -- purely vertical or mixed vertical-azimuthal field, 
in both the Braginskii and collisionless limits -- the final stages of mode evolution 
are  similar to those seen in standard MHD, with the growing 
mode becoming close to a nonlinear solution (aside from the 
effects of compressibility).  We must therefore consider 
alternate means for the saturation of the MRI, in particular 3-D turbulence. This
conclusion is supported by the fully 
kinetic simulations of \citet{Riquelme:2012kz} and \citet{Kunz:2014hm}, in which the 2-D KMRI was seen to grow to 
very large amplitudes.

In this section, we are concerned with how a 
growing mode  breaks up at large amplitudes into 3-D turbulence. 
This problem is somewhat separate from the study of the turbulent 
state itself (which we do not consider here), 
and has been studied by a variety of authors 
in terms of ``parasitic modes'' (e.g., \citealt{Goodman:1994dd,Pessah:2009uk,Latter:2009br}).
These are 3-D secondary instabilities
that feed off the large gradients in the growing MRI channel 
mode, acting to disrupt the mode and seed its transition into turbulence.
While the relevance of parasitic modes to  
  transport in the turbulent saturated state has been controversial (e.g., \citealt{Bodo:2008fr,Pessah:2009uk,Latter:2009br,Longaretti:2010ha}),
 they are nevertheless a helpful theoretical tool for understanding the
 initial saturation phase.

 The question we address here is whether one should expect any 
 striking differences (compared to MHD) in this initial saturation phase of the KMRI because of 
 differences in the behaviour of parasitic modes brought about by pressure anisotropy. Our conclusion -- within 
 the limitations of the LF model (\S\ref{sec:equations etc.}) -- is that there are \emph{not} significant differences. 
We also argue
 that the observations of larger transient channel amplitudes in \citetalias{Sharma:2006dh} are explained through the modes'
 increase in wavelength at large amplitudes due to 1-D pressure-anisotropy nonlinearities  (see \S\ref{sec: 1D vertical}).

Because of this null result, and given the simplifications inherent to our fluid-based model, we keep our discussion
 brief.  We do, however, reach these conclusions through two separate methods: (i) a 
 study of the effect of a mean $\Delta p$ on linear parasitic-mode growth rates in a sinusoidal channel mode, and (ii) 
 3-D nonlinear simulations using the modified version of the {\sc Zeus} code from \citetalias{Sharma:2006dh}. 
 We thus  feel that the general conclusions reached here are relatively robust. 
That being said, due to the variety of other effects that may be present in a true 
collisionless kinetic plasma, as well as 
the strong dependence of 
MHD MRI turbulence on microphysics (e.g., magnetic Prandtl number; \citealt{Fromang:2007cy,Meheut:2015it}),
we  do not necessarily claim that the initial stages of 3-D KMRI saturation should be similar to its 
MHD counterpart. Rather, our conclusion is more modest:  there
are no significant differences due to 
pressure anisotropy and heat fluxes (i.e., those kinetic effects contained within the LF model).

%
%
\begin{figure}
\begin{center}
\includegraphics[width=1.0\textwidth]{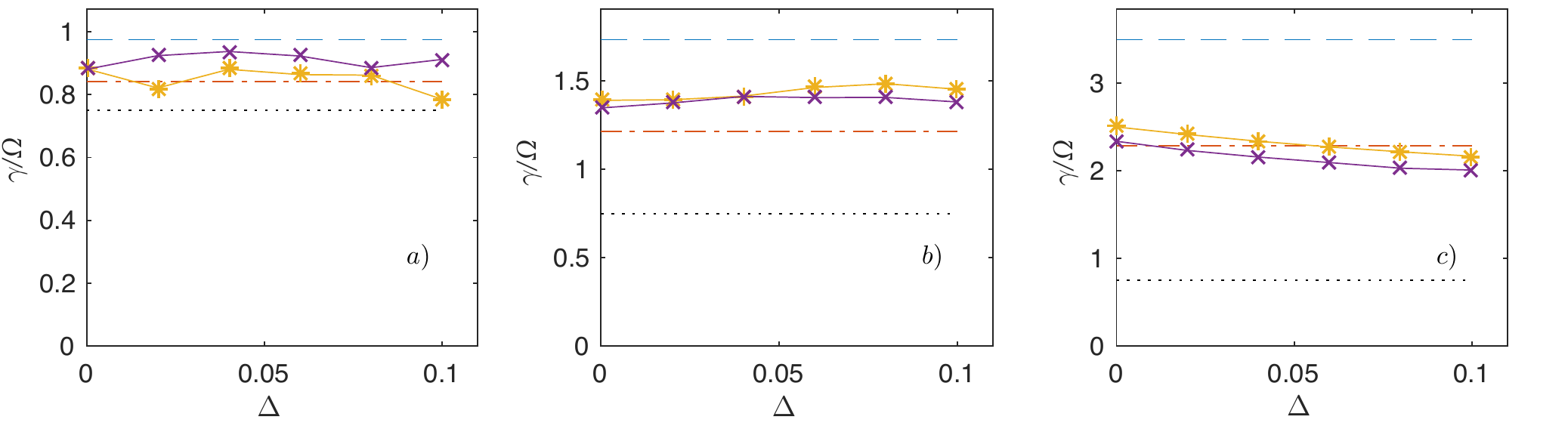}
\caption{Maximum parasitic growth rates $\gamma/\Omega$ as a function of $\Delta = (\delta p_{\perp0}-\delta p_{\|0})/p_{0} $ for $\beta_{0} \approx 90$ ($B_{0z}/\sqrt{4 \upi \rho_{0}} = 0.15c_{s}$) for (a) $\delta B_{0}/\sqrt{4 \upi\rho_{0}} =0.5c_{s} \approx 3.3 B_{0z}$, (b) $\delta B_{0}/\sqrt{4 \upi\rho_{0}} = c_{s} \approx 6.7 B_{0z}$, $\delta B_{0} /\sqrt{4 \upi\rho_{0}}= 2c_{s} \approx 13.3 B_{0z}$.  
Note that $\Delta  =0.1$ would 
correspond to a plasma fixed at the mirror limit in a constant background field $B_{0} \approx \sqrt{8\upi \Delta p_{0}}\approx 0.45 c_{s }\sqrt{4 \upi\rho_{0}}$ (but note that the channel mode field varies sinusoidally). In each figure the yellow 
stars illustrate the growth rate in KMHD without heat fluxes (i.e., CGL, $q_{\perp }=q_{\parallel}=0$), while
purple crosses illustrate KMHD growth rates including the simple model Eq.~\eqref{eq:simp heat fluxes} for the heat 
fluxes. For comparison, the dashed blue line shows the incompressible MHD result and the dash-dotted red line shows the 
isothermal compressible  MHD result (each at $\Delta= 0)$. The dotted black line is the MRI channel mode
growth rate $\gamma/\Omega = 3 /4$. Evidently, the variation of $\gamma$ with $\Delta$ is modest, and is probably  
too small to be of much consequence to MRI saturation. }
\label{fig:Parasites15}
\end{center}
\end{figure}

%
%
\begin{figure}
\begin{center}
\includegraphics[width=1.0\textwidth]{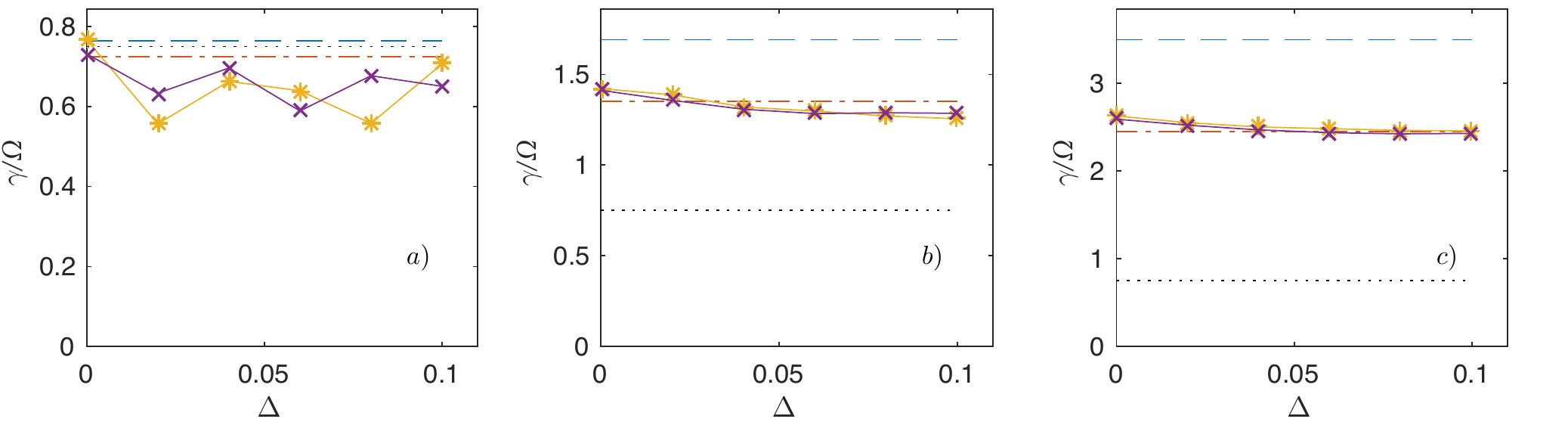}
\caption{As in figure \ref{fig:Parasites15} but with a background field $\beta_{0} \approx 800$  ($B_{0z}/\sqrt{4 \upi \rho_{0}} = 0.05c_{s}$). 
Although a growing MRI mode would have a shorter wavelength at this $B_{0z}$, which will make the parasitic 
modes more unstable at a given amplitude due to the larger gradients, we choose to keep the same $k = 2\upi/L_{z}$ 
as figure \ref{fig:Parasites15} to provide a direct comparison (recall also from \S\ref{sec:1D}  that the mode wavelength can increase during evolution). }
\label{fig:Parasites05}
\end{center}
\end{figure}

\subsection{Linear parasitic mode growth rates}\label{sub:Parasitics}

In this section we directly calculate parasitic mode growth rates for MRI and KMRI channel modes in a vertical background magnetic field.  We do not consider a mixed azimuthal-vertical background field configuration here primarily 
because the 1-D results of \S\ref{sec: 1D azim} suggested that such modes are always relatively disordered when they reach large amplitudes anyway, due to their strong nonlinear disruption (from long wavelengths to short wavelengths) when $\delta B_{y}\sim B_{y}$ (see, e g., figure~\ref{fig:By NL}b). Thus the very idealized linear problem, based on purely sinusoidal background profiles, is presumably much less relevant for this case (we rectify this omission in the 3-D simulations below; see figure~\ref{fig:ZEUS runs}b).

Motivated by previous MHD studies \citep{Goodman:1994dd,Pessah:2009uk,Latter:2009br,Latter:2010iz},
we consider 3-D linear perturbations, $f(\bm{x}) = f_{k_{x}k_{y}}(z)\exp(\imag k_{x}x+\imag k_{y}y)$ for $f = \{\bm{u}',\,\bm{B}',\,\rho',\,p_{\perp,\parallel}'\}$, evolving on top of
 a channel-mode background ($\delta \bm{u}$ and $\delta \bm{B}$ from \eqref{eq:channel profile},
with $\delta B_{0}$ a free parameter). That is, we decompose the fields as 
\begin{alignat}{5}
\bb{u} &{}= 		-Sx\ey 	&&{}-{} 	\sqrt{\frac{3}{5}} \frac{\delta B_0}{\sqrt{4\upi\rho}} \sin(kz) (\ex + \ey)	&&{}+{} \bb{u}', \nonumber\\*
\bb{B} &{}= 		B_{0} \ez 	&&{}-{} 	\delta B_0 \cos(kz) (\ex-\ey) 								&&{}+{} \bb{B}',\nonumber\\*
\rho &{}= 			\rho_0 	&&{}+{} 	0 													&&{}+{} \rho',\nonumber\\*
p_\perp &{}= 		p_0 		&&{}+{} 	\delta p_{\perp 0} 										&&{}+{} p'_{\perp},\nonumber\\*
p_\parallel &{}= 	p_0 		&&{}+{} 	\delta p_{\parallel 0} 										&&{}+{} p'_{\parallel} ,
\end{alignat}
with $k = 2\upi/L_{z}$, and linearize \eqref{eq:KMHD rho}--\eqref{eq:GL heat fluxes ql} in $\bm{u}'$, $\bm{B}'$, $\rho'$, and $p_{\perp,\parallel}'$. (For simplicity, we ignore spatial variation in $\delta p_{\perp,\parallel 0}$;
see discussion below.)
The resulting equations are solved numerically in a box with dimensions\footnote{The fastest-growing parasitic modes generally
have a wavelength several times that of the channel, necessitating a wide box \citep{Bodo:2008fr,Pessah:2009uk}.} $(L_{x},L_{y},L_{z})=(4,4,1)$, on a $16\times 16$ grid of Fourier modes in the homogenous $x$ and $y$ directions, and with a pseudopectral 
method and $64$ modes in the inhomogenous $z$ direction. Hyperviscous 
damping is used to remove energy just above the grid scale. We initialize with random noise 
in all variables, and evolve in time until $t=10\Omega^{-1}$ (by which time the most-unstable parasitic eigenmode mostly dominates). Fitting an exponential to the energy evolution at later times yields  the growth 
rate $\gamma$ of the least-stable parasitic mode.    Intuitively, a large parasitic mode
growth rate should be associated with rapid collapse of the channel mode  into MRI turbulence, because 
the parasitic modes will quickly  ``overtake'' the mode itself, with their 
3-D structure acting as a seed for the turbulence. Further, as the MRI mode
grows (i.e., as $\delta B_{0}$  increases) the parasitic growth rates should increase, 
since there are stronger gradients of $\delta\bm{u}$ and $\delta\bm{B}$ to feed the instabilities.

 To assess the impact of the pressure anisotropy, we apply a spatially constant background pressure anisotropy $\Delta p_{0}=\delta p_{\perp 0}-\delta p_{\parallel 0}$
in the KMHD models and calculate the resulting change in the maximum parasitic growth 
rate with $\Delta p_{0}$. 
A strong variation in   growth rate with $\Delta p_{0}$ would  indicate 
 that the saturation into turbulence is likely to depend sensitively on the self-generated pressure anisotropy, 
and thus differ strongly between collisional and collisionless plasmas.  
Of course, as discussed in \S\ref{sec:1D}, there will be spatial variation 
in $\Delta p$ at large $\delta B_{0}$, so the study here should be considered an approximation to the full problem, considering only the simplest effects. Similarly,  we neglect the 
influence of the background shear on the parasitic modes (this is common in previous MHD analyses), because without this simplification
the resulting time dependence of $k_{y}$ implies that an analysis in terms  of eigenmodes
is incorrect (one should consider transient, or
nonmodal, growth; \citealt{SCHMID:2007bz,Squire:2014es,Squire:2014cz}) This assumption is
not truly valid except at very large mode amplitudes when $\delta \bm{u}$  dominates strongly over the
mean shear, although we expect to capture the correct qualitative trends when the 
parasitic growth rate is  larger than the shearing rate.  
As mentioned above, given that the pressure anisotropy appears to cause little change to growth rates, 
we deliberately keep this section brief, postponing to possible future work the  detailed 
study of the mode structure and morphology (e.g.,  Kelvin-Helmholtz vs.~tearing-mode instability) or the variation of growth rates with $k_{x}$ and $k_{y}$ \citep{Goodman:1994dd,Latter:2009br,Prajapati:2010kh}.

Figures \ref{fig:Parasites15} and \ref{fig:Parasites05} illustrate representative examples of $\gamma$ as a function of 
$\Delta=(\delta p_{\perp0}-\delta p_{\|0})/p_{0}$ and $\delta B_{0}$.
In each case the chosen mode energy  is of order the thermal energy $\delta B_{0}/\sqrt{4 \upi\rho_{0}}\sim c_{s}$ 
and is  larger than the background $B_{0z}$ field. 
Because of this large $\delta B_{0}$, the parasitic growth rates are mostly larger than
 the MRI mode growth rate (dotted line), as required for a parasite to cause 
the channel mode to break up  into turbulence. 
The maximum of $\Delta$ plotted ($\Delta=0.1$) corresponds to a 
plasma that is everywhere at the mirror limit in a constant field of $B_{0} \approx \sqrt{8\upi \Delta p_{0}}\approx 0.45 c_{s}\sqrt{4 \upi\rho_{0}}$,
which is larger than the background $B_{0z}$ in each case 
(but smaller than the maximum $\delta B_{0}$ studied).
The first feature that is evident in figures \ref{fig:Parasites15} and \ref{fig:Parasites05}  is the  
suppression in parasitic-mode growth rates in the kinetic models (solid lines) and compressible 
fluids (red dash-dotted line), in comparison to incompressible fluids (blue dashed line). This property was also noted 
for compressible MHD 
in \citet{Latter:2009br}\footnote{Although this might be expected to lead to
larger saturation amplitudes in compressible MHD in comparison to incompressible MHD, the difference is  
offset by the more angular channel mode profiles that develop in compressible MHD. Because these have 
larger gradients (see  figure \ref{fig:Bz MRIs}{\it d}), this increases
the parasitic growth rate and  approximately 
cancels in the decrease in growth rate due to compressibility, leading to similar saturation behaviour. See the appendix of \citet{Latter:2009br} for more details.}.
Aside from this, we see that the differences between the kinetic models (both with and without heat fluxes)
and MHD are relatively modest\footnote{The apparent scatter at lower $\delta B_{0}$ is  
caused by the random initial conditions and relatively small time ($t=10$) at which
we calculate the growth rate. There can be several modes with similar growth rates (particularly 
$(k_{x}=1,k_{y}=0)$ and $(k_{x}=2,k_{y}=0)$ in figure \ref{fig:Parasites05}{\it b}), which contribute varying amounts
depending on the initial conditions, and thus lead to some scatter in the measured $\gamma$. This 
goes away at higher $\delta B_{0}$ because the much faster growth rates lead to stronger dominance 
of a single mode by  later times.}.  While there is there is a slight tendency for  $\gamma$ to decrease with $\Delta p$ at large mode amplitudes, 
changes in $\gamma$ of this magnitude are unlikely to
 make any notable differences in a nonlinear simulation. Furthermore, 
 there does not appear to be any significant change in behaviour at even higher $\delta B_{0}$ (not shown), 
 which leads us to conclude that  parasitic modes are broadly unaffected by the kinetic effects contained within
 the CGL and LF models. 
   
 Obviously, the results shown in figures \ref{fig:Parasites15} and \ref{fig:Parasites05} cover only a small portion 
 of parameter space in a rather idealized setting. In addition to 
 the results shown, we have calculated growth rates across a much wider parameter
 space in $B_{0z}$, $p_{0}$, $\delta B_{0}$, $\Delta $,  $k$ (the channel mode wavelength), small-scale dissipation coefficients (hyperviscosity, hyperresistivity, and their ratio), and  box dimensions.
 In addition, we have varied the ratio of $\delta \bm{u}$ and  $\delta \bm{B}$ away 
 from that of the fastest growing channel mode (i.e., differently from \eqref{eq:channel profile}),
 as might be caused, for example, by the effects of the self-generated  pressure anisotropy on the mode. 
 Finally, we have also considered parasitic mode evolution on more angular compressible profiles, similar to those 
 shown in figure \ref{fig:Bz MRIs}{\it d} \citep{Latter:2009br}.
In all cases, we have  failed to find
 any significant difference between standard 
 compressible MHD and the CGL or LF models, and
 so we refrain from presenting these results in any detail here. 
Of course, these studies have all assumed a spatially
constant $\Delta $ and $\rho_{0}$ profile, which will certainly change  results quantitatively
in some regimes. It is also possible that there 
are  modes in other regimes (e.g., much longer or shorter wavelength than the KMRI mode), that have not been 
captured by these studies. Nevertheless, we feel that the general conclusion 
that the parasitic modes are mostly insensitive to background pressure anisotropy 
should be robust,
given the wide range of parameter space for which this conclusion appears to hold.

\subsection{Nonlinear simulation}\label{sub: zeus sims}

Our second method for probing parasitic mode behaviour is to simply observe the evolution of a nonlinear 
KMRI channel mode in 3D. The maximum amplitude that such a mode reaches
before breaking up into turbulence should give a simple indication of 
the effectiveness of the parasitic modes. 
We use the modified version of the {\sc Zeus} code described in \citetalias{Sharma:2006dh}, which
 solves \eqref{eq:KMHD rho}--\eqref{eq:KMHD pl} with the LF closure \eqref{eq:GL heat fluxes qp}--\eqref{eq:GL heat fluxes ql} and pressure-anisotropy limiters.

This method is complementary to that described in the previous section:
although it cannot provide  detailed information on individual modes or variation with 
parameters, it is free from some of the caveats of the linear parasitic mode study (for example, the
assumption of spatially homogenous background density  profiles). It also allows us to  consider  the
mixed azimuthal-vertical field KMRI in a moderately realistic settling (as mentioned above, the stronger effects 
of 1-D nonlinearities in this case suggest that an idealized parasitic-mode study is not very worthwhile for the azimuthal-field KMRI).
Most importantly, 3-D simulations  directly address the issue we most care about: is the nonlinear saturation 
significantly different between the kinetic and MHD models?
In  \citetalias{Sharma:2006dh} the authors noted that there \emph{was} a significant difference, 
with MRI modes in kinetic 
 models growing to significantly larger amplitudes before being disrupted, even though
the turbulence itself maintained a similar level of angular momentum transport.
While this
 may appear to be at odds with our findings from the previous section, here we 
argue that this difference is primarily a consequence of the 
1-D effects described in \S\ref{sec:1D}.
Specifically, the positive pressure anisotropy can increase the wavelength of the mode well before 
it reaches saturation amplitudes. This effect  was
caused  by the choice of $\Delta \approx 7/\beta$ for the mirror instability limit in \citetalias{Sharma:2006dh}, which allows 
the anisotropies to have a strong dynamical effect before the mirror limit is enforced. For a given mode amplitude, these
longer wavelength modes are attacked more slowly by the parasites, due to the smaller gradients of
 $\delta\bm{u}$ and $\delta\bm{B}$ \citep{Goodman:1994dd}, thus 
 leading to a larger transient amplitude before the transition into turbulence.
 
 Our studies have deliberately  used a  setup that is similar to \citetalias{Sharma:2006dh}. We initialize with 
 random noise in all variables and a vertical field with $\beta_{0}=400$ in a box of dimension $(L_{x},L_{y},L_{z})=(1,2\upi,1)$. We take $k_{L}\approx 14$, which corresponds to capturing
 Landau damping correctly ($k_{\parallel}\approx k_{L}$) for low-amplitude modes with a wavelength
 of about half of the size of the box, and  limit 
 the positive pressure anisotropy at $p_{\perp}/p_{\parallel} -1 <\xi_{\mathrm{Mir}}/\beta_{\perp}$ (this 
 is $\Delta < \xi_{\mathrm{Mir}}/\beta $ for $\Delta p \ll p_{0}$). With either a purely vertical background field, 
 or a mixed azimuthal-vertical background field ($B_{0y}=B_{0z}$), we compare the mode saturation between MHD, the LF model with $\xi_{\mathrm{Mir}}=7$, and the LF model with $\xi_{\mathrm{Mir}}=1$. The vertical-field LF model runs are identical to runs \emph{Zl}4 and \emph{Zl}5 of \citetalias{Sharma:2006dh}.
 
%
%
\begin{figure}
\begin{center}
\includegraphics[width=0.49\columnwidth]{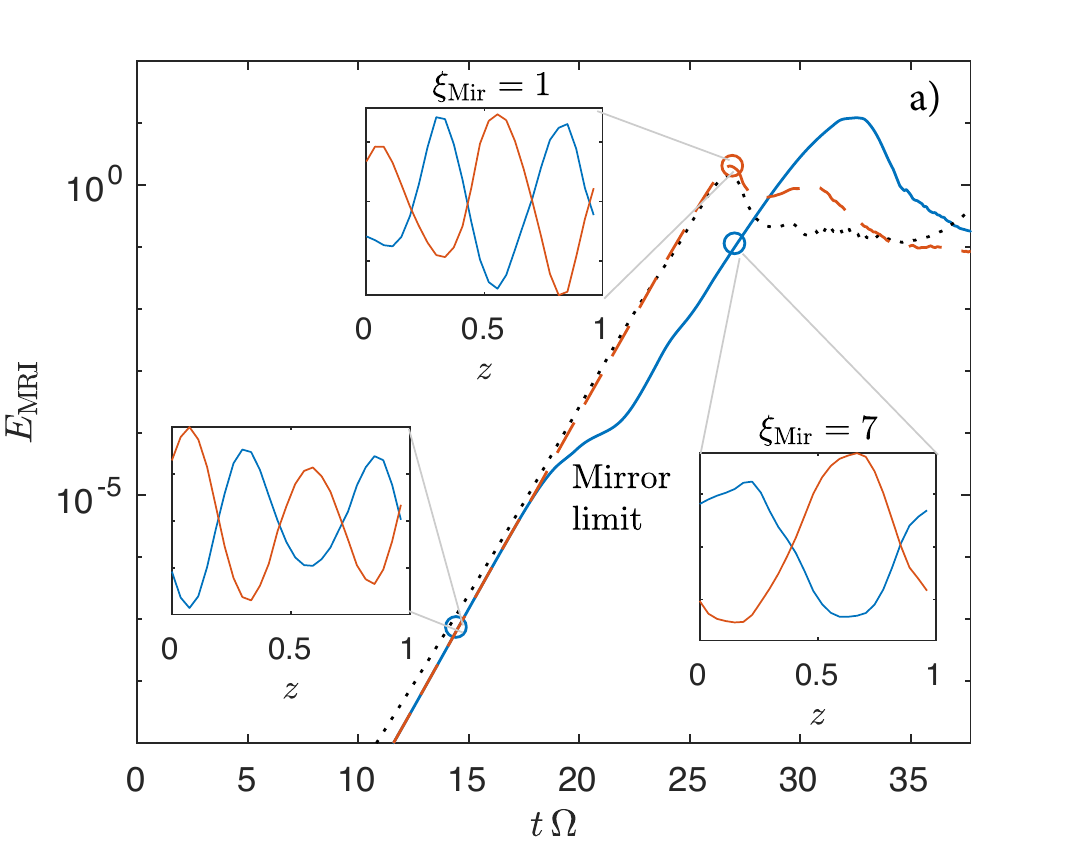}~~\includegraphics[width=0.49\columnwidth]{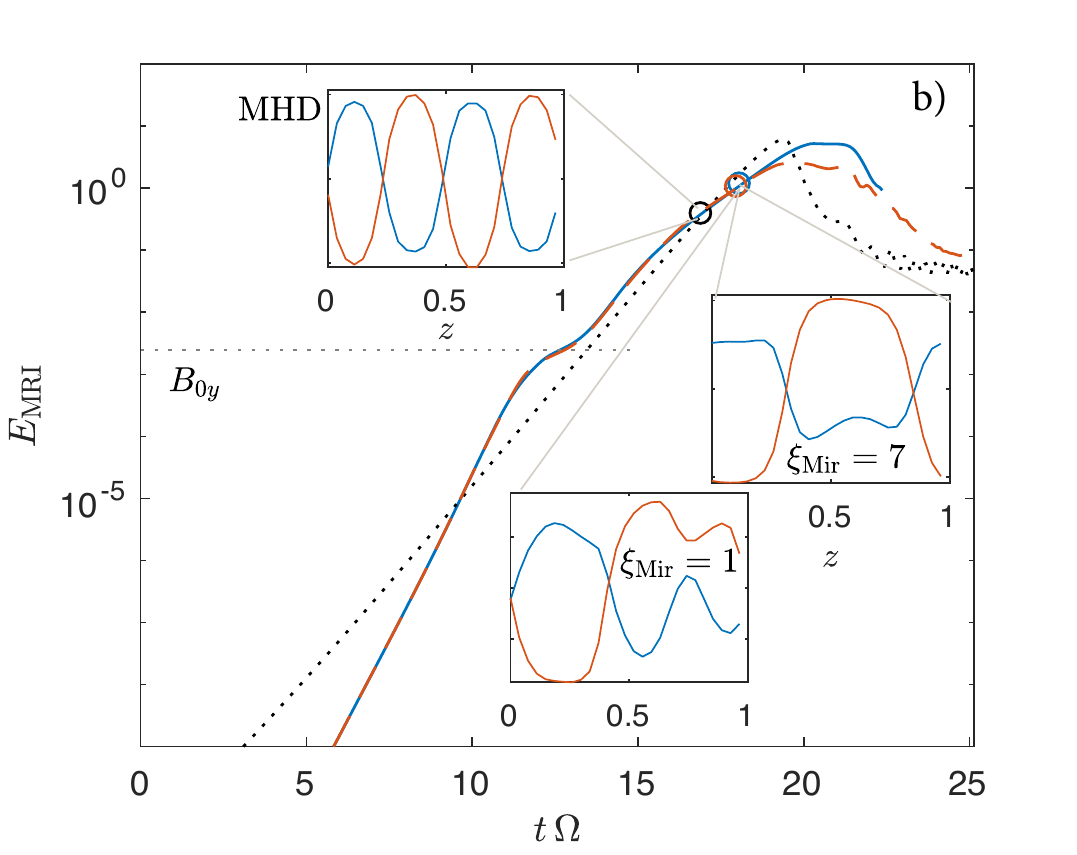}
\caption{Energy of the MRI perturbation, $E_{\mathrm{MRI}}=\int dz\,( \rho\, \delta u^{2}/2+\delta B^{2}/8\upi)$, as a function of time for a set of  3-D {\sc Zeus} simulations at $\beta=400$. We compare the evolution of the LF model \eqref{eq:KMHD rho}--\eqref{eq:GL heat fluxes ql} with 
 mirror limiter $\Delta=7/\beta$ as used in \citetalias{Sharma:2006dh} (blue  solid lines), the LF model with mirror
limiter $\Delta=1/\beta$ (red dashed lines), and standard MHD (black dotted lines).  The insets show the vertical mode structure ($\delta B_{x}$, blue; $\delta B_{y}$, red) at the times indicated by the circles. 
Panel (a) shows the case with a purely vertical background magnetic field ($B_{0y}=0$).
This illustrates how an  artificially high mirror limit ($\Delta=7/\beta$; blue solid line) causes the mode 
to move to longer wavelengths after it reaches the mirror limit at $t\approx 17$, which subsequently causes the mode to reach a very large amplitude before saturation. Panel (b) shows simulations 
with an azimuthal background magnetic field ($B_{0y}=B_{0z}$; the dotted line shows the  energy  of 
 $B_{0y}$); in this case, all three modes saturate
into turbulence at similar amplitudes. Given the relatively disordered mode structure in the kinetic runs (the insets compare the late-time structures of all three cases, as labeled), this behavior is consistent with the idea that there are not major differences between the parasitic modes' properties in the kinetic (LF) model and MHD (see text for further discussion).  Note that the time scale of the MHD case in panel (b) has been shifted to the left, so that all three modes 
reach saturation amplitudes at a similar time (the linear growth of the KMRI mode is faster, see figure \ref{fig:linear}).   }
\label{fig:ZEUS runs}
\end{center}
\end{figure}

 Our findings are illustrated in figure~\ref{fig:ZEUS runs}, which plots the modes' energy 
 evolution  in both  LF cases and standard MHD, with a purely vertical field (left-hand panel a) and with a mixed azimuthal-vertical field (right-hand panel b).
 The key result of this figure is the larger ($\sim$factor $10$) overshoot of the vertical-field KMRI mode (panel a) with $\xi_{\mathrm{Mir}}=7$ 
 (compared to MHD), which disappears at $\xi_{\mathrm{Mir}}=1$ 
 (i.e., there is effectively no difference between MHD and the LF model with the $\xi_{\mathrm{Mir}}=1$ mirror limiter).
As mentioned above, this leads us to attribute the differences between MHD and the LF model saturation
to the difference in the large-amplitude wavelength of the MRI modes.
Specifically, the strength of the vertical magnetic field for $\beta_0=400$ is such that modes with $k_{z} = 4\upi /L_{z}$ dominate
during the linear phase, but, with the artificially high mirror boundary $\xi_{\mathrm{Mir}}=7$, $\Delta$ becomes large enough to cause the 
KMRI mode to increase in scale to the largest in the box by later times (see insets). This does not occur 
with the standard mirror limit  $\xi_{\mathrm{Mir}}=1$.  While not unexpected, these results do show that there are not inherent differences in the parasitic modes' properties between the kinetic (LF) model and MHD for the vertical-field MRI.
Examination of mode evolution at a variety of other values for  $\beta_0$ and $\xi_{\mathrm{Mir}}$ (not shown) has produced  results 
that are generally consistent with this idea.

In  figure~\ref{fig:ZEUS runs}b, showing a mixed azimuthal-vertical background field configuration, we see that the saturation amplitudes of all three cases (the MHD model, and the $\xi_{\mathrm{Mir}}=1$ and $\xi_{\mathrm{Mir}}=7$ LF models) are similar. 
This seems to be because even at large amplitudes, the KMRI modes are relatively disordered and each have both 
$k = 2\upi/L_{z}$ and $k=4\upi/L_{z}$ components, while the MHD mode is nearly a pure   $k=4\upi/L_{z}$ mode. This more disordered KMRI state is expected based on the 1-D analysis of \S\ref{sec: 1D azim}: the mode is strongly 
disrupted as $\delta B_{y}$ surpasses $B_{0y}$, and has not had time to ``pick out'' the fastest growing mode (see  also figure~\ref{fig:By NL}b). Thus 
although the MHD mode might be more easily attacked by the parasitic modes (given its smaller scale), the more disordered KMRI modes are further from being nonlinear solutions, and thus  more easily evolve into turbulence. The net result is that they all saturate at approximately the same amplitude. As further evidence for this scenario, we see that the $k = 2\upi/L_{z}$ component of the $\xi_{\mathrm{Mir}}=7$ mode is  larger   than that of  the $\xi_{\mathrm{Mir}}=1$ mode (as expected because $\Delta p$ is larger, increasing the effective magnetic tension), explaining its slightly higher saturation amplitude.  Thus, these mixed azimuthal-vertical field KMRI results also suggest that there is little difference between the parasitic mode properties in the kinetic (LF) model and MHD. Again, we
have examined mode evolution at various other values for $\beta_{0}$ and $\xi_{\mathrm{Mir}}$ (not shown) and 
seen similar results; however, to truly study this case in detail would require much higher resolution simulations (so as to allow higher $\beta_{0}$; see figure \ref{fig:By NL}), which are beyond the scope of this work.

Overall, we see that all of our calculations -- both of linear parasitic-mode growth rates in idealized settings 
and 3-D  studies using the full nonlinear LF model -- are consistent 
with the idea that parasitic modes are \emph{not strongly affected} by the 
kinetic effects contained within the fluid models considered in this work. 
This seems to be the case for both the vertical-field KMRI and the mixed-azimuthal-vertical field KMRI (although 
we did not study the parasitic modes directly for the mixed-field case).
An obvious caveat of this conclusion is that we have considered only the LF model in this work, and future studies with fully kinetic methods (in particular, those
that can correctly capture plasma microinstability saturation) are needed to  shed light 
on whether our conclusions  also hold for truly collisionless plasmas.

\section{Kinetic effects not included in our models}\label{sub: extra kinetics}

Our approach throughout this paper has been to consider only the simplest kinetic effects, in
particular those arising from a self-generated gyrotropic $\Delta p$. Further, 
the Landau fluid models used for much of the kinetic modeling do not correctly capture
the all-important firehose and mirror microinstabilities, and we have resorted to applying phenomenological 
hard-wall limits as commonly used in previous works (\citetalias{Sharma:2006dh}; \citealt{Sharma:2007cr,SantosLima:2014cn}).
In this section we briefly outline some physical effects that are not included in our model, most of which must
be examined, in one way or another, through fully kinetic simulations \citep[e.g.][]{2016PhRvL.117w5101K}.
\begin{description}
\item[Mirror and firehose evolution: ] {Recent  kinetic simulations and analytical calculations  \citep{Kunz:2014kt,Hellinger:2015en,Rincon:2015mi,Melville:2015tt} paint an interesting picture of how the firehose and mirror instabilities saturate, with 
each going through a regime where fluctuations grow secularly with little particle scattering, followed by a saturated regime
in which the microinstabilities strongly scatter particles due to sharp small-scale irregularities in the magnetic field. The mirror instability 
is particularly interesting, both because it is more important than the firehose for  MRI dynamics (since ${\rm d}B/{\rm d}t>0$ quite generally), and because  
it grows secularly over macroscopic time scales  (i.e., for $t \sim |\grad \bm{u}|^{-1}$) before saturating and
scattering particles \citep{Kunz:2014kt,Rincon:2015mi,2015ApJ...800...27R,Melville:2015tt}.
 This may add another time-scale and nonlinear feature
into the 1-D MRI evolution described in \S\ref{sec:1D}, whereby the effective collisionality 
is strongly enhanced some time $t\sim\Omega^{-1}$ after $\Delta p $ initially reaches the mirror limit. 
It is unclear if there will be a significant change in macroscopic behaviour 
with the onset of particle scattering, and  fully kinetic MRI simulations with large scale separations $\Omega/\Omega_{i}\gg1$ are needed to address this issue (see \S\ref{sec:implications} for further discussion).}
\item[Other kinetic instabilities: ]{There are a variety of other pressure-anisotropy-induced kinetic
 instabilities that we have ignored throughout this
work. For the ion dynamics, the most important of these is likely the ion-cyclotron instability (see, e.g., \citealt{1993JGR....98.3963G}). While  general theoretical analysis (\citealt{Gary:1997dp}; \citetalias{Sharma:2006dh}) and solar-wind 
observations/theory  \citep{2002GeoRL..29.1839K,Bale:2009de,Verscharen:2016sw} suggest that the ion-cyclotron instability is less important than the mirror instability when $T_{e}\sim T_{i}$ and $\beta_{0}\gtrsim 1$,  
it may become more important at lower $T_{e}/T_{i}$ (as expected in low-luminosity accretion flows). In particular, the works of \citet{2015ApJ...800...88S} and \citet{2015ApJ...800...89S} suggest that there is a transition around $T_{e}/T_{i}\lesssim 0.2 $ (or somewhat lower when $\beta_{i}\gtrsim 30$) below which the ion-cyclotron instability dominates over the mirror instability in regulating pressure anisotropy (this behavior is at least partially accounted for by the linear effects of electron pressure anisotropy; see \citealt{2000JGR...105.2393P,2013JGRA..118..785R}). 
At least for lower $\beta_{i}$ plasmas, this will modify the threshold at which the pressure anisotropy is nonlinearly regulated (see \S2.3 of  \citetalias{Sharma:2006dh}) and change the microphysical mechanism through which this regulation occurs \citep{2015ApJ...800...88S}.}
\item[Non-thermal distributions: ]{Having focused on fluid models, we cannot address the many interesting questions surrounding
non-thermal particle distributions that might develop. Strong non-thermal distributions have been seen in a variety of kinetic 
simulations  \citep{Riquelme:2012kz,2015PhRvL.114f1101H,2016PhRvL.117w5101K}, perhaps
due to magnetic reconnection.}
\item[Electrons: ]{We have completely neglected any discussion of electron dynamics throughout this work. This can be loosely justified either when 
 the  electrons are (significantly) colder than ions, or in a weakly collisional regime, when ions  dominate the anisotropic stress in the momentum equation due to the higher electron collisionality.
 However, even  in such regimes, where the anisotropic stress due to elections is nominally subdominant to that of the ions, there may be additional subtleties induced by their thermodynamics. For example, the ion-cyclotron instability 
increases in importance compared to the mirror instability when $T_{e}\ll T_{i}$ (see above, \citealt{2015ApJ...800...88S}). Further, the induced electron-pressure-anisotropy stress  will presumably not be efficiently regulated by ion-scale instabilities, potentially allowing the anisotropic  electron stress to grow to dynamically important values even if $T_{e}\ll T_{i}$, and/or exciting electron instabilities (e.g., the electron whistler instability; \citealp{2017JGRA..122.4410K,2017arXiv170807254R}).
In addition to the possible influence of electron-anisotropy instabilities and stresses, there  are a variety of important questions
to explore related to the proportion of viscous heating imparted to ions and electrons in RIAFs (e.g., see \citealt{Sharma:2007cr,2015MNRAS.454.1848R,2015ApJ...800...89S,2017arXiv170807254R}).}
\item[Non-gyrotropic effects: ]{As the temporal and spatial  scales of the MRI approach the gyroscale, the 
approximation of gyrotropy -- that the pressure tensor is symmetric about the magnetic-field lines -- breaks down. 
In this case, either more complex fluid models \citep{Passot:2012go} or fully kinetic treatments
are needed  to understand any key differences due to non-gyrotropic effects. While such effects are unlikely 
to be astrophysically important in current-epoch disks (where the separation between macroscopic scales and
the gyroscale is often $\sim$$10^{10}$ or more), they may be important 
 for understanding the amplification of very weak ($\beta \gtrsim \Omega_i/\Omega$) fields in the high-redshift universe \citep{Heinemann:2014cl,Quataert:2015dv}. In addition, numerical simulations
will always have limited scale separations (in order to resolve both the macroscopic scales and the gyroradius),
and knowledge of such effects could be important for the complex task of characterizing the limitations of 
kinetic simulations (see \S\ref{sub: FLR implications}).  }
\end{description}

\section{Implications for the design of kinetic MRI turbulence simulations}\label{sec:implications}

In light of the above caveats concerning detailed kinetic effects absent in our models, it is clear that continued efforts to more rigorously simulate KMRI turbulence are needed. In this section, we leverage the results of this paper, as well as those from existing kinetic simulations of the KMRI and of Larmor-scale velocity-space instabilities, to provide some guidance for such efforts. Driving the discussion is a recognition that achieving a healthy scale separation in a kinetic simulation of magnetorotational turbulence is numerically expensive, perhaps prohibitively so. We thus focus primarily on non-asymptotic behaviour that might occur when $\Omega/\Omega_i$ is not sufficiently small, and provide some estimates for what $\Omega/\Omega_i$ must be smaller than in order to capture the effects predicted in this paper. We stress that the asymptotic regime focused on in this paper is likely the most astrophysically relevant one, even if it is difficult to realize in fully kinetic simulations.

\subsection{Pressure-anisotropy overshoot due to finite scale separation}\label{sub: delta p overshoot}

First, in order for the mirror instability to effectively regulate the positive pressure anisotropy during the growth of KMRI, the growth of the mirrors must deplete the anisotropy faster than it is being adiabatically replenished by the KMRI. This requires $\gamma_{\rm m} / \gamma_{\rm kmri} > 1$, where $\gamma_{\rm m}$ and $\gamma_{\rm kmri}$ are the growth rates of the mirror instability and KMRI, respectively. The maximum growth rate of the mirror instability is given by $\gamma_{\rm m} \sim \Omega_i \Lambda^2_{\rm m}$, where $\Lambda_{\rm m} \doteq \Delta - 1/\beta_\perp$ is positive when the plasma is mirror unstable \citep{Hellinger:2007}. Thus, we require $\Lambda_{\rm m} > ( \gamma_{\rm kmri} / \Omega_i )^{1/2}$ for the mirror instability to outpace the KMRI-driven production of positive pressure anisotropy. For the vertical-field case, $\gamma_{\rm kmri} = S/2$ at maximum growth, and so we require $\Lambda_{\rm m} > (S / 2\Omega_i)^{1/2}$.\footnote{See the inset of Fig.~6 in \citet{Kunz:2014kt} for a numerical demonstration of the scaling $\textrm{max}(\Lambda_{\rm m}) \propto (S/\Omega_i)^{1/2}$, where $S$ is the shear rate at which pressure anisotropy is driven. See also Fig.~1{\it c} in \citet{2016PhRvL.117w5101K} for a demonstration that $\Lambda_{\rm m} \simeq \Delta \simeq 0.12 \approx (S/2\Omega_i)^{1/2}$ when the mirror modes begin to first deplete the  pressure anisotropy driven by the vertical-field KMRI. (Note that our definition for $\Delta$ is actually twice as large as the quantity $(p_\perp - p_\parallel)/p_0$ plotted in Fig.~1{\it c} of \citet{2016PhRvL.117w5101K}, which temporarily peaks at $\simeq$$0.07$ before the mirrors are able to drive the pressure anisotropy towards marginal mirror stability. The factor of 2 difference is because of the additional thermal pressure in $p_0$ from the electrons.)} When there is an azimuthal magnetic field present, $\gamma_{\rm kmri} \approx (2S\Omega)^{1/2}$ at maximum growth, and so we require $\Lambda_{\rm m} >  (S/\Omega_i)^{1/2} \, (2\Omega/S)^{1/4} \approx (S/\Omega_i)^{1/2}$. Of course, in this case we must also contend with the firehose instability in regions where $\Delta \propto \delta B_y < 0$ (see (\ref{eq:By linear dP})), for which the instability criterion is $\Lambda_{\rm f} \doteq \Delta + 2 / \beta_\parallel < 0$. With $\gamma_{\rm f} \sim \Omega_i |\Lambda_{\rm f}|^{1/2}$ as the growth rate for the fastest-growing oblique firehose \citep{Yoon:1993of,2000JGR...10510519H}, we require $|\Lambda_f| > (S/\Omega_i)^2 \, (2\Omega/S) \simeq (S/\Omega_i)^2$ for the firehose instability to outpace the KMRI-driven production of negative pressure anisotropy. The parallel-propagating firehose has $\gamma_{\rm f} = \Omega_i |\Lambda_{\rm f}|$ as its maximum growth rate \citep[e.g.][]{1968PhFl...11.2259D,Rosin:2011er}, and thus grows slower than its oblique counterpart for $|\Lambda_{\rm f}| \lesssim  1$.

We now determine whether these criteria place prohibitive constraints on kinetic simulations. For the vertical-field KMRI, $\Delta \sim (\delta B/B_0)^2$ (see (\ref{eq:DeltaLF})), and so the mirror instability will grow fast enough to deplete the pressure anisotropy $\Lambda_{\rm m} \rightarrow 0^+$ when
\begin{equation}\label{eqn:zmri_mirror}
\left(\frac{\delta B}{B_0}\right)^2 \gtrsim \frac{1}{\beta_0} + \left(\frac{S}{\Omega_i}\right)^{\!1/2},
\end{equation}
beyond which the plasma is kept marginally mirror stable (and the results of this paper then follow). The asymptotic limit $(S/\Omega_i) \beta^2_0 \rightarrow 0$ was taken throughout this paper to obtain $(\delta B/B_0)^2 \sim 1/\beta_0$ at the mirror instability threshold. The additional factor of $(S/\Omega_i)^{1/2}$ due to overshoot beyond this threshold may be quite appreciable in a contemporary kinetic simulation of the MRI, perhaps $\sim$$0.1$ \citep[e.g.][]{2016PhRvL.117w5101K} or even more \citep[e.g.][]{Riquelme:2012kz,2015PhRvL.114f1101H}, and thus might be comparable to, if not larger than, $1/\beta_0$. The situation will, of course, improve as the amplification of the magnetic-field strength by the KMRI increases $\Omega_i$ and decreases $\beta$.  Thus, in the final, turbulent saturated state, the effect of  finite scale-separation will presumably be less severe than during the early, weakly nonlinear phases that have been the focus of this work; nonetheless,  one should at least be aware of the impact of insufficient scale separation on the early phase of the KMRI.

In a mixed azimuthal-vertical guide field, the KMRI-driven pressure anisotropy is linear in the mode amplitude (see \eqref{eq:By linear dP}). Because, in this case, it is the pressure anisotropy which provides the azimuthal torque to transport angular momentum and drive the instability (rather than the magnetic tension), it matters all the more how efficiently the pressure anisotropy is regulated. If the lack of scale separation allows the pressure anisotropy to grow much beyond the mirror threshold, the instability's behaviour once $\delta B_{y}\gtrsim \beta_{0}^{-2/3}B_{0y}$ may be completely different. Let us be quantitative, assuming $B_{0y}\approx B_{0z}$ and that $\beta_{0}$ is sufficiently high ($\beta_{0z}\gtrsim 10^{3}$ at least) that the scalings derived in Appendix \ref{app: AKMRI linear properties}  hold. Then, the maximum growth rate of  the azimuthal KMRI, $\gamma \approx (2S\Omega)^{1/2}$,  occurs at wavenumbers satisfying $|k| v_{Az} / \Omega \approx 2\beta_{0}^{-1/6}$ (see equation \ref{eq:app: max k MVI}), and the driven pressure anisotropy (\ref{eq:By linear dP}) is
\begin{equation}
\Delta \approx \sqrt{\frac{\upi S}{ \Omega}} \beta_{0}^{-1/3} \frac{\delta B_y(t)}{B_0} \left(\frac{ 2\beta_0^{-1/6}\Omega}{v_{Az}|k| }\right) ,
\end{equation}
where the final term in parentheses is order unity. Thus, the mirror instability will grow fast enough to deplete the excess positive pressure anisotropy when
\begin{equation}\label{eqn:ymri_mirror}
\left(\frac{\delta B_y}{B_0}\right)^2 \gtrsim \frac{\Omega}{\upi S} \left( \frac{1}{\beta^{2/3}_0} + \frac{S^{1/2}\beta^{1/3}_0 }{\Omega^{1/2}_i} \right)^2 ,
\end{equation}
which may be readily compared to (\ref{eqn:zmri_mirror}). The asymptotic limit $(S/\Omega_i) \beta^2_0 \rightarrow 0$ was taken throughout this paper to obtain $\delta B_y / B_0 \sim \beta^{-2/3}_0$ at the mirror instability threshold. The additional factor of $(S/\Omega_i)^{1/2} \beta^{1/3}_0$ is due to the necessary overshoot beyond this threshold, which, again, may not be all that small in a contemporary kinetic simulation. For the firehose, the negative pressure anisotropy will be efficiently depleted when
\begin{equation}
\left( \frac{\delta B_y}{B_0} \right)^2 \gtrsim  \frac{4\Omega }{\upi S} \left( \frac{1}{\beta^{2/3}_0} + \frac{S^2\beta^{1/3}_0}{2\Omega^2_i} \right)^2 ,
\end{equation}
which will occur {\em after} the mirror criterion (\ref{eqn:ymri_mirror}) is satisfied because of its more forgiving threshold (as long as $(S/\Omega_i)^{1/2}$ is small compared to $1/\beta_0$). 

\subsection{The impact of pressure-anisotropy overshoot on the KMRI}\label{sub: impact of overshoot}

One notable impact of these pressure-anisotropy overshoots is on the fastest-growing KMRI mode wavelength. For the vertical-field case, this wavelength will increase due to the nonlinear pressure anisotropy by an amount $\approx$$(3/2 + \beta \Lambda_{\rm m}/2)^{1/2}$. With $\Lambda_{\rm m} \gtrsim (S/\Omega_i)^{1/2}$ required for the mirrors to outpace the production of anisotropy, this could easily be a factor of several increase unless adequate scale separation is used. For example, having $\beta = 400$ and $S/\Omega_i = 0.01$ would result in a pressure-anisotropy-driven increase in the KMRI wavelength by a factor of $\approx$$5$. In a computational box with vertical extent $L_z \equiv c_s / \Omega$, this means that a maximally growing mode with $k v_A \simeq \Omega$ and $\gamma \simeq S/2$ would shift from having $\lambda / L_z \approx 2\upi (2/\beta)^{1/2} \approx 0.4$ to $\lambda/L_z \approx 2$, bigger than the box. The mode's wavelength would, of course, stop increasing once $\lambda / L_z = 1$. But if, at that point, $\Lambda_{\rm m} \ge (S/\Omega)/2\upi^2 - 3/\beta$, then all the KMRI modes in the box would be stabilized by the effectively increased magnetic tension. Since $\Lambda_{\rm m}$ must grow to $\sim$$(S/\Omega_i)^{1/2}$ before the mirrors can efficiently relax the pressure anisotropy and thereby remove some of this excess tension, we find that values of $S/\Omega_i \ge [ (S/\Omega)/2\upi^2 - 3/\beta ]^2$ will ultimately stabilize the KMRI\footnote{That is, unless $\beta$ decreases (or $\Omega_i$ increases) by an appreciable amount due to the KMRI-driven field amplification by the time that $\lambda$ reaches the box size. In this case, we would have $\delta B/B_0 \gg 1$  and  other nonlinear effects could become important also (e.g., compressibility).}. It is, however, likely that this stabilization would be transient:
 even if the KMRI stops growing, the mirrors will continue relaxing $\Delta p$ (albeit rather slowly), and at some point $\Delta p$ would be sufficiently low so as to render the KMRI unstable again.
Finally, we note that if the box does continue to support unstable KMRI modes on the largest scales $\lambda/L_z = 1$, one must ensure that the horizontal extent of the box is large enough to fit the parasitic modes (\S\ref{sub:Parasitics}).

In the azimuthal-vertical-field case, the fastest-growing mode occurs on scales satisfying $k v_{Az} / \Omega \sim \beta_{0}^{-1/6}$, or $\lambda / L_z \sim \beta_{0}^{-1/3}$. These scales are larger than those of the standard MRI and the vertical-field KMRI. We have predicted that, as the mirror and firehose instabilities kick in and regulate the pressure anisotropy, the influence of the $\delta p_\perp$ and $\delta p_\parallel$ perturbations on the mode evolution is suppressed and the KMRI reverts back to its standard, MHD-like behaviour. This involves a suppression of long-wavelength MRI modes (i.e., $\gamma_{\rm kmri}$ decreases for $k v_A \lesssim 1$) and a transition phase in the nonlinear evolution (``Region 2'' in figure \ref{fig:By NL}{\it a}), in which the mode becomes more MHD-like at smaller scales with the kinetic and magnetic energies in approximate equipartition. If the mirror regularisation is especially sluggish due to inadequate scale separation, this phase might be skipped altogether and a $\lambda/L_z = 1$ mode will take the place of what is instead seen in figure \ref{fig:By NL}{\it b}. It is also worth noting that the large value of $\beta_{0} = 5000$ used in \S\ref{sec: 1D azim} to accentuate the different regions of evolution would require an especially small value of $S/\Omega_i$ in a kinetic simulation.

\subsection{The microphysics of the firehose and mirror instabilities}\label{sub: firehose mirror beta constraints}

A further constraint on the choice of $S/\Omega_i$ concerns the means by which mirror/firehose instabilities regulate the pressure anisotropy. In order for these instabilities to efficiently keep the anisotropy near the instability thresholds via the anomalous pitch-angle scattering of particles, the scattering rate must be $\sim$$S\beta$, and this number must be smaller than $\Omega_i$. 

In the case of the firehose instability, when $S/\Omega_i \ll 1/\beta$ the firehose fluctuations saturate at a mean level $\langle|\delta\bb{B}/B|^2\rangle \sim (\beta S/\Omega_i)^{1/2}$ in a time $\sim$$[\beta/(S\Omega_i)]^{1/2} \ll S^{-1} \sim \gamma^{-1}_{\rm kmri}$ \citep{Kunz:2014kt,Melville:2015tt}. This is achieved via pitch-angle scattering of the particles off Larmor-scale firehose fluctuations, which precludes the adiabatic production of pressure anisotropy. In this limit, since local shear in a macroscopic plasma flow will change in time at the rate comparable to the shear itself, one can safely consider the anomalous collisionality associated with the firehose fluctuations to turn on instantaneously, in line with the macroscopic modeling assumption used in this paper. At sufficiently high $\beta$ and/or $S$ such that $\beta \gtrsim \Omega_i/S$, however, this is no longer true and the firehose fluctuations saturate at an order-unity level independent of either $\beta$ or $S$, after growing secularly without scattering particles for one shear time \citep{Melville:2015tt}. Thus, for kinetic simulations of MRI turbulence to reliably demonstrate the anomalous-scattering model of pressure-anisotropy regulation used in this work and others \citep[e.g.][]{Sharma:2006dh,SantosLima:2014cn,Mogavero:2014io,Chandra:2015bh,Foucart:2015gy}, parameters must be chosen such that $S/\Omega_i < 1/\beta$, preferably by a decade or more. Again, satisfying this inequality becomes easier as the KMRI grows the magnetic-field strength. But, for the early phases of evolution, this cautions against setting $\beta_0$ too high, because this will impose stiff constraints  on an acceptable value of $S/\Omega_{i0}$.

In the arguably more relevant case of the mirror instability, when $S/\Omega_i \ll 1/\beta$ the mirror fluctuations saturate at a mean level $\langle (\delta B/B)^2 \rangle \sim 1$ in a time $\sim$$S^{-1} \sim \gamma^{-1}_{\rm kmri}$ \citep{Kunz:2014kt,2015ApJ...800...27R,Melville:2015tt}. While marginal stability and mirror saturation is ultimately achieved via pitch-angle scattering of the particles off Larmor-scale bends at the ends of the magnetic mirrors, there is a long ($\sim$$S^{-1}$) phase in which the pressure anisotropy is held marginal {\em without} appreciable particle scattering. This is achieved by corralling an ever increasing number of particles into the deepening magnetic wells, in which these particles become trapped, approximately conserve their $\mu$ as they bounce to and fro, and perceive no average change in $B$ (and thus no generation of net pressure anisotropy) along their bounce orbits. The increase in the large-scale $B$ is offset by the decrease in the intra-mirror $B$. During this phase of evolution, the mirror fluctuations grow at the rate required to offset the production of pressure anisotropy by the KMRI-driven growth in the large-scale magnetic-field strength. A few things must be satisfied for kinetic simulations of KMRI growth to produce results similar to those predicted in this paper. First, the hard-wall limiter on the pressure anisotropy that we (and others) use must be an adequate (if not complete) representation of the otherwise more complicated mirror-driven regulation. This is particularly true during the $\mu$-conserving phase of the mirror instability: does it matter to the large (i.e., KMRI) scales that an ever-increasing population of trapped particles are ignorant of the KMRI-driven magnetic-field growth during this phase? If not, then fine; simply limit the pressure anisotropy at the mirror instability using an enhanced collisionality, nothing more sophisticated being necessary. But, if so, then an effort must be made in the kinetic simulation to ensure that its results, if different from those predicted in this paper, are truly asymptotic. Namely, since the $\mu$-conserving phase of the mirror instability lasts just one shear time, whereas the KMRI growth phase typically lasts several shear times, there must be enough scale separation so that the mirrors can saturate {\em before} $\delta B/B_0$ of the KMRI enters into the nonlinear phases we predicted in \S\ref{sec:1D}. Secondly, do the Larmor-scale mirror distortions in the magnetic-field direction greatly affect the efficacy of the heat flow? An answer in the affirmative is suggested by \citet{2016MNRAS.460..467K,2016ApJ...824..123R,2017arXiv170803926R}. But, if the effect of these distortions is simply a reduction in the magnitude of the heat flow, then the footnote in \S\ref{sec:eqns} applies: our results are not strongly affected. The reason is that, by the time the mirror instability is triggered, the heat flows have already spatially smoothed the pressure anisotropy on the scale of the KMRI mode, fulfilling their main role in influencing the mode evolution.

\subsection{Finite-Larmor-radius effects}\label{sub: FLR implications}

There is additional physics that enters when $S/\Omega_i$ is not sufficiently small, which might complicate the evolution of the KMRI beyond that envisaged herein. First, gyroviscous effects become important when $\beta \gtrsim 4\Omega_i/\Omega$ \citep{2007ApJ...662..512F,Heinemann:2014cl}. For such $\beta$, the polarity of the mean magnetic field influences the stability and MRI growth rates. Without a good scale separation between $\Omega$ and $\Omega_i$, finite-Larmor-radius effects might therefore artificially modify the KMRI, even at modest $\beta$. Secondly, note that an equilibrium particle distribution function in a strongly magnetized differentially rotating disc can be quite different than an equilibrium distribution function in an unmagnetized disc. In the latter, a tidal anisotropy of the in-plane thermal motions of the particles is enforced by epicyclic motion in the rotational supported plasma (see \S 3.1 of \citealt{Heinemann:2014cl}); this effect goes away in a strongly magnetized plasma, where gyrotropy of the distribution function about the magnetic field is enforced. This is important because, if the initial magnetic field is inclined with respect to the rotation axis (i.e., $b_y \ne 0$, $b_z \ne 1$) and $S/\Omega_i$ is not sufficiently small, this tidal anisotropy can function as a field-biased pressure anisotropy and potentially drive mirror fluctuations {\em in the equilibrium state}. To wit, the equilibrium field-biased pressure anisotropy is $\Delta \approx (3/2) (S/\Omega_i) ( b^2_y - 1/3)/b_{z}$ to leading order in $S/\Omega_i$ when $b_z \ne 0$. A disc with $b_y = b_z = 1/\sqrt{2}$ would be mirror unstable from the tidal anisotropy alone if $S/\Omega_i \gtrsim 2\sqrt{2}\beta^{-1}$.
 (A plasma with, say, $\beta_0 = 400$ and $S/\Omega_i = 0.01$ would thus be mirror-unstable, even without the KMRI-driven pressure anisotropy.) If the background field is {\em exactly} azimuthal, then the field-biased pressure anisotropy $\Delta = S / (2\Omega) > 0$, and any large-$\beta$ plasma would be trivially mirror-unstable, no matter how strongly the plasma is magnetized.
 
 \subsection{The saturated state}\label{sub: saturated state kinetic sims}

Finally, one must be cognizant of the physical constraints and computational demands not only during the early stages of the KMRI, but also in the saturated state. In going, say, from $\beta_0 \approx 10^3$ to $\beta \approx 4$, as often seen in a typical magnetorotationally turbulent saturated state \citep[e.g.][]{2006PhRvL..97v1103P}, the ion Larmor radius might shrink by a factor between $\approx$$4$ (if $\mu$ is somehow conserved during this process) and $\approx$$16$ (if $\mu$ isn't). Increased scale separation is, of course, a good thing, but only to a point. If $\rho_i$ decreases so much that it falls under the numerical grid, then the anomalous particle scattering from  ion-Larmor-scale magnetic structures that plays a regulatory role for the pressure anisotropy will be attenuated, fundamentally changing the physics of the mirror and firehose instabilities.

%

%
%
%
\section{Conclusions}\label{sec:discussion}

A persistent feature of high-$\beta$ collisionless plasmas  
is the appearance of  nonlinearity due to pressure anisotropy in regimes 
where similar dynamics in a collisional (MHD) plasma are linear (e.g., \citealt{Schekochihin:2006tf,Mogavero:2014io,Squire:2016ev}).
Such behaviour generically arises because, for similar values of the magnetic field, 
the mechanisms that generate a pressure anisotropy  are proportional to the total pressure (e.g., ${\rm d}\Delta p / {\rm d}t \sim p_0 \, {\rm d}\ln B/{\rm d}t$ in the collisionless case), while
the momentum stresses induced by this anisotropy (i.e., its dynamical effects) 
depend on $\Delta p$ itself (i.e., not on $\Delta p/p_{0}$). Thus a 
larger background pressure leads to larger anisotropic stresses, which 
 dominate the Lorentz force by a factor $\sim$$\beta$. This implies that
nonlinear effects can become important for very small changes in magnetic field strength.  
However, as is well known, once the anisotropy grows beyond $\Delta p \sim \pm B^{2}$, the 
firehose and mirror instabilities grow rapidly at ion gyroscales and  
limit any further growth in $\Delta p$. We are then left 
with the question of whether the resulting dynamics on large scales 
are effectively MHD-like (as occurs if the microinstability-limited $\Delta p$
is dynamically unimportant), or whether there are strong differences compared to MHD.
In either case, we can expect that  linear instabilities will be nonlinearly
modified for amplitudes well below where such modifications occur in MHD.

This work has considered the influence of this physics on the collisionless (kinetic) and weakly collisional (Braginskii) magnetorotational instability (KMRI),
focusing on the characteristics of the instability at high $\beta $before its saturation into turbulence.
Our general motivations have been to:
\begin{enumerate}
\item {Understand if there are 
any alternate (pressure-anisotropy related)
means for the linear KMRI to saturate in various regimes. Such a mechanism
could significantly alter expected angular-momentum transport properties of kinetic MRI turbulence.}
\item{Inform current and future kinetic numerical simulations of the KMRI -- which are complex, computationally expensive, and difficult to analyze -- on some key differences and similarities as compared to well-known MHD results.  }
\end{enumerate}
Our main finding is that  the KMRI at large amplitudes 
behaves quite similarly to the standard (MHD) MRI. In fact, in some cases -- in particular, the MRI in a mean azimuthal-vertical field (also known as the magnetoviscous instability) -- the 
MRI transitions from kinetic to MHD-like behaviour as its amplitude increases. 
Furthermore, in all cases studied we have seen the channel mode ($k_{x}=k_{y}=0$ MRI mode) emerge  as an \emph{approximate nonlinear
solution of the kinetic equations} at large amplitudes, in the same way as occurs in MHD \citep{Goodman:1994dd}. This is because the 
mirror-limited pressure anisotropy has the same form as the Lorentz force (since $\Delta p \propto B^{2}$), and this vanishes identitically for an MRI channel mode. 
This points to an interesting robustness of the channel-mode solution in collisionless plasmas that had not been previously fully appreciated.

The similarity between the nonlinear physics of the KMRI and the MHD MRI is certainly not a given;
for example, the nonlinear dynamics of shear-Alfv\'en waves, which are  related to the MRI \citep{Balbus:1998tw}, differ very significantly between collisional and kinetic plasmas \citep{Squire:2016ev,Squire:2016ev2}.
Further, there remain a variety of 1-D nonlinear effects that cause modest differences when compared to standard MHD, 
and these could be important for the difficult task of designing and interpreting 3-D fully kinetic simulations. For example, 
depending on the level of overshoot of the pressure anisotropy above the mirror instability threshold  
(as would occur if there were insufficient scale separation between the large-scale dynamics and the gyroscale; see \S\ref{sec:implications}), 
the MRI mode may migrate to longer wavelengths at moderate amplitudes, or (in extreme cases) be completely 
stabilized. A more detailed overview of the most relevant 1-D results is given in \S\ref{sub:1-D conclusions}.

Motivated by the finding that there are no viable 1-D mechanisms 
for halting the growth of the kinetic MRI, as also found in previous numerical simulations (\citetalias{Sharma:2006dh}; \citealt{Sharma:2007cr,Riquelme:2012kz,Kunz:2014hm,2015PhRvL.114f1101H}), 
we are left with the conclusion that 3-D effects must govern the collapse of a KMRI
channel mode into a turbulent-like state.
Following previous  MHD studies \citep{Goodman:1994dd,Pessah:2009uk,Latter:2009br}, we have 
considered 3-D mechanisms for mode saturation in terms of \emph{parasitic modes}, secondary 
instabilities that feed off the large field and flow gradients of the channel mode, acting to disrupt it and
cause its collapse into turbulence. Using both linear studies of parasitic modes and
3-D nonlinear simulations (with the modified {\sc Zeus} version of \citetalias{Sharma:2006dh}), we 
have found very little difference between the behaviour of parasitic modes in kinetic and MHD models.
We have further shown that the observations of  \citetalias{Sharma:2006dh} of larger saturation 
amplitudes in kinetic models as compared to MHD may be straightforwardly explained by the migration 
of kinetic channel modes to longer wavelengths due to the mean pressure anisotropy (i.e., 1-D effects).
This suggests that MHD results may be used to give simple, zeroth-order estimates of the 
expected amplitude at which a KMRI channel mode should saturate into turbulence. Similar conclusions have also been found in global Braginskii MHD simulations of accretion disks \citep{2017MNRAS.470.2240F}.

Although our results suggest that the breakdown into turbulence occurs in a similar way in kinetic theory and MHD, 
this does not necessarily imply that the saturated state of the turbulence is similar. Indeed,
even following the pioneering 3-D nonlinear kinetic  MRI simulations of \citet{2015PhRvL.114f1101H} and \citet{2016PhRvL.117w5101K}, many properties of the  saturated state of the KMRI -- i.e., the
turbulence -- remain largely unknown. The zero-net-flux simulation of \citet{2016PhRvL.117w5101K} found a level of turbulence that was comparable to high-Prandtl-number turbulence in MHD. However, there are some notable differences; for instance, a greater prevalence of coherent flows, and the fact that (in contrast to MHD) a large proportion of this transport 
 arises from the pressure anisotropy directly. Some similar results were found in \citetalias{Sharma:2006dh}, \citet{Sharma:2007cr}, and \citet{2017MNRAS.470.2240F} using  fluid closures.
 However, these current results have explored only small regions of parameter space (e.g., the case of zero net flux), and it remains unknown how kinetic MRI turbulence relates (if at all) to 
 MHD MRI turbulence.
 Both the strength of kinetic MRI turbulence and the different heating processes involved (in particular, the relative level of ion versus electron heating; \citealp{1998ApJ...500..978Q,Quataert:1999gn}), remain crucial unknowns in constraining phenomenological disk models.

\begin{acknowledgments}
We would like to thank Prateek Sharma for kindly providing the modified version of the ZEUS code, which formed the basis for the  nonlinear results of \S\ref{sub: zeus sims}.  
J.S.~was funded in part by the Gordon and Betty Moore Foundation through Grant GBMF5076 to Lars Bildsten, Eliot Quataert and E.~Sterl Phinney. 
E.Q.~was supported by Simons Investigator awards from the Simons Foundation and  NSF grants AST 13-33612 and AST 17-15054. 
M.W.K.~was supported in part by NASA grant NNX17AK63G and US DOE Contract DE-AC02-09-CH11466.
Some of the numerical calculations presented in this work were run on Caltech's  Zwicky cluster (NSF MRI award PHY-0960291) 
\end{acknowledgments}


\appendix

\label{APPENDIX BEGINS HERE}

\section{Linear properties of the KMRI with a background azimuthal-vertical field}\label{app: AKMRI linear properties}

In this appendix, we  derive various properties of the KMRI in the general case when the
background field has a mixed azimuthal-vertical configuration. We focus on modes 
with $k_{x}=k_{y}=0$ as in \S\ref{sec:linear MRI}. 

\subsection{The fastest-growing wavenumber}
An important input to the nonlinear arguments put forth in \S\ref{sec: 1D azim} is the
scaling of the fastest-growing wavenumber (and growth rate) with $\beta_{0}$.
Our starting point is the Landau-fluid dispersion relation, obtained through the characteristic polynomial 
of the matrix resulting from the 
linearization of \eqref{eq:KMHD rho}--\eqref{eq:GL heat fluxes ql} (with $S/\Omega=3/2$ and $\nu_{c}=0$).
We wish to find $k_{\mathrm{max}}$, the wavenumber that maximizes the growth rate $\gamma=\Im(\omega)$, 
as a function of $\beta_{0z}=8\upi p_{0}/B_{0z}^{2}$ and $\alpha \equiv B_{0y}/B_{0z}$, assuming $\beta_{0z}\gg1$ 
(because we consider only vertical modes, it is most straightforward to work with $\beta_{0z}$ and $v_{Az}$, as opposed to quantities defined with $B_{0}^{2}=B_{0y}^{2}+B_{0z}^{2}$).
Anticipating the scaling $k_{\mathrm{max}}v_{Az}/\Omega\sim \beta_{0}^{-1/6}$, $\omega- i \sqrt{3} \sim -\beta_{0}^{-1/3} $
we insert the ansatz $\beta_{0z}= \epsilon^{-6}\bar{\beta}_{0z}$, $k =\epsilon {k}_{0}\bar{\beta}_{0z}^{-1/6}$, $\omega = i\sqrt{3}+ i \epsilon^{2}\gamma^{(1)}$ and expand the resulting expression in $\epsilon$.
This yields the solution
\begin{equation}
\frac{\gamma}{\Omega} \approx i\sqrt{3} + \gamma^{(1)}\approx i\sqrt{3} - \frac{\alpha^{2}k_{0}^{3}+12(1+\alpha^{2})^{3/2}}{6\sqrt{\upi}\alpha^{2}k_{0}\beta_{0z}^{1/3}},\label{eq:app: correction to MVI gamma}\end{equation}
which is an approximate KMRI dispersion relation, valid at high $\beta$ near the peak growth rate. Maximizing \eqref{eq:app: correction to MVI gamma} over $k_{0}$, we find,
\begin{equation}
\frac{k_{\mathrm{max}}v_{Az}}{\Omega}\approx \left(\frac{12}{\upi\alpha}\right)^{1/6}\left(\frac{1}{\alpha}+\alpha\right)^{1/2}\beta_{0z}^{-1/6}, \label{eq:app: max k MVI}
\end{equation}
and 
\begin{equation}
\frac{\gamma_{\mathrm{max}}}{\Omega}\approx i\sqrt{3} - \frac{3^{5/6}}{(2\upi\alpha)^{1/3}}\left(\frac{1}{\alpha}+\alpha\right)\beta_{0z}^{-1/3},\label{eq:app: max gamma MVI}
\end{equation}
for the maximum growth rate, $\gamma_{\mathrm{max}} = \gamma(k_{\mathrm{max}})$.

%
%
%
\begin{figure}
\begin{center}
\includegraphics[width=0.6\columnwidth]{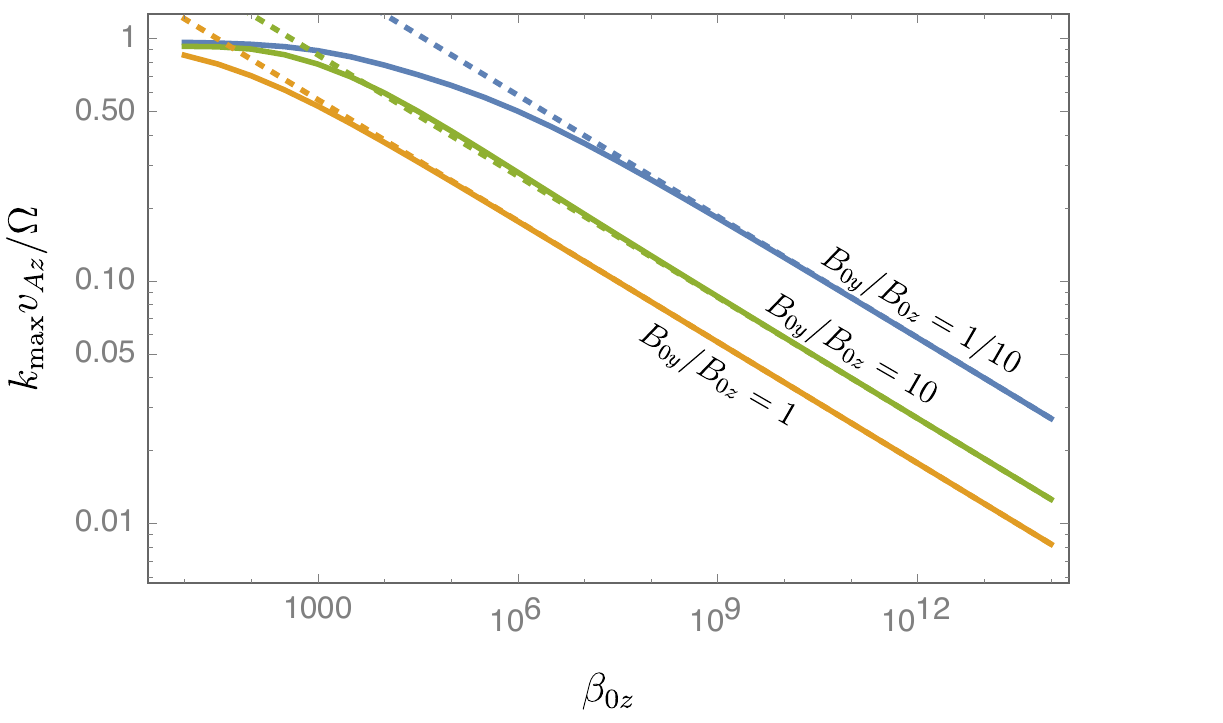}
\caption{Scaling of $k_{\mathrm{max}}$, the wavenumber of the fastest-growing mode, 
as a function of $\beta_{0z} = 8\upi p_{0}/B_{0z}^{2}$ for different choices of $\alpha \equiv B_{0y}/B_{0z}$. Solid 
lines show results from the numerical solution of the LF dispersion relation; dashed lines show 
the asymptotic result $k_{\mathrm{max}}v_{A}/\Omega\sim \beta^{-1/6}$ (see \eqref{eq:app: max k MVI}). }
\label{fig: maximizing k for MVI}
\end{center}
\end{figure}
%
%
%

Unsurprisingly (given that we carried out an expansion in $\beta_{0}^{-1/6}$) the expressions \eqref{eq:app: max k MVI}--\eqref{eq:app: max gamma MVI} are  accurate only at very high $\beta_{0}$, particularly when $\alpha\neq 1$. A comparison
with the true $k_{\mathrm{max}}$, obtained by numerically maximizing the numerically computed dispersion relation,  is illustrated in figure~\ref{fig: maximizing k for MVI}. We see 
 that, very approximately, the asymptotic result \eqref{eq:app: max k MVI} is valid when 
it predicts $k_{\mathrm{max}}v_{Az}/\Omega\lesssim 1$ (as should be expected, since $k_{\mathrm{max}}v_{Az}/\Omega\sim 1$ is the fastest-growing wavelength of the standard MRI). This suggests 
that the results \eqref{eq:app: max k MVI} and \eqref{eq:app: max gamma MVI} are applicable when $\beta_{0z}\gg (12/\upi) \alpha^{-4}$ for $\alpha\ll1$, or when $\beta_{0z}\gg (12/\upi) \alpha^{2}$ for $\alpha \gg 1$. For $\beta_{0z}$ lower than these 
estimates, $k_{\mathrm{max}}$ is less than the prediction \eqref{eq:app: max k MVI} (there are also minor deviations above the prediction \eqref{eq:app: max k MVI} when $\alpha>1$, see figure~\ref{fig: maximizing k for MVI}).
It is also worth noting that the dispersion relation around $k\approx k_{\mathrm{max}}$ is not very strongly 
peaked (see, e.g., figure \ref{fig:linear} in the main text). This implies that the fastest-growing mode grows only slightly faster than those with a similar 
wavelength, and it is unlikely to completely dominate by the time it reaches nonlinear amplitudes (see \S\ref{sec: 1D azim} and \S\ref{sub: zeus sims}). 

\subsection{The fastest-growing mode}

The structure of the fastest-growing KMRI mode is also relevant to the nonlinear arguments of \S\ref{sec: 1D azim}. This can be found by inserting $k_{\mathrm{max}}$ and $\omega = \gamma_{\mathrm{max}}$ into 
the matrix resulting from the 
linearization of \eqref{eq:KMHD rho}--\eqref{eq:GL heat fluxes ql}, and solving for the
amplitudes of each component $\delta u_{x}$, $\delta B_{x}$, etc., in terms of $\delta p_{\perp}$.
To lowest order in $\beta_{0z}^{-1}$, this yields $\delta p_{\parallel}\approx -\alpha^{2} \delta p_{\perp}$, 
as well as the following relations for the fastest-growing KMRI mode:
\begin{gather}
\frac{\delta B_{x}}{B_{0z}}\approx-\frac{1}{2}\left({\frac{1 }{12 \upi^{2}}}\right)^{1/6}\frac{ \alpha^2+1 }{\alpha^{1/3}  } \beta_{0z}^{2/3} \frac{\delta p_{\perp}}{p_{0}},\label{eq:app: Bx mode amp MVI}\\ \frac{\delta B_{y}}{B_{0z}}\approx\left({\frac{1 }{12\upi^{2}}}\right)^{1/3}\frac{ (\alpha^2+1)^{2} }{\alpha^{5/3}  }\beta_{0z}^{1/3}\frac{\delta p_{\perp}}{p_{0}},\label{eq:app: By mode amp MVI}\\ 
\frac{\delta u_{x}}{v_{Az}}\approx\frac{\imag}{2}\left({\frac{3 }{16\upi}}\right)^{1/6} \alpha^{1/3}(\alpha^{2}+1)^{1/2}\beta_{0z}^{5/6} \frac{\delta p_{\perp}}{p_{0}},\\  \frac{\delta u_{y}}{v_{Az}}\approx\frac{\imag}{4}\left({\frac{81 }{16\upi}}\right)^{1/6} \alpha^{1/3}(\alpha^{2}+1)^{1/2}\beta_{0z}^{5/6} \frac{\delta p_{\perp}}{p_{0}}.\label{eq:app: uy mode amp MVI}
\end{gather}
We see that $\delta B_{x}/B_{0}\sim \beta_{0}^{1/3}\delta B_{y}/B_{0} $, \emph{viz.,} the mode is dominated by the radial magnetic field. 

A more intuitive way of understanding the structure of the KMRI mode is through the  relations $\delta p_{\perp}/p_{0}\approx \delta \rho/\rho_{0} -\imag \sqrt{\upi }\xi \delta B/B_{0}$, and $\delta p_{\parallel}/p_{0}\approx \delta \rho/\rho_{0} +\imag \sqrt{\upi }\xi \delta B/B_{0}$,
where $\xi = \omega/(\sqrt{2} \hat{\bm{b}}\cdot \bm{k}\, c_{s})$ 
and $\delta B/B_{0} = \delta B_{y} B_{0y}/B_{0}^{2}$ is the perturbation to the field strength. 
These relations are straightforwardly derived from the linear $\delta p_{\perp}$ and $\delta p_{\parallel}$ equations  by balancing the  the production of pressure anisotropy against the smoothing action of the heat fluxes; see \citetalias{Quataert:2002fy}. For $\alpha=B_{0y}/B_{0z}\approx 1$, these lead to equation \eqref{eq:By linear dP}, which is used in \S\ref{sec: 1D azim} and \S\ref{sec:implications} to estimate the amplitude at which the KMRI mode reaches the firehose and mirror limits
(inserting $k_{\mathrm{max}}$, one can also obtain \eqref{eq:app: By mode amp MVI}).

\section{The form of the nonlinearity in growing KMRI modes}\label{app: MRI nonlinearity}

In this appendix, we derive, using asymptotic expansions, the form of the nonlinearity in
growing KMRI modes. The method used is almost identical to that in the appendices of \citet{Squire:2016ev2}, and the results are very similar, yielding few surprises. However, 
the results do serve to more formally justify some of the claims made in the main text, in particular
those relating to the smoothing effects of the heat fluxes in \S\S\ref{sub: general comparison}, \ref{sec: 1D vertical}. 
They also allow one to explicitly calculate the form of the nonlinearity that causes the changes to KMRI mode shape illustrated in figure \ref{fig:Bz MRIs}.

We consider three cases -- a double-adiabatic model, a collisionless LF model, and a Braginskii MHD model -- each with a purely vertical field (see \S\S\ref{sec: 1D vertical}--\ref{sec: 1D brag} and figure \ref{fig:Bz MRIs}). While the double-adiabatic model is not considered in the main text (neglect of the heat fluxes is never a good approximation at high $\beta$), it provides a nice illustration of the importance of heat fluxes for high-$\beta$ KMRI dynamics. We treat only the early nonlinear behaviour, that is, when pressure anisotropy first becomes important
at low mode amplitudes. We also do not treat the $B_{0y} \ne 0$ KMRI (\S\ref{sec: 1D azim}), since such modes stay close to linear until $\Delta p$ reaches the firehose and mirror limits, at which point there are strong nonlinear modifications that cannot be captured with this type of asymptotic method (see figure \ref{fig:By NL}).

\subsection{Equations and method}\label{app: equations}

Our method here is nearly identical to that used in \cite{Squire:2016ev2} to study shear-Alfv\'en waves, with only 
minor modifications to the equations to account for the rotation and shear flow. We consider a mode of wavelength $2\upi/k_{\|}$ in a background plasma with density $\rho_{0}$, thermal pressure $p_{0}$, and vertical magnetic field $\bm{B}_{0}=B_{0}\hat{\bm{z}}$. For simplicity, we normalize length scales to $k^{-1}_{\|}$, velocities to $v_{A0} \equiv B_0 / \sqrt{4\upi\rho_0}$, time scales to $\omega^{-1}_{A}\equiv(k_{\|}v_{A0})^{-1}$, densities to $\rho_{0}$, pressures to $p_{0}$, and magnetic fields to $B_{0}$. Splitting the velocity $\bb{u}$ into its equilibrium ($\bb{U}_0 = -Sx\ey$) and fluctuating ($\delta\bb{u}$) parts, equations \eqref{eq:KMHD rho}--\eqref{eq:GL heat fluxes ql}  become, respectively,
\begin{gather}
\D{t}{\rho} = - \rho\grad\bcdot\delta\bm{u},\label{eq:NDMHD rho}\\[2ex]
{\rho} \left( \D{t}{\delta\bm{u}} + 2\overline{\Omega}\hat{\bm{z}}\btimes \delta\bm{u} - \overline{S} \delta u_{x}\hat{\bm{y}} \right) = -\grad\left(\frac{\beta_{0}}{2} {p}_{\perp} + \frac{{B}^{2}}{2}\right)+{\grad} \bcdot \left[ \hat{\bm{b}}\hat{\bm{b}}\left(\frac{\beta_{0}}{2}\Delta + {B}^{2} \right)\right], \label{eq:NDMHD u}\\*
\D{t}{\bb{B}} +\overline{S} B_{x} \hat{\bm{y}} = {\bm{B}}\bcdot {\grad} {\delta\bm{u}} - {\bm{B}}{\grad} \bcdot {\delta\bm{u}},\label{eq:NDMHD B}\\*
\D{t}{{p}_{\perp}} + p_{\perp}\hat{b}_{x}\hat{b}_{y}\overline{S} =  - \beta_{0}^{\,1/2} \left[ {\grad} \bcdot ({q}_{\perp}\hat{\bm{b}}) + {q}_{\perp}{\grad} \bcdot \hat{\bm{b}}\right] + {p}_{\perp}\hat{\bm{b}}\hat{\bm{b}}\dbldot  {\grad} {\delta\bm{u}} - 2{p}_{\perp}{\grad} \bcdot {\delta\bm{u}} -\overline{\nu}_{c}\Delta,\label{eq:NDMHD pp}\\*
\D{t}{{p}_{\parallel}}  - 2p_{\|}\hat{b}_{x}\hat{b}_{y}\overline{S} = - \beta_{0}^{\,1/2} \left[ {\grad} \bcdot ({q}_{\parallel}\hat{\bm{b}}) -2 {q}_{\perp}{\grad} \bcdot \hat{\bm{b}}\right] -2{p}_{\parallel}\hat{\bm{b}}\hat{\bm{b}}\dbldot  {\grad} {\delta\bm{u}} - {p}_{\parallel}{\grad} \bcdot {\delta\bm{u}} + 2\overline{\nu}_{c}\Delta,\label{eq:NDMHD pl}\\*
{q}_{\perp} = -\sqrt{\frac{ {p}_{\parallel}}{\upi {\rho}}} \frac{1}{|k_{\parallel}| +\overline{\nu}_{c}  (\beta_{0} \upi {p}_{\parallel}/{\rho} )^{-1/2}} \left[ {\rho}   {\nabla}_{\parallel} \left(\frac{{p}_{\perp}}{{\rho} }\right)  - {p}_{\perp}\left(1-\frac{{p}_{\perp}}{{p}_{\parallel}} \right)\frac{{\nabla}_{\parallel} {B}}{{B}}  \right], \label{eq:NDMHD qp}\\*
{q}_{\parallel} = -2\sqrt{\frac{ {p}_{\parallel}}{\upi {\rho} }} \frac{1}{|k_{\parallel}| +(3\upi/2-4)\overline{\nu}_{c}  (\beta_{0} \upi {p}_{\parallel}/{\rho} )^{-1/2}} \,{\rho}  {\nabla}_{\parallel} \left(\frac{{p}_{\parallel}}{{\rho} }\right),\label{eq:NDMHD ql}
\end{gather}
where $\overline{\nu}_{c}\equiv\nu_{c}/\omega_{A}$ and $\overline{\Omega}\equiv\Omega/\omega_{A}$. The heat fluxes $q_{\perp,\|}$ are normalized using the sound speed $c_{s}\equiv\beta^{1/2}_0v_{A0} = \sqrt{2 p_{0}/\rho_{0}}$ (note that we have changed the definition of $c_{s}$ from the main text here, so as to remove various inconvenient factors of $2$ from equations~\eqref{eq:NDMHD rho}--\eqref{eq:NDMHD ql}). As in the main text, $\Delta\equiv p_{\perp}-p_{\parallel}$ denotes the dimensionless pressure anisotropy, $\beta_{0}\equiv8\upi p_{0}/B_{0}^{2}$, and
\begin{equation}
\D{t}{} = \pD{t}{} - \overline{S} x \pD{y}{} + \delta\bm{u}\bcdot\grad
\end{equation}
is the convective derivative. 

Following \S\ref{sec: 1D vertical}, we focus on the nonlinear evolution of a 1D (in $z$) channel mode. This involves an asymptotic expansion of \eqref{eq:NDMHD rho}--\eqref{eq:NDMHD ql}, which is constructed as follows. Figure \ref{fig:linear} shows that the $k_\parallel =k_z$ KMRI mode grows fastest for $k_{\|}v_{A}/\Omega \sim 1$, and so we order $\overline{\Omega}\sim 1$. Following \citet{Squire:2016ev2}, we order the (dimensionless) MRI mode amplitude $\delta\bm{B}_{\perp} \sim \delta\bm{u}_{\perp} \sim \epsilon \ll 1$ and the equilibrium plasma beta parameter $\beta_{0}$ such that the 
effect of the pressure anisotropy $\Delta$ is as important as that of the linear terms, {\it viz.}, $\beta_{0 }\Delta \sim 1$.  Because the growth rate of the fastest-growing MRI mode is $\gamma \sim \Omega \sim \omega_{A}$, we order the (dimensionless) spatial and temporal derivatives to be $\sim$$\mathcal{O}(1)$. This ordering captures nonlinear effects on the MRI mode just before it drives the pressure anisotropy to the mirror limit $\beta_{0}\Delta\approx 1$, in both the collisionless and weakly collisional (Braginskii) cases.

In what follows, we use $\langle f \rangle$ to denote a spatial ($z$) average of some function $f$ and $\widetilde{f}\equiv f-\langle f\rangle$ to denote the spatially varying part of $f$.

\subsection{Collisionless limit: Double-adiabatic closure}\label{app:sub: double-adiabatic}

Because the pressure anisotropy is generated proportional to the change in $B$, in the double-adiabatic  
case $\Delta$ scales as $\sim$$\delta B_{\perp}^{2}$. Thus we order $\beta_{0}\sim \epsilon^{-2}$. 
The ordering of all variables is then as follows (cf.~\citealt{Squire:2016ev2}):
\begin{subequations}
\begin{alignat}{5}
\bb{u}_\perp 	&= -\overline{S}x\ey 				&&+{} \epsilon\, \delta\bm{u}_{\perp1} 	&&+{} \epsilon^{2}\,\delta\bm{u}_{\perp2}	&&+{} \dots, 	&&{} \label{eq:Bx expansion 1}\\
u_z 			&= 0							&&+{} 0							&&+{} \epsilon^{2} \,\delta u_{z2} 		&&+{} \epsilon^{3} \,\delta u_{z3}	&&+{} \dots \\
\bb{B}_\perp 	&= 0 						&&+{} \epsilon\, \delta\bm{B}_{\perp1} 	&&+{} \epsilon^{2}\,\delta\bm{B}_{\perp2}	&&+{} \dots				,	&&{} \\
\rho 			&= 1							&&+{} 0							&&+{} \epsilon^{2}\,\delta\rho_{2} 		&&+{}\epsilon^{3}\,\delta\rho_{3} 	&&+{}\dots,\\
p_{\perp} 		&= 1							&&+{} 0							&&+{} \epsilon^{2} \,\delta p_{\perp 2}	&&+{} \epsilon^{3} \,\delta p_{\perp 3}&&+{} \dots,\\ 
p_{\parallel} 	&=1							&&+{} 0							&&+{} \epsilon^{2} \,\delta p_{\parallel 2}	&&+{} \epsilon^{3} \,\delta p_{\parallel 3} &&+{} \dots,\label{eq:pl expansion 1}
\end{alignat}
\end{subequations}
where the numerical subscripts denote the order in $\epsilon$. Equations \eqref{eq:Bx expansion 1}--\eqref{eq:pl expansion 1} are inserted into the MRI equations \eqref{eq:NDMHD rho}--\eqref{eq:NDMHD pl} with $q_{\perp}=q_{\|}=0$ and the result is examined order by order in $\epsilon$. 

\paragraph{\emph{Order $\epsilon^{0}$.}} Only the $z$ component of \eqref{eq:NDMHD u} contributes at this order, giving $\partial_{z}\delta p_{\|2}=0$ or $\widetilde{\delta p_{\|2}}=0$. This condition expresses parallel pressure balance. 

\paragraph{\emph{Order $\epsilon^{1}$.}} 
The parallel component of the momentum equation \eqref{eq:NDMHD u} gives $ \widetilde{\delta p_{\|3}}=0$. The perpendicular components of the momentum and induction equations \eqref{eq:NDMHD u}--\eqref{eq:NDMHD B} provide evolution equations for the linear MRI:
\begin{gather}
\partial_{t}{\delta\bm{u}_{\perp1}} + 2\overline{\Omega}\hat{\bm{z}}\btimes \delta\bm{u}_{\perp1} - \overline{S} \delta u_{x1 }\hat{\bm{y}} =  \partial_{z}\left[ \delta\bm{B}_{\perp1} \left(1+ \frac{\beta_0}{2} \Delta_{2}\right)\right],\label{app:eq: u eqn CGL} \\
\partial_{t}{\delta\bm{B}_{\perp1}} + \overline{S}\delta B_{x1}\hat{\bm{y}} = \partial_{z} \delta\bm{u}_{\perp1}.\label{app:eq: B eqn CGL}
\end{gather}
To close this system, we require $\Delta_{2}=\delta p_{\perp2}-\delta p_{\|2}$ as a function of $\delta \bm{u}_{\perp2}$ and $\delta \bm{B}_{\perp2}$, which is obtained at next order. 

\paragraph{\emph{Order $\epsilon^{2}$.}} At this order, we require only the pressure equations \eqref{eq:NDMHD pp}--\eqref{eq:NDMHD pl} to obtain $\Delta_{2}$ for use in \eqref{app:eq: u eqn CGL}. Expanding $\hat{\bm{b}}\hat{\bm{b}}\dbldot\grad \delta\bm{u}=\hat{b}^2_{z}\partial_{z}\delta u_{z}+\hat{\bm{b}}\hat{\bm{b}}\dbldot\grad \delta\bm{u}_{\perp}$, \eqref{eq:NDMHD pp}--\eqref{eq:NDMHD pl} become
\begin{gather}
\partial_t \delta p_{\perp 2} + \partial_z{\delta u_{z2}} = \delta \bm{B}_{\perp1}\bcdot \partial_{z}{\delta \bm{u}_{\perp1}} - \overline{S}\delta B_{x1}\delta B_{y1} = \frac{1}{2}\partial_{t}{\delta B_{\perp 1}^{2}} ,\label{eq:pp CGL expansion}\\
\partial_t \delta p_{\parallel 2} + 3 \partial_{z}{\delta u_{z2}} = -2\delta \bm{B}_{\perp1}\bcdot \partial_{z}{\delta \bm{u}_{\perp1}} + 2\overline{S}\delta B_{x1}\delta B_{y1} = -\partial_{t}{\delta B_{\perp 1}^2}, \label{eq:pl CGL expansion}
\end{gather}
where $\delta B^2_{\perp1}\equiv \delta B_{x1}^{2}+\delta B_{y1}^{2}$; the final equalities in these equations follow from \eqref{app:eq: B eqn CGL}. We can then solve for $\partial_{z}\delta u_{z2} = \widetilde{\partial_{z}\delta u_{z2}}$ using \eqref{eq:pl CGL expansion} and insert this into \eqref{eq:pp CGL expansion} to find
\begin{equation}\label{eq:Delta2 CGL}
\partial_{t}\Delta_{2} = \frac{5}{6}\partial_{t}\delta B_{\perp1}^{2} + \frac{2}{3} \partial_{t}\langle \delta B_{\perp1}^{2}\rangle,
\end{equation}
If we then assume that the mode starts growing from vanishingly small initial conditions, \eqref{eq:Delta2 CGL} may be straightforwardly integrated to obtain
\begin{equation}\label{eq:Delta2}
\Delta_{2}=\frac{5}{6}\delta B_{\perp1}^{2} + \frac{2}{3}\langle \delta B_{\perp1}^{2}\rangle.
\end{equation}
This expression may be inserted into \eqref{app:eq: u eqn CGL} to yield a simple nonlinear equation for the growing MRI mode:
\begin{align}
&\partial^2_{t} \delta \bm{B}_{\perp 1} + 2\overline{\Omega} \ez\btimes \partial_{t} \delta \bm{B}_{\perp 1} - 2\overline{S} \overline{\Omega} \delta B_{x1} \ex \nonumber\\*
\mbox{} &\quad = \partial^2_z \left\{ \delta \bm{B}_{\perp 1} \left[ 1 + \frac{\beta_0}{2} \left( \frac{5}{6} \delta B^2_{\perp 1} + \frac{2}{3} \langle \delta B^2_{\perp 1} \rangle \right) \right] \right\} ,
\end{align}
where we have grouped all nonlinear terms on the right-hand side.

This rather simple expression of $\Delta_{2}$ (equation \eqref{eq:Delta2}) arises because, while parallel pressure balance enforces $\partial_{z}\delta p_{\|}\approx 0$ in the growing mode, there is no equivalent pressure balance condition for $\delta p_\perp$. The same result can also be obtained by projecting 
the driving of $\Delta p$ due to the MRI mode onto the eigenmodes of the double-adiabatic equations; this agrees with 1-D nonlinear simulations (not shown). Because the
spatial variation in the anisotropy is comparable to its mean (i.e., $\langle \Delta_{2}\rangle\sim \widetilde{\Delta_{2}}$) the double-adiabatic model will cause nonlinear modifications to the mode shape as it approaches the mirror limit (similar to the Braginskii model; see figure \ref{fig:Bz MRIs}{\it c} and \S\ref{app:sub: Braginskii}).

\subsection{Collisionless: Landau-fluid closure}\label{app:sub: LF}

In this section, we repeat the calculation detailed in \S\ref{app:sub: double-adiabatic} but include the heat fluxes $q_{\perp}$ \eqref{eq:NDMHD qp} and $q_{\parallel}$ \eqref{eq:NDMHD ql} with $\overline{\nu}_{c} = 0$. In a high-$\beta$ plasma with Alfv\'{e}nic fluctuations, such flows rapidly smooth pressure perturbations and, as a result, lead to a $\Delta_{2}$ that is smooth, {\it viz.},  $\widetilde{\Delta_{2}}=0$.  The ordering is the same as that used in the double-adiabatic case (\S\ref{app:sub: double-adiabatic}), but with the addition of the heat fluxes,
\begin{align}
q_{\perp} &= \epsilon^{2} \frac{1}{\sqrt{\upi}} \frac{\partial_{z}}{|k_{z}|}( \delta p_{\perp 2} -  \delta \rho_{2} ) + \epsilon^{3} \frac{1}{\sqrt{\upi}}\frac{\partial_{z}}{|k_{z}|}(  \delta p_{\perp 3} -  \delta \rho_{3} ) + \mathcal{O}(\epsilon^{4}), \\*
q_{\parallel} &= \epsilon^{2} \frac{2}{\sqrt{\upi}}\frac{\partial_{z}}{|k_{z}|}(  \delta p_{\parallel 2} -  \delta \rho_{2} ) + \epsilon^{3} \frac{2}{\sqrt{\upi}}\frac{\partial_{z}}{|k_{z}|}(  \delta p_{\parallel 3} -  \delta \rho_{3} ) + \mathcal{O}(\epsilon^{4}).
\end{align}
The $\grad\bcdot (q_{\perp,\parallel}\hat{\bm{b}})$ contributions to the pressure equations \eqref{eq:NDMHD pp}--\eqref{eq:NDMHD pl} then simplify to 
$\epsilon^{2}\partial_{z} q_{\perp,\|2} + \epsilon^{3}\partial_{z} q_{\perp,\|3} + \mathcal{O}(\epsilon^{4})$, i.e., 
heat flows along $\bm{B}_{0}=\hat{\bm{z}}$ to lowest order. 

The equations up to order $\epsilon^{1}$ are identical to those found in \S\ref{app:sub: double-adiabatic}, aside from additional contributions from the $\grad\bcdot (q_{\perp,\parallel}\hat{\bm{b}})$ terms in the pressure equations \eqref{eq:NDMHD pp}--\eqref{eq:NDMHD pl}, namely,
\begin{gather}
\upi^{-1/2} \beta_0^{1/2}|k_{z}| (\delta p_{\perp 2}-\delta \rho_{2}) =0,\label{eq:lowest order p2 condition 1} \\*
2\upi^{-1/2} \beta_0^{1/2} |k_{z}| (\delta p_{\parallel 2}-\delta \rho_{2}) =0\label{eq:lowest order p2 condition 2},\end{gather}
where we have used $\partial_{z}^{2}/|k_{z}| = -|k_{z}|$ to simplify the nonlocal diffusion operators. Combining \eqref{eq:lowest order p2 condition 1}--\eqref{eq:lowest order p2 condition 2} with the continuity equation \eqref{eq:NDMHD rho} and parallel pressure balance 
$\delta \widetilde{p_{\|2}} = 0$, we obtain 
\begin{equation}
\widetilde{\delta p_{\perp 2}}  =  \widetilde{\delta \rho_{2}} = \widetilde{\delta u_{z2}} =\widetilde{\Delta_{2}} =0.\label{eq: pressure 2 is smooth}
\end{equation}
This formally justifies the statements in \S\S\ref{sub: general comparison},\ref{sec: 1D vertical} that $\Delta$ is spatially constant to lowest order. 

At order $\epsilon^{2}$, the pressure equations \eqref{eq:NDMHD pp}--\eqref{eq:NDMHD pl} are 
\begin{gather}
\partial_{t}\delta p_{\perp 2} + \partial_{z} \delta u_{z2}  + \upi^{-1/2} \beta_0^{1/2}|k_{z}| (\delta p_{\perp 3}-\delta \rho_{3}) = \frac{1}{2}\partial_{t}\delta B_{\perp1}^{2},\label{eq:pp Landau expansion}\\*
\partial_{t}\delta p_{\parallel 2} + 3\partial_{z} \delta u_{z2} + 2\upi^{-1/2}\beta_0^{1/2}|k_{z}| (\delta p_{\parallel 3}-\delta \rho_{3})= -\partial_{t}\delta B_{\perp1}^{2}.\label{eq:pl Landau expansion}
\end{gather}
Spatially averaging these equations, using $\widetilde{\delta p_{\perp 2}}=\widetilde{\delta p_{\| 2}}=0$,
and again assuming that the mode growth starts from negligibly small amplitudes (i.e., $\delta B_{\perp1}^{2}(t=0)=0$), we find
\begin{equation}
\Delta_{2} = \frac{3}{2}\langle \delta B_{\perp1}^{2}\rangle.\label{app:eq: LF delta MRI}
\end{equation}
This can be inserted into \eqref{app:eq: u eqn CGL}--\eqref{app:eq: B eqn CGL}
to obtain the following evolution equation for $\delta \bm{B}_{\perp1}$:
\begin{equation}\label{eq:Bprp1_LF}
\partial^2_{t} \delta \bm{B}_{\perp 1} + 2\overline{\Omega} \ez\btimes \partial_{t} \delta \bm{B}_{\perp 1} - 2\overline{S} \overline{\Omega}\, \delta B_{x1} \ex
= \partial^2_z \left[ \delta \bm{B}_{\perp 1} \left( 1 + \frac{3\beta_0}{4} \langle \delta B^2_{\perp 1} \rangle \right) \right] ,
\end{equation}
which remains valid until $\Delta_{2}$ exceeds the mirror threshold (at which point its growth should be limited by the unresolved mirror instability, as discussed in \S\ref{sec: 1D vertical}).

As expected, the presence of such strong heat fluxes has  rendered the KMRI equations  \eqref{app:eq: u eqn CGL}--\eqref{app:eq: B eqn CGL} much simpler 
than in the double-adiabatic case.\footnote{Interestingly,  a similar saturation mechanism also arises for the standard (MHD) MRI in a non-periodic system when it is near the marginal-stability condition \citep{2015RSPSA.47140699V}.} 
Physically, the spatial average inside the nonlinear term in \eqref{eq:Bprp1_LF} implies that, if a KMRI mode is initially sinusoidal, it will remain so even as it becomes nonlinear (see figure \ref{fig:Bz MRIs}{\it b} for a demonstration of this property). We can then use \eqref{eq:Bprp1_LF} to write down an ordinary differential equation for the amplitude evolution of a single MRI mode $\delta \bm{B}_{\perp} = \delta \bm{B}_\perp(t)\erm^{\imag k_{\|}z}$, $\delta\bm{u}_\perp = \delta\bm{u}_\perp(t)\erm^{\imag k_{\|} z}$:
\begin{gather}
\DD{t}{\delta\bm{B}_{\perp 1}} + 2\overline{\Omega}\ez\btimes \D{t}{\delta\bm{B}_{\perp 1}} - 2 \overline{S}\overline{\Omega} \, \delta B_{x1} \ex = -k^2_\parallel \delta\bm{B}_{\perp 1} \left( 1 + \frac{3\beta_0}{8} \delta B^2_{\perp 1} \right) ,\label{app:eq: MRI single 1} \\*
\delta\bm{u}_{\perp 1} = - \frac{\imag}{k_\parallel} \left( \D{t}{\delta\bm{B}_{\perp 1}} + \overline{S} \delta B_{x1} \ey \right) ,\label{app:eq: MRI single 2}
\end{gather}
%
where we have used $\langle \sin^{2}(k_{\|}z)\rangle=1/2$.
Solutions to \eqref{app:eq: MRI single 1}--\eqref{app:eq: MRI single 2} correctly reproduce the  change in relative amplitudes of $\delta \bm{B}_\perp$ and $\delta \bm{u}_\perp$ as seen in figure \ref{fig:Bz MRIs}{\it b} (e.g., the relative increase in $\delta u_{y}$ and relative decrease of $\delta u_{x}$). Of course, if there is more than one growing mode, the pressure-anisotropy nonlinearity 
\eqref{app:eq: LF delta MRI} does couple the modes, which could allow, for example, a larger wavelength mode to ``take over'' due to the positive pressure anisotropy (see \S\ref{sec: 1D vertical}).

If one were so inclined, a continuation of the expansion to $\mathcal{O}(\epsilon^{3})$ would  yield equations for the spatial variation in $\Delta$. However, because the expected effect on the mode is very small for $\beta_{0}\gg 1$, we do not do this here (see, e.g., App.~A.3.1 of \citealt{Squire:2016ev2}).

\subsection{Weakly collisional: Braginskii closure}\label{app:sub: Braginskii}

In this section, we treat the weakly collisional, Braginskii MHD limit. As discussed in \S\S\ref{sec:equations etc.},\ref{sec: 1D brag},\ref{subsub: Braginskii with By}, the Braginskii 
regime is relevant when $\overline{\nu}_{c}\equiv\nu_{c}/\omega_{A}\sim \nu_{c}/\Omega \gg 1$, with the corrections to the MRI 
due to pressure anisotropy becoming unimportant when $\overline{\nu}_{c}\gtrsim \beta$ (see \citealt{Sharma:2003hf}). Within the relevant range $1\ll \overline{\nu}_{c}\ll \beta$, one may obtain a variety of behaviours depending upon whether or not the heat fluxes play a 
significant role in the evolution of $\Delta p$. If $\overline{\nu}_{c}\gg \beta^{1/2}$, the heat fluxes are suppressed
by the collisionality and do not strongly influence $\Delta p$;  if instead $\overline{\nu}_{c}\lesssim \beta^{1/2}$, the heat fluxes  smooth $\Delta p$ in space\footnote{Specifically, the relative magnitude of the rate of heat-flux smoothing of $\Delta p$ compared to its rate of creation (via ${\rm d}\ln B/{\rm d}t$)
 varies between $\beta^{1/2}$ (as in the collisionless case),  when $\overline{\nu}_{c}\ll \beta^{1/2}$, and $1$, when $\overline{\nu}_{c}\sim \beta^{1/2}$.} on a timescale shorter than that over which $\Delta p$ is produced by the 
changing $B$ (see  \citealt{Squire:2016ev2} for further discussion). We present here only the former limit ($\overline{\nu}_{c}\gg \beta^{1/2}$), which leads to 
the closure used in the main text, equation \eqref{eq:Brag closure}; in the $\overline{\nu}_{c}\ll \beta^{1/2}$ limit, the heat fluxes smooth out $\Delta p$ near the nodes of 
$\delta B$ and so $\Delta p$ is almost spatially constant as the mirror limit is approach (see the discussion in  \S\ref{sec: 1D brag}). In the intermediate case $\overline{\nu}_{c} \sim \beta^{1/2}$, a valid closure for $\Delta p$ has been obtained by \citet{Squire:2016ev2} -- see their equations (B12)--(B15).

The ordering introduced above, $\delta \bm{B}_{\perp}\sim \delta\bm{u}_{\perp} \sim \epsilon$ with $\beta_{0 }\Delta \sim 1$, coupled with the Braginskii pressure anisotropy $\Delta \sim \overline{\nu}_{c}^{-1} {\rm d}\ln B/{\rm d}t$, suggests
that we order $\overline{\nu}_{c}\sim \epsilon^{2}\beta_{0}$. The simultaneous 
requirement that $\overline{\nu}_{c}\gg \beta_{0}^{1/2}$ for the ``high-collisionality'' regime where heat-fluxes
are collisionally suppressed then implies the ordering $\overline{\nu}_{c}\sim \mathcal{O}(\epsilon^{-4})$, $\beta_{0}\sim \mathcal{O}(\epsilon^{-6})$. For the other variables, we adopt the orderings $p_{\perp}\sim p_{\|}\sim \rho \sim 1+\mathcal{O}(\epsilon^{6})$ and $\delta u_{z}\sim\mathcal{O}(\epsilon^{6})$.
The $\mathcal{O}(1)$ and $\mathcal{O}(\epsilon)$ equations are then almost the same 
as in \S\ref{app:sub: double-adiabatic}: the parallel momentum equation gives $\partial_{z}\delta p_{\|6}=0$, and the perpendicular momentum and induction equations are identical to \eqref{app:eq: u eqn CGL}--\eqref{app:eq: B eqn CGL} but with $\Delta_{2}$ replaced by $\Delta_{6}$.
At $\mathcal{O}(\epsilon^{2})$, the pressure equations \eqref{eq:NDMHD pp} and \eqref{eq:NDMHD pl} may be combined to give
\begin{equation}
\overline{\nu}_{c} \Delta_{6} =  \delta \bm{B}_{\perp1}\bcdot \partial_{z}\delta \bm{u}_{\perp1} - \overline{S} \delta B_{x1} \delta B_{y1} = \frac{1}{2}\partial_{t} \delta B^2_{\perp 1} ;
\end{equation}
the heat-flux terms appear at $\mathcal{O}(\epsilon^{4})$. This is exactly as was anticipated (cf.~\eqref{eq:Brag closure}).
The nonlinear equation for the growing mode is then simply
\begin{equation}\label{app:eq: B eqn Brag}
\partial^2_{t} \delta \bm{B}_{\perp 1} + 2\overline{\Omega} \ez\btimes \partial_{t} \delta \bm{B}_{\perp 1} - 2\overline{S} \overline{\Omega}\, \delta B_{x1} \ex
= \partial^2_z \left[ \delta \bm{B}_{\perp 1} \left( 1 + \frac{\beta_0}{4\overline{\nu}_c} \partial_t \delta B^2_{\perp 1} \right) \right] ,
\end{equation}
%
%
Because the nonlinearity in \eqref{app:eq: B eqn Brag} depends on $\delta B^2_{\perp 1}(z)$ (rather than $\langle \delta B^2_{\perp 1}\rangle$), it will 
 distort the shape of an initially sinusoidal mode, as discussed in \S\ref{sec: 1D brag} and exhibited in figure \ref{fig:Bz MRIs}{\it c}. Thus, we cannot reduce \eqref{app:eq: B eqn Brag} to an ordinary differential equation for a single mode, as in the collisionless (LF) derivation (see \eqref{app:eq: MRI single 1}).

\bibliographystyle{jpp}
\bibliography{fullbib,bib_extrapapers}

\end{document}